\documentclass[aps,twocolumn,nofootinbib,groupedaddress,superscriptaddress,longbibliography]{revtex4-1}

\usepackage{amsmath,amssymb}
\usepackage[normalem]{ulem}
\usepackage{graphicx}
\usepackage{multirow,makecell}
\usepackage{bm}
\usepackage{mathtools}
\usepackage{dsfont}
\usepackage{amsfonts}
\usepackage{xcolor}
\usepackage{url}
\usepackage{rotating}
\usepackage{wasysym}
\usepackage{diagbox}
\usepackage[colorlinks=true,linkcolor=blue,citecolor=blue,urlcolor=blue,hypertexnames=false]{hyperref}
\newcommand{\nocontentsline}[3]{}
\newcommand{\tocless}[2]{\bgroup\let\addcontentsline=\nocontentsline#1{#2}\egroup}
\def\ba#1\ea{\begin{align}#1\end{align}}
\def\bg#1\eg{\begin{gather}#1\end{gather}}
\def\bpm{\begin{pmatrix}}
\def\epm{\end{pmatrix}}

\newcommand{\nn}{\nonumber \\ }
\newcommand{\td}[1]{\widetilde{#1}}
\newcommand{\mrm}[1]{{\mathrm #1}}
\newcommand{\mns}{\mrm{-}}

\newcommand{\hf}{\frac{1}{2}}
\newcommand{\mhf}{\mns \frac{1}{2}}
\newcommand{\mch}[2]{\mc{C}_{#1}^{#2}}
\newcommand{\mchd}{\mc{C}_m^{x\cm{y}}}
\newcommand{\mchdp}{\mc{C}_m^{xy}}

\newcommand{\bb}[1]{{\mathbf #1}}
\newcommand{\mbb}[1]{{\mathbf #1}}
\newcommand{\bx}{\bb x}
\newcommand{\bk}{\bb k}
\newcommand{\bR}{\bb R}

\newcommand{\cm}{\overline}
\newcommand{\mc}[1]{\mathcal{#1}}

\newcommand{\der}{\partial}
\newcommand{\dg}{\dagger}
\newcommand{\om}{\omega}
\newcommand{\sg}{\sigma}
\newcommand{\vph}{\varphi}

\newcommand{\ket}[1]{|#1\rangle}
\newcommand{\bra}[1]{\langle#1|}
\newcommand{\brk}[2]{\langle#1|#2\rangle}
\newcommand{\kbr}[2]{|#1\rangle\langle#2|}

\newcommand{\ketr}[1]{|#1\rangle}
\newcommand{\brar}[1]{\langle#1|}
\newcommand{\brkr}[2]{\langle#1|#2\rangle}

\newcommand{\Rf}[1]{Ref.~\onlinecite{#1}}

\newcommand{\eq}[1]{Eq.~\eqref{#1}}
\newcommand{\eqs}[1]{Eqs.~\eqref{#1}}
\newcommand{\fig}[1]{Fig.~\ref{#1}}
\newcommand{\figs}[1]{Figs.~\ref{#1}}

\newcommand{\kSubgroupsmag}{{\ttfamily k-Subgroupsmag}}
\newcommand{\MWYCKPOS}{{\ttfamily MWYCKPOS}}

\newcommand{\SUPERGROUPS}{{\ttfamily SUPERGROUPS}}
\newcommand{\mxybo}{\td{M}_{x\cm{y}}}
\newcommand{\mxyb}{M_{x\cm{y}}}
\newcommand{\mxyo}{\td{M}_{xy}}
\newcommand{\mxy}{M_{xy}}
\newcommand{\tmx}{TG_x}
\newcommand{\tmy}{TG_y}
\newcommand{\Tg}{T_G}
\newcommand{\bG}{\cm{\Gamma}}
\newcommand{\bGp}{\cm{\Gamma'}}
\newcommand{\bM}{\cm{M}}
\newcommand{\bX}{\cm{X}}
\newcommand{\bY}{\cm{Y}}

\newcommand{\psm}{\psi_\pm(M)}
\newcommand{\bkp}{\bk_\perp}
\newcommand{\mbk}{\mathbf k}
\newcommand{\mbkp}{\mbk_\perp}
\newcommand{\cyo}{\td{C}_{2_1 y}}
\newcommand{\sfig}{SFig.}
\newcommand{\sfigs}{SFigs.}
\newcommand{\stable}{Supplementary Table}
\newcommand{\sn}{Supplementary Note}
\newcommand{\sns}{Supplementary Notes}
\newcommand{\ourtitle}{
Magnetic wallpaper Dirac fermions and topological magnetic Dirac insulators
}
\allowdisplaybreaks

\begin{document}
\title{\textbf{\ourtitle}}

\author{Yoonseok \surname{Hwang}}
\thanks {These authors contributed equally to this work.}
\affiliation{Center for Correlated Electron Systems, Institute for Basic Science (IBS), Seoul 08826, Korea}
\affiliation{Department of Physics and Astronomy, Seoul National University, Seoul 08826, Korea}
\affiliation{Center for Theoretical Physics (CTP), Seoul National University, Seoul 08826, Korea}

\author{Yuting \surname{Qian}}
\thanks {These authors contributed equally to this work.}
\affiliation{Center for Correlated Electron Systems, Institute for Basic Science (IBS), Seoul 08826, Korea}
\affiliation{Department of Physics and Astronomy, Seoul National University, Seoul 08826, Korea}

\author{Junha \surname{Kang}}
\affiliation{Center for Correlated Electron Systems, Institute for Basic Science (IBS), Seoul 08826, Korea}
\affiliation{Department of Physics and Astronomy, Seoul National University, Seoul 08826, Korea}
\affiliation{Center for Theoretical Physics (CTP), Seoul National University, Seoul 08826, Korea}

\author{Jehyun \surname{Lee}}
\affiliation{Center for Correlated Electron Systems, Institute for Basic Science (IBS), Seoul 08826, Korea}
\affiliation{Department of Physics and Astronomy, Seoul National University, Seoul 08826, Korea}
\affiliation{Center for Theoretical Physics (CTP), Seoul National University, Seoul 08826, Korea}

\author{Dongchoon \surname{Ryu}}
\affiliation{Center for Correlated Electron Systems, Institute for Basic Science (IBS), Seoul 08826, Korea}
\affiliation{Department of Physics and Astronomy, Seoul National University, Seoul 08826, Korea}
\affiliation{Center for Theoretical Physics (CTP), Seoul National University, Seoul 08826, Korea}

\author{Hong Chul \surname{Choi}}
\email{chhchl@snu.ac.kr}
\affiliation{Center for Correlated Electron Systems, Institute for Basic Science (IBS), Seoul 08826, Korea}
\affiliation{Department of Physics and Astronomy, Seoul National University, Seoul 08826, Korea}

\author{Bohm-Jung \surname{Yang}}
\email{bjyang@snu.ac.kr}
\affiliation{Center for Correlated Electron Systems, Institute for Basic Science (IBS), Seoul 08826, Korea}
\affiliation{Department of Physics and Astronomy, Seoul National University, Seoul 08826, Korea}
\affiliation{Center for Theoretical Physics (CTP), Seoul National University, Seoul 08826, Korea}
%%%%%%%%%%%%%%%%%%%%%%

%%%%%%%%%%%%%%%%%%%%%%
\begin{abstract}
Topological crystalline insulators (TCIs) can host anomalous surface states which inherits the characteristics of crystalline symmetry that protects the bulk topology.
Especially, the diversity of magnetic crystalline symmetries indicates the potential for novel magnetic TCIs with distinct surface characteristics.
Here, we propose a topological magnetic Dirac insulator (TMDI), whose two-dimensional surface hosts fourfold-degenerate Dirac fermions protected by either the $p'_c4mm$ or $p4'g'm$ magnetic wallpaper group.
The bulk topology of TMDIs is protected by diagonal mirror symmetries, which give chiral dispersion of surface Dirac fermions and mirror-protected hinge modes. We propose candidate materials for TMDIs including Nd$_4$Te$_8$Cl$_4$O$_{20}$ and DyB$_4$ based on first-principles calculations, and construct a general scheme for searching TMDIs using the space group of paramagnetic parent states.
Our theoretical discovery of TMDIs will facilitate future research on magnetic TCIs and illustrate a distinct way to achieve anomalous surface states in magnetic crystals.
\end{abstract}
%%%%%%%%%%%%%%%%%%%%%%

\maketitle

\let\oldaddcontentsline\addcontentsline
\renewcommand{\addcontentsline}[3]{}
%%%%%%%%%%%%%%%%%%%%%%

%%%%%%%%%%%%%%%%%%%%%%
\section{Introduction}
\label{sec:intro}
%%%%%%%%%%%%%%%%%%%%%%
The surface states of topological insulators (TIs) have anomalous characteristics that are unachievable in ordinary periodic systems~\cite{hasan2010rmp}.
A representative example is the twofold-degenerate gapless fermion on the surface of three-dimensional (3D) TIs protected by time-reversal symmetry (TRS)~\cite{bernevig2006quantum,fu2007topological,fu2007inversion,hsieh2008topological}.
Contrary to the case of ordinary two-dimensional (2D) crystals with TRS in which
gapless fermions appear in pairs, a single gapless fermion can exist on the surface of TIs through its coupling to the bulk bands. 
Such a violation of fermion number doubling~\cite{nielsen1981no,haldane1988model,ahn2019failure} is a representative way in which the anomalous characteristics of surface states are manifested at the boundary of TIs.

In topological crystalline insulators (TCIs)~\cite{fu2011topological,chiu2016rmp}, crystalline symmetries enrich the ways in which anomalous surface states are realized. 
For example, in systems with rotation symmetry and TRS, variants of the fermion doubling theorem enabled by symmetries can be anomalously violated on the surface of TCIs~\cite{fang2019new}.
Additionally, in the case of mirror-protected TCIs~\cite{hsieh2012topological}, although the number of surface gapless fermions can be even, the surface band structure exhibits a chiral dispersion along mirror-invariant lines such that anomalous chiral fermions appear in the one-dimensional (1D) mirror-resolved subspace of the 2D surface Brillouin zone (BZ).
More recently, studies showed that in crystals with glide mirrors, the anomalous surface states can have an hourglass-type band connection~\cite{wang2016hourglass}.
Moreover, when the surface preserves two orthogonal glide mirrors, a single fourfold-degenerate Dirac fermion~\cite{young2015dirac} was shown to be achievable as an anomalous surface state~\cite{wieder2018wallpaper}.

In magnetic crystals, there is great potential to achieve a new type of magnetic TCI with distinct anomalous surface states~\cite{watanabe2018structure,elcoro2021magnetic,peng2022topological,okuma2019topological,xu2020high,bouhon2021topological,shiozaki2022classification} because there are abundant magnetic crystalline symmetries described by 63 magnetic wallpaper groups (MWGs) and 1421 magnetic space groups (MSGs)~\cite{belov19564color,litvin2013magnetic}, which are overwhelmingly larger than the 17 wallpaper groups and 230 space groups of nonmagnetic crystals~\cite{bradlyn2017topological, po2017symmetry,kruthoff2017topological,song2018quantitative,khalaf2018symmetry}.
Very recently, exhaustive studies of magnetic topological phases and their classification have been performed~\cite{watanabe2018structure,elcoro2021magnetic,peng2022topological}, and various novel magnetic topological phases have been systematically categorized.
However, as far as we can tell, all the surface states of magnetic TCIs reported up to now appear in the form of twofold-degenerate gapless fermions, whose detailed band connection depends on the surface symmetry.
%

%%%%%%%%%%%%%%%%%%%%%%
\begin{figure*}[t!]
\centering
\includegraphics[width=0.98\textwidth]{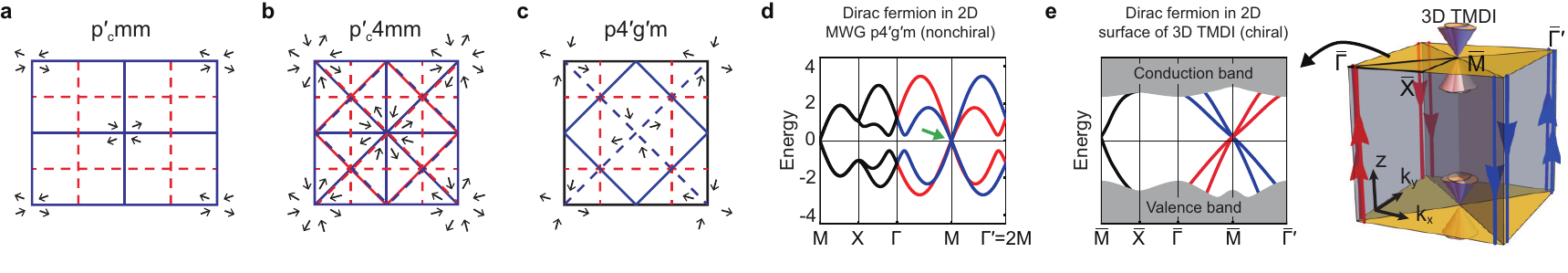}
\caption{
{\bf MWGs and fourfold-degenerate Dirac fermions.}
(a)-(c) MWGs that protect fourfold degeneracy.
Type-IV MWGs (a) $p'_cmm$ and (b) $p'_c4mm$, and (c) Type-III MWG $p4'g'm$.
Black arrows represent spin configurations located at generic positions.
The styles of lines indicate the types of symmetry elements: glides (blue dashed lines), antiunitary glides (red dashed lines), mirrors (blue solid lines), and antiunitary mirrors (red solid lines).
(d) Typical band structure of 2D crystals belonging to MWG $p4'g'm$.
$\Gamma=(0,0)$, $X=(\pi,0)$, $M=(\pi,\pi)$, and $\Gamma'=(2\pi,2\pi)$ denote high-symmetry points.
A fourfold-degenerate Dirac fermion appears at $M$ near $E=0$ (green arrow).
Red (blue) lines denote states with eigenvalue $+i$ ($-i$) of the diagonal mirror.
Along $\Gamma$-$M$-$\Gamma'$, each mirror eigenvalue sector of the Dirac fermion has a nonchiral dispersion.
(e) Left: a Dirac fermion appearing as an anomalous surface state of a topological magnetic Dirac insulator (TMDI) with a chiral dispersion in each mirror sector.
Right: schematic depiction of a TMDI, which hosts a fourfold-degenerate Dirac fermion on the (001) surface, and mirror-protected hinge modes on the sides invariant under diagonal mirrors.
}
\label{fig1}
\end{figure*}
%%%%%%%%%%%%%%%%%%%%%%

Here, we propose a magnetic TCI with fourfold-degenerate gapless fermions on the surface, coined the topological magnetic Dirac insulator (TMDI).
A fourfold-degenerate gapless fermion, a Dirac fermion for short hereafter, can appear on the surface of a magnetic insulator when the MWG of the surface is one of the three MWGs $p4'g'm$, $p'_cmm$, and $p'_c4mm$, among 63 possible MWGs.
Contrary to the surface Dirac fermion in nonmagnetic crystals protected by two orthogonal glides, our surface Dirac fermion is protected by symmorphic symmetries combined with either an antiunitary translation symmetry or an antiunitary glide mirror.

In particular, in magnetic crystals whose (001)-surface MWG is either $p4'g'm$ or $p'_c4mm$, the bulk topology is characterized by the mirror Chern number (MCN) $\mchd$ about the diagonal mirror planes normal to either the [110] or $[1\bar{1}0]$ direction.
Because of this, in TMDIs, the way in which the surface anomaly is realized is different from the case of the nonmagnetic Dirac insulator~\cite{wieder2018wallpaper} and more similar to the case of mirror-protected nonmagnetic TCIs~\cite{hsieh2012topological}.
Namely, along the mirror-invariant line on the surface BZ, the Dirac fermion develops a chiral dispersion relevant to the MCN.
Moreover, the MCN of TMDIs also induces hinge modes at open boundaries along the $x$ and $y$ directions, which respect diagonal mirrors.

Using first-principles calculations, we propose candidate materials for TMDIs, including Nd$_4$Te$_8$Cl$_4$O$_{20}$ and DyB$_4$.
Since the database for magnetic materials only has a limited number of materials, we construct a systematic way to find candidate magnetic materials for TMDIs using the space group of paramagnetic parent compounds.
%%%%%%%%%%%%%%%%%%%%%%

%%%%%%%%%%%%%%%%%%%%%%
\section{Dirac fermions and magnetic wallpaper groups}
\label{sec:MWG}
%%%%%%%%%%%%%%%%%%%%%%
In 2D magnetic crystals, Dirac fermions with fourfold-degeneracy can be symmetry-protected at the BZ corner, $ M=( \pi,\pi)$,
by \textit{three MWGs}, i.e., Type-III MWG $p4'g'm$ and Type-IV MWGs $p'_cmm$ and $p'_c4mm$ [see \figs{fig1}(a)-(c)].
Here, we use the notation of Belov and Tarkhova (BT)~\cite{belov19564color} for denoting MWGs and the notation of Belov, Neronova, and Smirnova (BNS)~\cite{belov1957Shubnikov} for denoting MSGs.
Note that Type-III MWGs have antiunitary spatial symmetries combining TRS $T$ with spatial symmetries, while Type-IV MWGs have antiunitary translation symmetries combining $T$ and fractional lattice translations.
All three MWGs have mirror-invariant lines, whose normal directions are $\hat{x}$, $\hat{y}$, or $\hat{x} \pm \hat{y}$.

In 2D systems belonging to the Type-III MWG $p4'g'm$ described in \fig{fig1}(c), a Dirac fermion is protected at $M$ by twofold rotation about the $z$-axis $C_{2z}$, antiunitary glide mirror $\tmy \equiv T \{m_y| \hf,\hf \}$, and off-centered diagonal mirror $\mxybo = \{ m_{x\cm{y}} | \hf, \mhf \}$.
Here, the notation $\{g|\mathbf{t}\}$ denotes the point group symmetry $g$ followed by a partial lattice translation $\mathbf{t}$.
$m_{x \cm{y},y}$ are mirror symmetries that act on real-space coordinates as $m_{x \cm{y}}: (x,y,z) \to (y,x,z)$ and $m_y: (x,y,z) \to (x,-y,z)$. %
[See the conventions in Supplementary Note (SN) 1.]
As detailed in SN 3, the fourfold degeneracy is formed by four states $\psi_\pm$, $\tmy\psi_\pm$, $\mxybo \psi_\pm$, and $\tmy \mxybo \psi_\pm$, where $\psi_\pm$ is an energy eigenstate with $C_{2z}$ eigenvalue $\pm i$.

In contrast, 2D systems belonging to the Type-IV MWGs $p'_cmm$ and $p'_c4mm$, described in \figs{fig1}(a) and (b), respectively, have common symmetry elements, i.e., antiunitary translation $\Tg = \{T |\hf,\hf \}$ and two mirrors $M_x=\{m_x|\bb{0}\}$ and $M_y=\{m_y|\bb{0}\}$, where $m_x: (x,y,z) \to (-x,y,z)$.
At $M$, these symmetry elements anticommute with each other, and $\Tg^2=-1$.
These relations protect the fourfold degeneracy formed by $\psi_\pm$, $\Tg \psi_\pm$, $M_y \psi_\pm$, and $\Tg M_y \psi_\pm$, where $\psi_\pm$ has $M_x$ eigenvalue $\pm i$ (see SN3).
Note that the same symmetry representation was also studied in Ref.~\cite{young2017filling}.

A typical band structure supported by MWG $p4'g'm$ is shown in \fig{fig1}(d).
Since MWG $p4'g'm$ has diagonal mirror $\mxybo$, the energy bands can be divided into two different mirror eigensectors along the $\Gamma M \Gamma'$ direction.
Here, $\Gamma=(0,0)$, and $\Gamma'=(2\pi,2\pi)$.
Focusing on the band structures in each mirror sector, we find that the numbers of upward (chiral) and downward (antichiral) bands crossing the Fermi level [$E=0$ in \fig{fig1}(d)] at $M$ are the same.
Otherwise, the mirror-resolved band structure cannot be periodic along $\Gamma M \Gamma'$.
Hence, a Dirac fermion in 2D crystals belonging to MWG $p4'g'm$ is nonchiral in each mirror sector.
Similar phenomena also occur in MWGs $p'_cmm$ and $p'_c4mm$.
In general, the mirror-resolved dispersion of Dirac fermion in 2D crystals protected by MWGs is nonchiral.
Although the local dispersion near the Dirac point may exhibit either chiral or nonchiral dispersion, the full band dispersion along the mirror invariant line is always nonchiral in 2D crystals.
%%%%%%%%%%%%%%%%%%%%%%

%%%%%%%%%%%%%%%%%%%%%%
\begin{figure*}[t!]
\centering\includegraphics[width=0.95\textwidth]{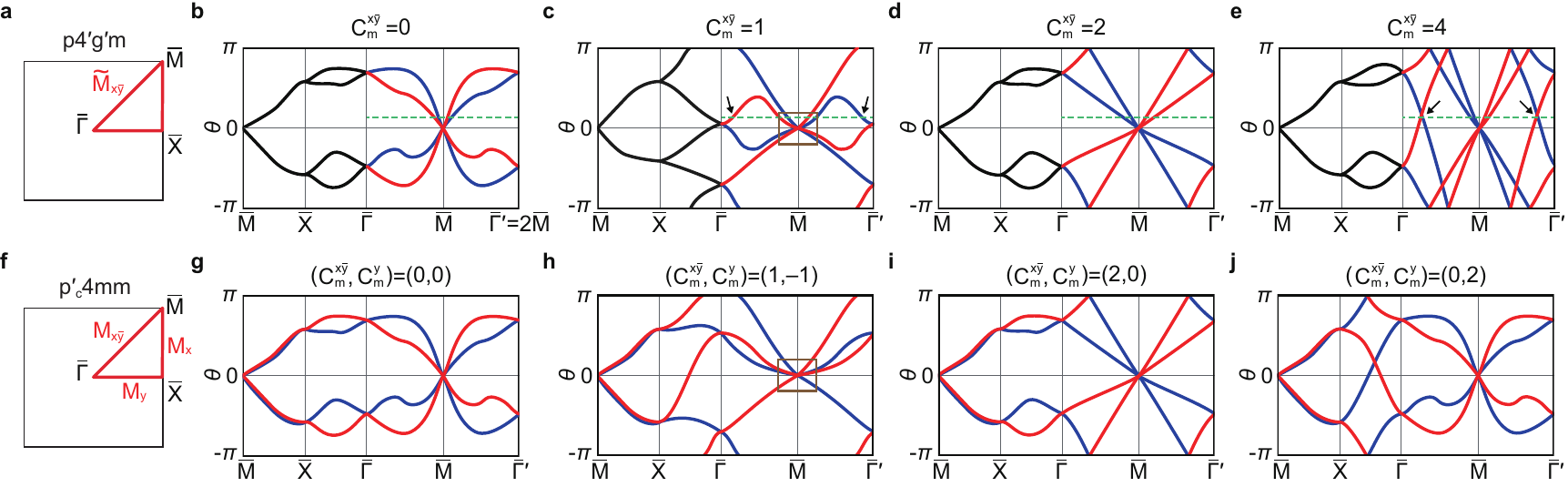}
\caption{
{\bf Wilson loop spectra of 3D magnetic TCIs with (001)-surface MWGs $p4'g'm$ and $p'_c4mm$.}
Classification of Wilson loop spectra, whose connectivity is equivalent to the (001)-surface band structure, based on the MCNs $\mchd$ and $\mc{C}_m^y$.
(a)-(e) Type-III TMDIs with MWG $p4'g'm$ on the (001) surface.
(a) (001)-surface BZ. $\mxybo$ is the diagonal mirror used to define $\mchd$.
(b)-(e) Wilson loop spectra corresponding to (b) $\mchd=0$, (c) $\mchd=1$, (d) $\mchd=2$, and (e) $\mchd=4$.
For convenience, the position of the Dirac fermion at $\bM$ is adjusted to be located at $\theta(\bk)=0$.
The red (blue) lines correspond to Wilson bands with mirror eigenvalue $+i$ ($-i$).
The green dashed lines are the reference lines used to count the MCN $\mchd$.
In (c)-(e), in which $\mchd \neq0$, the dispersion of the Dirac fermion can be chiral in each mirror sector along $\bG$-$\bM$-$\bGp$.
Note that $\bGp=(2\pi,2\pi)$.
In (b), where $\mchd=0$, the Dirac fermion is nonchiral.
In (d), where $\mchd=2$, the reference line is crossed by two chiral (upward) modes with mirror eigenvalue $+i$.
In (c), where $\mchd=1$, the dispersion is chiral along the entire $\bG$-$\bM$-$\bGp$ line but locally looks nonchiral near the Dirac point.
The Wilson loop structure in (c) can be deformed into that in (h) by pushing the twofold crossing at $\bG$ upward, which gives a locally chiral dispersion at $\bM$.
When $|\mchd|>2$, a Dirac fermion must appear with other gapless surface states along the $\bG$-$\bM$-$\bGp$ line [black arrows in (e)].
(f)-(j) Type-IV TMDIs with MWG $p'_c4mm$ on the (001) surface.
(f) (001)-surface BZ.
(g)-(j) Wilson loop spectra corresponding to (g) $(\mchd,\mc{C}_m^y)=(0,0)$, (h) $(\mchd,\mc{C}_m^y)=(1,-1)$, (i) $(\mchd,\mc{C}_m^y)=(2,0)$, and (j) $(\mchd,\mc{C}_m^y)=(0,2)$.
Note that $\mchd=\mc{C}_m^y$ (mod 2) holds for insulators.
}
\label{fig2}
\end{figure*}
%%%%%%%%%%%%%%%%%%%%%%

%%%%%%%%%%%%%%%%%%%%%%
\section{Chiral surface Dirac fermions}
\label{sec:bulk}
%%%%%%%%%%%%%%%%%%%%%%
A 2D Dirac fermion, which is nonchiral in 2D systems,
can be chiral on the surface of 3D magnetic TCIs, as illustrated in \fig{fig1}(e).
Here, we systematically search for 3D magnetic insulators that can host a Dirac fermion on the (001) surface. 
As a 2D fourfold-degenerate Dirac fermion can be protected by one of the three MWGs $p4'g'm$, $p'_cmm$, and $p'_c4mm$,
we focus on the MSGs whose (001) surface has one of these three MWGs. 
By studying the MSG symbols and the detailed surface symmetries, we find that there are at least 31 MSGs that can be generated from such MWGs and additional generators compatible with the MWGs.
(See \stable~1.)

All 31 MSGs have mirror planes whose normal vectors are orthogonal to the (001) direction.
Thus, the corresponding MCNs can give chiral dispersions along the mirror-invariant lines on the (001) surface.
First, MWG $p4'g'm$ has off-centered diagonal mirrors~\cite{yang2017topological} $\mxybo=\{ m_{x\cm{y}} | \hf, \mhf \}$ and $\mxyo=\{ m_{xy} | \hf, \hf \}$ [see \fig{fig1}(c)].
For the 11 MSGs relevant to MWG $p4'g'm$, we define four MCNs $\mch{\pm}{k_x=k_y}$ and $\mch{\pm}{k_x=-k_y}$, which are defined in the $k_x=k_y$ and $k_x=-k_y$ planes, respectively.
Here, the $\pm$ sign denotes the mirror eigenvalues of occupied bands.
All the MCNs are equivalent up to sign because of the symmetry relations among $\mxybo$, $\mxyo$, and $TC_{4z}$.
Hence, the bulk topology can be classified by $\mchd \equiv \mch{+}{k_x=k_y}=-\mch{-}{k_x=k_y}$.
Similarly, we can define MCNs for the 5 MSGs related to MWG $p'_c4mm$, which have four mirrors, $M_x$, $M_y$, $\mxyb$, and $\mxy$.
Among them, only two MCNs, $\mchd$ and $\mc{C}_m^y \equiv \mch{+}{k_y=0}=- \mch{-}{k_y=0}$, are independent, and serve as bulk topological invariants.
Finally, for the 15 MSGs related to MWG $p'_cmm$, the relevant MCNs are $\mc{C}_m^x\equiv\mch{+}{k_x=0}=-\mch{-}{k_x=0}$ and $\mc{C}_m^y\equiv\mch{+}{k_y=0}=-\mch{-}{k_y=0}$.
For more detailed discussions on the MCNs, see SN4.
%%%%%%%%%%%%%%%%%%%%%%

%%%%%%%%%%%%%%%%%%%%%%
%\subsection{Wilson loop and surface band structure}
%\label{subsec:WL}
%%%%%%%%%%%%%%%%%%%%%%
Now, we classify the Wilson loop spectra~\cite{yu2011equivalent,alexandradinata2014wilson,alexandradinata2016topological,bouhon2019wilson,bradlyn2019disconnected} according to the MCNs.
We consider the $k_z$-directed Wilson loop $\mc{W}_z(\mbkp)$,
\ba
\mc{W}_z(\mbkp)_{nm}
&= \bra{u_n(\mbkp, \pi)} \prod^{\pi \leftarrow -\pi}_{k_z} P_{\rm occ}(\mbkp, k_z) \ket{u_m(\mbkp, -\pi)},
\label{eq:WLdef}
\ea
where $P_{\rm occ}(\mbk) \equiv \sum_{n=1}^{n_{\rm occ}} \ket{u_n(\mbk)} \bra{u_n(\mbk)}$ is a projection operator for occupied bands $\ket{u_n(\mbk)}$ and $\mbkp=(k_x,k_y)$.
Since the Wilson loop is unitary, its eigenvalue can be collectively denoted as $\{e^{i \theta(\mbkp)}\} =\{e^{i \theta_a(\mbkp)}| \theta_a(\mbkp) \in (-\pi,\pi], a=1,\dots,n_{\rm occ}\}$.
Then, $\{\theta(\mbkp)\}$ defines the Wilson loop spectrum, or equivalently, the Wilson bands.
Wilson loop spectra and surface band structures have the same spectral features~\cite{alexandradinata2016topological,fidkowski2011model}.
Thus, the band structure on the (001) surface
can be systematically classified based on Wilson loop analysis.
(The details on tight-binding notation and Wilson loop is provided in SN2 and SN5.)

First, let us consider the MSGs related to MWG $p4'g'm$ [see \figs{fig2}(a)-(e)].
At $\bM=(\pi,\pi)$, four Wilson bands form a fourfold degeneracy, which can be identified as a Dirac fermion on the (001) surface.
The connectivity of Wilson bands is classified by the MCN $\mchd$, which is encoded in the slope of Wilson bands in each mirror-sector crossing a horizontal reference line [a green dashed line in \fig{fig2}(b)] along the $\bG$-$\bM$ direction.
In \fig{fig2}(d), for example, as two Wilson bands with mirror eigenvalue $-i$ intersect the reference line with a negative slope, we obtain $\mchd=2$.
See SN5 for the details on the counting rules for $\mchd$.

Now, we compare the Wilson loop spectra of topological phases with nonzero $\mchd$ and the trivial phase with zero $\mchd$ by focusing on the region near the fourfold degeneracy at $\bM$.
Along the $\bG$-$\bM$-$\bGp$ line, the four bands are divided into two different mirror sectors.
When $\mchd=0$, as in \fig{fig2}(b), the dispersion in each mirror sector is nonchiral, similar to that of Dirac fermions in 2D crystals in \fig{fig1}(d).
In \fig{fig2}(b), chiral and antichiral modes in the same mirror sector (e.g., the $+i$ sector) cross the green dashed reference line with opposite signs of the group velocity.
In contrast, their numbers are not equal in \fig{fig2}(d), where $\mchd=2$.

The Dirac fermions in \figs{fig2}(c) and (e), which correspond to $\mchd=1$ and 4, respectively, appear with additional surface states (black arrows).
When $\mchd=1$, the dispersion is chiral along the entire $\bG$-$\bM$-$\bGp$ line but locally looks nonchiral near the Dirac point.
However, if the dispersion along $\bG$-$\bM$-$\bGp$ is deformed such that the additional surface states near $\bG$ are pushed away from the Fermi level (which corresponds to $\theta=0$ in Wilson loop spectra), the Dirac fermion in \fig{fig2}(c) becomes chiral, as in \fig{fig2}(h).
In contrast, such a deformation is impossible in \fig{fig2}(e), where $\mchd=4$.
In general, one can show that a nonzero MCN $|\mchd| \le 2$ manifests as a chiral dispersion of the surface Dirac fermion along $\bG$-$\bM$-$\bGp$, provided that there is no additional surface state other than the Dirac fermion at the Fermi level.
In contrast, when $|\mchd|>2$, additional surface states always appear along $\bG$-$\bM$-$\bGp$.
Hence, the chiral dispersion of the Dirac fermion when $|\mchd| \le 2$ and the coexistence of additional surface states when $|\mchd| > 2$ are signatures of the nontrivial bulk topology of 3D TMDIs with nonzero $\mchd$.
An exact formulation of the relation among the chiral dispersion of the Dirac fermion, MCN $\mchd$, and number of additional surface states is given in SN6.

Similarly, one can analyze the Wilson loop spectra of the MSGs related to MWG $p'_c4mm$ [see \figs{fig2}(f)-(j)].
The MCN $\mchd$ ($\mc{C}_m^y$) can be determined by examining the $\mxyb$ ($M_y$) eigenvalues and the slopes of Wilson bands crossing a reference line along $\bG$-$\bM$ ($\bG$-$\bX$).
The relation between $\mchd$ and the chiral dispersion of the Dirac fermion is identical to the case of the MSGs with MWG $p4'g'm$.
The only additional feature is that $\mchd$ and $\mc{C}_m^y$ must be equivalent up to modulo 2, i.e., $\mchd = \mc{C}_m^y$ (mod 2), in insulating phases.

Finally, in the 15 MSGs related to MWG $p'_cmm$, the Wilson loop spectra can be classified by the MCNs on the $k_{x,y}=0$ planes, $\mc{C}_m^x(0)$ and $\mc{C}_m^y(0)$, related to $M_{x,y}$ mirrors.
The MCNs on the $k_{x,y}=\pi$ planes, $\mc{C}_m^x(\pi)$ and $\mc{C}_m^y(\pi)$, are always trivial, while surface Dirac fermions can be chiral only for nonzero $\mc{C}_m^{x,y}(\pi)$.
Thus, the surface Dirac fermion in the MSGs with MWG $p'_cmm$ is nonchiral and trivial.
%%%%%%%%%%%%%%%%%%%%%%

%%%%%%%%%%%%%%%%%%%%%%
\begin{figure}[t!]
\centering
\includegraphics[width=0.48\textwidth]{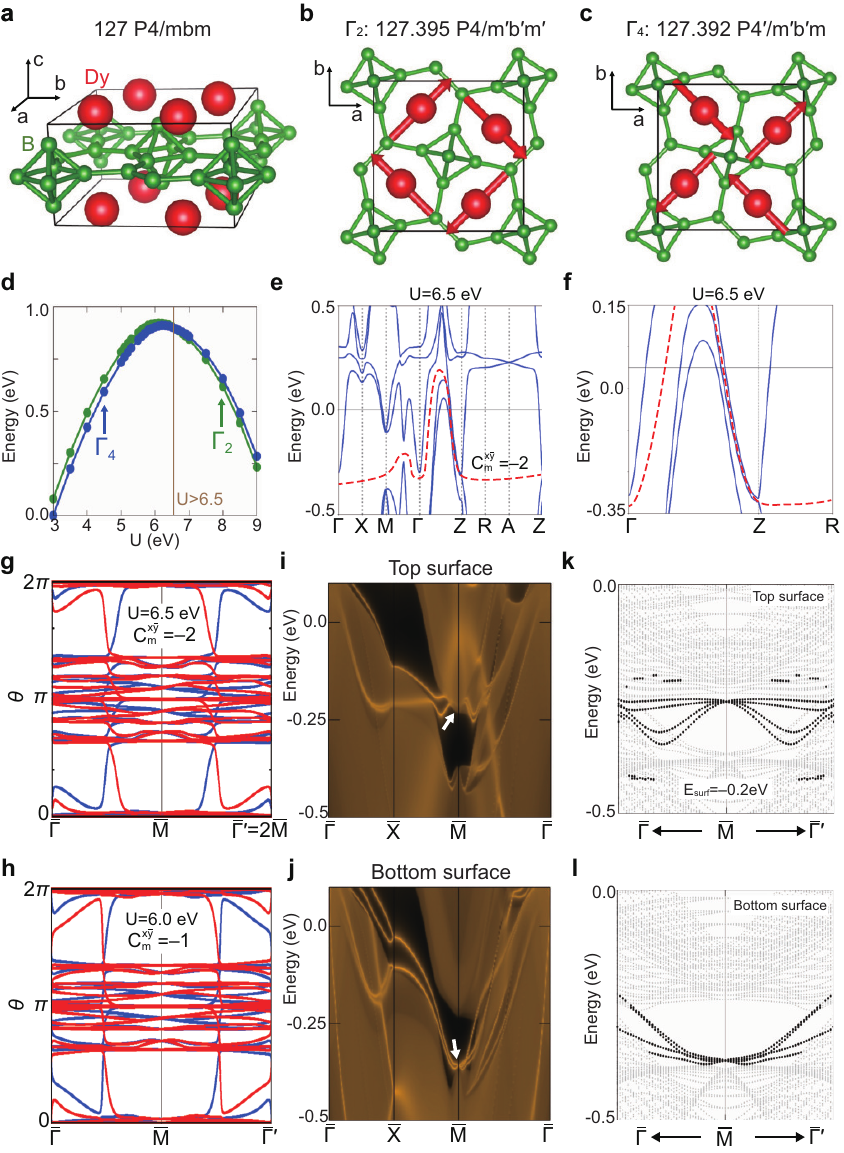}
\caption{
{\bf DyB$_4$, a Type-III TMDI candidate.}
(a) Tetragonal crystal structure of DyB$_4$.
(b)-(c) Magnetic spin structures in the (b) $\Gamma_2$ and (c) $\Gamma_4$ states. The
$\Gamma_4$ state corresponds to Type-III MSG 127.392 $P4'/m'b'm$ with (001)-surface MWG $p4'g'm$.
(d) Density functional theory (DFT) total energy differences between the $\Gamma_4$ and $\Gamma_2$ states. The absolute values are adjusted to be in the range between 0 eV and 1 eV. The magnetic phase transition is indicated by the brown solid line. The $\Gamma_4$ state has lower energy than the $\Gamma_2$ state when $U$ $\leq$ 6.5 eV.
(e)-(f) Bulk band structure from DFT+$U$ calculations ($U$=6.5 eV and $J$=1 eV) for the $\Gamma_4$ state.
The band gap near the Fermi level is indicated by the red dashed lines.
(g)-(h) Wilson loop spectra below the red dashed line along $\bG$-$\bM$-$\bGp$.
The winding structure exhibits (g) $\mchd=-2$ at $U$=6.5 eV and (h) $\mchd=-1$ at $U$=6.0 eV.
(i)-(j) (001)-surface spectra for the (i) top (B-terminated) and (j) bottom (Dy-terminated) surfaces obtained using the surface Green's function method.
The white arrows guide where surface states appear.
(k)-(l) (001)-surface band structures from a 60-layer slab calculation, which are drawn to show surface states indicated by the white arrows in (i) and (j).
Bulk and surface bands are represented by gray and black, respectively.
Surface Dirac fermions are moved inside the gap by applying a surface potential of -0.2 eV on the top.
(i-l) are calculated using the Wannierized tight-binding model from the DFT+$U$ calculation ($U$=6.5 eV).
}
\label{fig3}
\end{figure}
%%%%%%%%%%%%%%%%%%%%%%

%%%%%%%%%%%%%%%%%%%%%%
\section{Topological magnetic Dirac insulators}
\label{sec:hinge}
%%%%%%%%%%%%%%%%%%%%%%
According to the Wilson loop analysis, the bulk band topology of the 11 MSGs with Type-III MWG $p4'g'm$ and the 5 MSGs with Type-IV MWG $p'_c4mm$ can be characterized by $\mchd$ and $(\mchd,\mc{C}_m^y)$, respectively.
In these 16 MSGs, when $\mchd\neq0$, a Dirac fermion whose mirror-resolved dispersion is chiral can appear on the (001) surface.
Based on this, we define TMDIs as 3D magnetic TCIs with nonzero $\mchd$ hosting a 2D chiral Dirac fermion on the (001) surface.
Additionally, according to the (001)-surface MWG, TMDIs can be divided into Type-III and Type-IV TMDIs such that Type-III (Type-IV) TMDIs have (001)-surface MWG $p4'g'm$ ($p'_c4mm$).

Interestingly, TMDIs also exhibit higher-order topology~\cite{benalcazar2017quantized,benalcazar2017electric,langbehn2017reflection,schindler2018higher1,schindler2018higher2,khalaf2018higher,khalaf2018symmetry,trifunovic2019higher,fang2019new,lange2021subdimensional} with hinge modes at open boundaries along the $x$ and $y$ directions when the entire finite-size systems respect the diagonal mirror symmetries.
The MCN for the diagonal mirror $\mchd$ follows the higher-order bulk-boundary correspondence~\cite{langbehn2017reflection,schindler2018higher1,trifunovic2019higher}.
The dispersion of hinge modes can be both chiral and helical depending on the details of the systems.
Note that the number of chiral and antichiral hinge modes at each hinge can be changed by a mirror-symmetric and bulk-gap-preserving perturbation because such a perturbation can close and reopen a surface gap~\cite{schindler2018higher1,khalaf2018higher,geier2018second}.
However, the MCN $\mchd$ protects at least $|\mchd|$ hinge modes at each mirror-invariant hinge.
(See SN7 for more details.)
We provide a tight-binding model for a Type-III TMDI in SN10, which confirms the bulk-boundary correspondence described above. 
%%%%%%%%%%%%%%%%%%%%%%

%%%%%%%%%%%%%%%%%%%%%%
\section{Candidate materials}
\label{sec:DFT}
%%%%%%%%%%%%%%%%%%%%%%

%%%%%%%%%%%%%%%%%%%%%%
\begin{figure}[t!]
\includegraphics[width=0.48\textwidth]{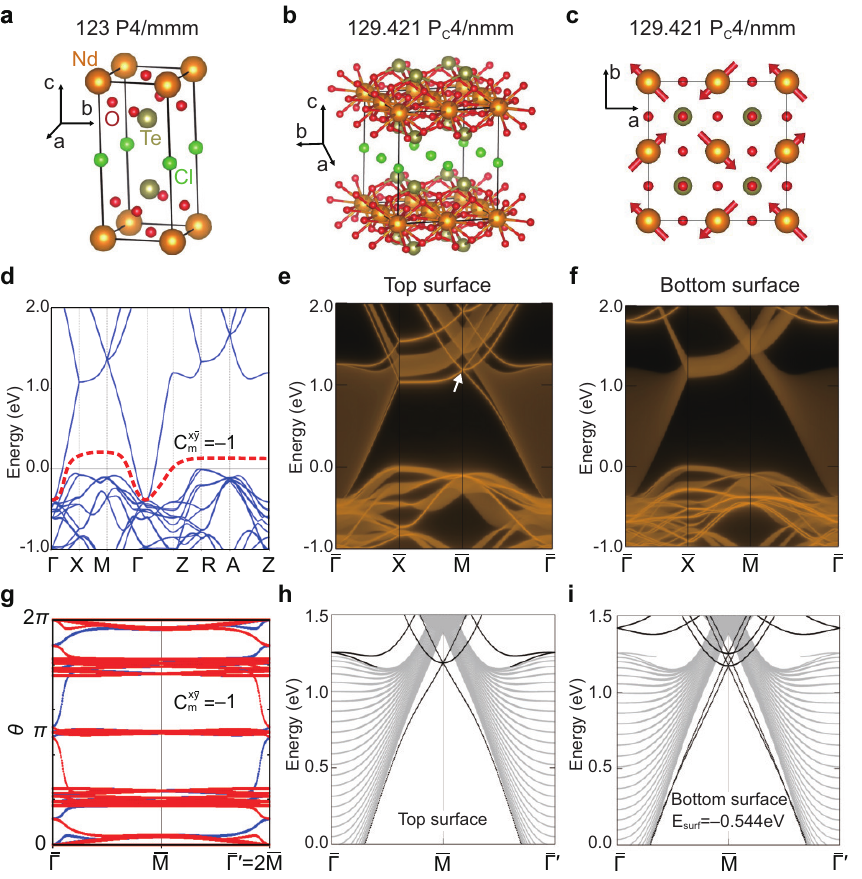}
\caption{
{\bf Nd$_4$Te$_8$Cl$_4$O$_{20}$, a Type-IV TMDI candidate.} 
(a)-(b) Crystal structures of NdTe$_2$ClO$_5$ and Nd$_4$Te$_8$Cl$_4$O$_{20}$ ($2 \times 2 \times 1$ supercell).
(c) Magnetic structure compatible with Type-IV MSG 129.421 $P_C4/nmm$ and (001)-surface MWG $p'_c4mm$.
(d) Bulk band structure from DFT+$U$ calculations ($U$=6 eV).
The band gap near the Fermi level is indicated by the red dashed line.
(e)-(f) (001)-surface spectra for the (e) top (O-terminated) and (f) bottom (Nd-terminated) surfaces obtained using the surface Green's function method.
The surface Dirac fermion is indicated by the white arrow in (e).
(g) $k_z$-directed Wilson loop spectrum along the diagonal path $\bG$-$\bM$-$\bGp$.
The winding structure exhibits $\mchd=-1$.
(h)-(i) (001)-surface for the (h) top and (i) bottom surfaces from a 30-layer slab calculation.
Bulk and surface bands are represented by gray and black, respectively.
The fourfold-degenerate Dirac fermions are located approximately 1 eV from the Fermi level on the top surface.
A surface potential of -0.544 eV is introduced to push the Dirac point on the bottom surface into the gap.
}
\label{fig4}
\end{figure}
%%%%%%%%%%%%%%%%%%%%%%

Using first-principles calculations, we propose DyB$_4$ and Nd$_4$Te$_8$Cl$_4$O$_{20}$ as candidate materials for a Type-III TMDI and a Type-IV TMDI, respectively, whose electronic and topological properties are summarized in \figs{fig3} and \ref{fig4}, respectively.
Although their band structure are metallic, as the systems have nonzero direct gap at all momenta, their mirror Chern numbers are well defined.
We note that the two candidate materials, DyB$_4$ and Nd$_4$Te$_8$Cl$_4$O$_{20}$, are not available in the existing material databases~\cite{ICSD,springermaterials, MaterialsProject,zhang2019catalogue, tang2019comprehensive, vergniory2019complete, xu2020high,gallego2016magndata,vergniory2022all}.
In general, searching for candidate magnetic materials is more challenging than searching for nonmagnetic materials because
the number of available materials in magnetic material databases~\cite{xu2020high,gallego2016magndata} is limited to approximately 1600~\cite{gallego2016magndata}, which is much smaller than the number of nonmagnetic materials.
To overcome this limitation, we will also propose a general scheme to systematically search for candidate materials for TMDIs, whose magnetic structures and MSGs are derived from their parent paramagnetic states.

Let us first consider DyB$_4$, a member of the rare-earth tetraborides family, whose crystal structure is shown in \fig{fig3}(a). 
The paramagnetic parent phase of DyB$_4$ has space group 127 $P4/mbm$.
In experiments~\cite{sim2016Spontaneous}, this material was reported to have two competing spin configurations that correspond to the $\Gamma_2$ (with MSG 127.395 $P4/m'b'm'$) and $\Gamma_4$ (with MSG 127.392 $P4'/m'b'm$) magnetic states, as shown in \figs{fig3}(b) and (c). 
Between them, the $\Gamma_2$ state was reported to be more favored~\cite{sim2016Spontaneous}.
Note that the $\Gamma_4$ phase can support a Type-III TMDI with (001)-surface MWG $P4'g'm$.
Here, we investigate the conditions in which the $\Gamma_4$ state becomes the magnetic ground state in DyB$_4$ based on DFT+$U$(where $U$ is the onsite Coulomb interaction) calculations.
We examine the total energies of the $\Gamma_2$ and $\Gamma_4$ states as a function of $U$ with fixed Hund's coupling $J$ = 1 eV.
As shown in Fig.~\ref{fig3}(d), DyB$_4$ undergoes a magnetic phase transition from the $\Gamma_4$ to $\Gamma_2$ state when $U>6.5$ eV.

An indirect gap near the Fermi level $E_{\rm F}$=0.0 eV exists in the DFT+ $U$ result ($U$=6.5 eV, $J$=1.0 eV), as shown in Fig.~\ref{fig3}(e).
Because the gap along the $\Gamma$-$Z$ line is tiny, as shown in \fig{fig3}(f), a topological phase transition between phases with different MCNs $\mchd$ can be easily induced by a small perturbation.
(See SN9 for the band structures between $U$=6.0 eV and 6.5 eV and SN12 for a detailed analysis of this topological phase transition.)
Figures~\ref{fig3}(g) and (h) show the $k_z$-directed Wilson loop spectra for $U$=6.5 eV and 6.0 eV, respectively.
Because of gap closing and reopening at $Z$, a topological phase transition from the $\mchd=-1$ state to the $\mchd=-2$ state occurs as $U$ increases from 6.0 eV to 6.5 eV.
Hence, we identify DyB$_4$ at $U$=6.5 eV (6.0 eV) as a Type-III TMDI with $\mchd=-2$ ($-1$).

More specifically, let us consider DyB$_4$ at $U$=6.5 eV.
Figures~\ref{fig3}(i)-(j) show the surface Green's function calculations for the top (B-terminated) and bottom (Dy-terminated) surfaces.
A surface Dirac fermion with fourfold degeneracy is clearly identified near $E=-0.35$ eV on the bottom surface [\figs{fig3}(j) and (l)].
In contrast, the top surface state is buried in the bulk conduction band and thus not visible in the gap.
However, the hidden surface state can be brought into the gap by applying a surface potential $E_{\rm surf}=-0.2$ eV, as shown in~\fig{fig3}(k).

The second material Nd$_4$Te$_{8}$Cl$_4$O$_{20}$ is a candidate for a Type-IV TMDI with $(\mchd,\mc{C}_m^y)=(-1,1)$.
Since its magnetic properties are unknown, this material is found by using our general scheme to search for candidate materials, as briefly explained below.
MSG 129.421 $P_C 4/nmm$ (123.16.1014 $P_P4/m'mm$ in the OG setting) is one of the MSGs that can host Type-IV TMDIs, which is derived from nonmagnetic space group (SG) 123 $P4/mmm$.
The crystal structure of NdTe$_2$ClO$_5$~\cite{ndteocl} in the paramagnetic phase has SG 123 $P4/mmm$, as shown in \fig{fig4}(a).
Introducing a $2 \times 2 \times 1$ supercell, Nd atoms can form a spin configuration compatible with MSG 129.421 $P_C 4/nmm$, as shown in \figs{fig4}(b) and (c).
Assuming the magnetic ordering corresponding to MSG 129.421 $P_C 4/nmm$, we study the electronic and topological properties of Nd$_4$Te$_{8}$Cl$_4$O$_{20}$.
The bulk band structure in \fig{fig4}(d) shows a semimetallic state with an indirect gap between $-0.5$ eV and 0.0 eV.
For the bands below this gap, we compute the $k_z$-directed Wilson loop spectrum, which exhibits a nonzero MCN, $\mchd=-1$, as shown in \fig{fig4}(g).
Hence, we identify Nd$_4$Te$_{8}$Cl$_4$O$_{20}$ as a candidate for a Type-IV TMDI with (001)-surface MWG $p'_c4mm$.

The surface Green's function calculations for the top (O-terminated) and bottom (Nd-terminated) surfaces are shown in \figs{fig4}(e) and (f).
Within a gap with a magnitude of 1.5 eV at $\bM$, a surface Dirac fermion with fourfold degeneracy is clearly identified near $E$=1 eV on the top surface.
As shown in \fig{fig4}(h), the four states of the surface Dirac fermion are split into two nondegenerate bands and accidentally nearly degenerate bands, in a similar way as in \fig{fig2}(h).
Although bottom surface states buried in bulk states are not visible in the gap [\fig{fig4}(f)], the hidden surface state can be pushed into the gap by applying a surface potential of $-0.544$ eV, as shown in~\fig{fig4}(i).

Our material search scheme used to find Nd$_4$Te$_{8}$Cl$_4$O$_{20}$
can generally be applied to all 16 MSGs (11 Type-III and 5 Type-IV MSGs) that can host TMDIs, as follows.
For each MSG, (i) determine the SG of paramagnetic phases, (ii) obtain the transformation matrix of basis vectors between the magnetic unit cell and crystal unit cell, and (iii) build up the magnetic unit cell from the paramagnetic unit cell according to the transformation matrix as well as restrict the spin configuration of magnetic atoms.
(See SN8 for details.)
Additionally, using this material search scheme, we studied
other candidate materials including two TMDI candidates with nonzero $\mchd$ (Ce$_2$Pd$_2$Pb, Ce$_2$Ge$_2$Mg) and four materials with $\mchd=0$ (HoB$_4$, FeSe, FeTe, AlGeMn, N$_3$TaTh).
(See SN9 for details.)
%%%%%%%%%%%%%%%%%%%%%%

%%%%%%%%%%%%%%%%%%%%%%
\section{Discussion}
\label{sec:dis}
%%%%%%%%%%%%%%%%%%%%%%
Let us discuss the symmetry-breaking effect on the 2D Dirac fermion, either in 2D bulk crystals or on the surface of 3D TMDIs.
When a perturbation that preserves diagonal mirrors in MWGs $p4'g'm$ and $p'_c4mm$ is applied, the Dirac fermion in 2D crystals becomes gapped while that on the surface of 3D TMDIs remains gapless because of the nonzero MCN.
In contrast, a mirror-breaking perturbation can induce various phase transitions in both cases because a 2D Dirac fermion corresponds to a multicritical point in magnetic systems~\cite{young2017filling}.
For example, MWG $p4'g'm$ is reduced to $p2g'g'$ when $\mxybo$ and $TC_{4z}$ are broken by tensile strain (see SN13).
In this case, a mass term is allowed; thus, the Dirac fermion can be gapped.
The Chern number of the resulting gapped phase changes by 2 when the sign of the mass is reversed.
This, in turn, indicates the appearance of chiral edge channels at domain walls between two gapped domains with opposite signs of the mass. 

Finally, we discuss the bulk topological responses of TMDIs.
Since mirror symmetry reverses the spatial orientation, it quantizes the axion angle $\theta$ to $0$ or $\pi$, which is related to the MCN by $\theta/\pi = \mchd$ (mod 2)~\cite{varjas2015bulk,peng2022topological}.
Hence, TMDIs with odd $\mchd$ correspond to an axion insulator exhibiting quantized magnetoelectric effects~\cite{qi2011rmp}.
In contrast, TMDIs with an even $\mchd$ have a vanishing axion angle. %
However, according to a recent theoretical proposal~\cite{lin2022spin}, even helical higher-order topological insulators (HOTIs) with $\theta=0$ can exhibit spin-resolved magnetoelectric effects.
In the case of 3D TMDIs, spin-resolved bands are ill-defined because of noncollinear magnetic orderings inherent in Type-III and -IV MSGs.
However, by combining the spin and sublattice degrees of freedom, the pseudospin-resolved magnetoelectric effect can be defined in 3D TMDIs with an even $\mchd$,
which is demonstrated using a tight-binding model in SN11.
A more systematic study of bulk topological responses in magnetic TCIs with noncollinear magnetic ordering is an important subject for future research.
%%%%%%%%%%%%%%%%%%%%%%
\vspace{0.3cm}

%%%%%%%%%%%%%%%%%%%%%%
\begin{center}
\textbf{Methods}
\end{center}

For the ab-initio calculation, the Vienna Ab initio Simulation Package (VASP) is employed with the projector augmented-wave method (PAW)~\cite{kresse1999ultrasoft}.
We employ the generalized gradient approximation (PBE-GGA) for exchange-correlation potential~\cite{perdew1998perdew}.
The default VASP potentials, an energy cutoff with 400 eV, and a $8 \times 8 \times 14$ Monkhorst-pack k-point mesh are used.
The spin-orbit coupling is considered because of the presence of the heavy rare-earth atoms in the unit cell.
On-site Coulomb interaction is taken into account with $U$ =6.5 eV and $J$ (Hund's coupling) = 1 eV in DyB$_4$ and $U$=6 eV and $J$=0 eV in NdTe$_2$ClO$_5$.
Wannier Hamiltonians were constructed by WANNIER90~\cite{mostofi2014updated} and symmetrized by WannSymm code~\cite{wannsymm}.
The WannierTools package~\cite{wu2018wanniertools} was used to produce the slab band structure and Wilson loop spectra based on the symmetrized Hamiltonian.
The experimental crystal structure of DyB$_4$~\cite{sim2016Spontaneous} are used without structural relaxation.
Details on the ab-initio calculation can be found in SN S9.
%%%%%%%%%%%%%%%%%%%%%%
\vspace{0.3cm}

%%%%%%%%%%%%%%%%%%%%%%
%\let\oldaddcontentsline\addcontentsline
%\renewcommand{\addcontentsline}[3]{}
\begin{acknowledgments}
We thank Benjamin Wieder and Aris Alexandradinata for fruitful discussions.
This work was supported by the Institute for Basic Science in Korea (Grant No. IBS-R009-D1),
Samsung Science and Technology Foundation under Project Number SSTF-BA2002-06,
the National Research Foundation of Korea (NRF) grant funded by the Korea government (MSIT) (No. 2021R1A2C4002773 and No. NRF-2021R1A5A1032996).
\end{acknowledgments}
%%%%%%%%%%%%%%%%%%%%%%
\vspace{0.3cm}

%%%%%%%%%%%%%%%%%%%%%%
\noindent \textbf{Data availability}
The data that support the findings of this study are available from the corresponding authors upon reasonable request.
%%%%%%%%%%%%%%%%%%%%%%
\vspace{0.3cm}

%%%%%%%%%%%%%%%%%%%%%%
\noindent \textbf{Code availability}
The numerical codes used in this paper are available from the corresponding authors upon reasonable request.
%%%%%%%%%%%%%%%%%%%%%%
\vspace{0.3cm}

%%%%%%%%%%%%%%%%%%%%%%
\noindent \textbf{Competing interests}
The authors declare no competing interests. 
%%%%%%%%%%%%%%%%%%%%%%
\vspace{0.3cm}

%%%%%%%%%%%%%%%%%%%%%%
\noindent \textbf{Author contributions}
Y.H. J.K. and J.L. performed the model and symmetry analyses.
Y.Q., D.R., and H.C.C. performed the ab initio calculations.
B.-J.Y. supervised the project.
Y.H., Y.Q., H.C.C., and B.-J.Y. analyzed the results and wrote the manuscript.
All authors contributed to the discussion and manuscript revision.
Y.H. and Y.Q. are the co-first authors contributed equally to this work.
%%%%%%%%%%%%%%%%%%%%%%
\vspace{0.3cm}

%%%%%%%%%%%%%%%%%%%%%%
\noindent \textbf{Correspondence} and requests for materials should be addressed to H.C.C. or B.-J.Y.
%%%%%%%%%%%%%%%%%%%%%%

%%%%%%%REFERENCES%%%%%%%%
%\bibliographystyle{apsrev}
\bibliography{Refs.bib}

%%%%%%%%APPENDIX%%%%%%%%
\let\addcontentsline\oldaddcontentsline
%%%%%%%%%%%%%%%%%%%%%%

%%%%%%%%%%%%%%%%%%%%%%
\clearpage
\onecolumngrid
\begin{center}
\textbf{\large Supplemental Material for ``\ourtitle"}
\end{center}
%%%%%%%%%%%%%%%%%%%%%%
\setcounter{section}{0}
\setcounter{figure}{0}
\setcounter{equation}{0}
\setcounter{table}{0}
\renewcommand{\thefigure}{S\arabic{figure}}
\renewcommand{\theequation}{S\arabic{equation}}
\renewcommand{\thesection}{S\arabic{section}}
\renewcommand{\thetable}{S\arabic{table}}
\tableofcontents
%%%%%%%%%%%%%%%%%%%%%%
\hfill \\
\onecolumngrid
%%%%%%%%%%%%%%%%%%%%%%

\begin{center}
\textbf{\large Supplementary Notes}
\end{center}

%%%%%%%%%%%%%%%%%%%%%%
\section{Conventions for symmetry elements}
\label{app:Conv}
%%%%%%%%%%%%%%%%%%%%%%
We summarize the conventions used in this work for denoting symmetries.
Unitary symmetries are collectively denoted as $\sg$.
Arbitrary unitary symmetry element can be written as $\sg=\{\sg_0|a,b,c\}=\{E|a,b,c\} \{\sg_0|\bb 0\}$.
Here $E$ is the identity and $\{E|a,b,c\}$ is the translation by $a \bb a_1 + b \bb a_2 + c \bb a_3$.
An antiunitary symmetry can be represented as a combination of time-reversal $T$ and unitary symmetry.
We denote antiunitary symmetries as $\sg'$ collectively.
The conventional/primitive lattice vectors are normalized as $\bb a_1=(1,0,0)$, $\bb a_2=(0,1,0)$, and $\bb a_3=(0,0,1)$.
The point group elements introduced in this work act on real-space coordinates $(x,y,z)$ as follows:
\bg
c_{4z}(x,y,z) = (-y,x,z), \,\, c_{2z}(x,y,z) = (-x,-y,z), \,\,
m_x(x,y,z) = (-x,y,z), \,\, m_y(x,y,z) = (x,-y,z), \nn
m_{x\cm{y}}(x,y,z) = (y,x,z), \,\, m_{xy}(x,y,z) = (-y,-x,z), \,\,
i(x,y,z)=(-x,-y,-z).
\eg
The $2\pi$ rotation of half-integer spin is denoted as $\cm{E}$ or simply $-1$.
For example, $m_x^2=\cm{E}$ in half-integer spin or double group representation.
For a definite description, let us consider a mirror $\mxyb=\{m_{x\cm{y}}|\bb 0\}$ and an antiunitary glide mirror $\tmy=T\{m_y|\hf,\hf,0\}$ as examples.
In our notation, $\mxyb$ acts as $\mxyb(x,y,z) = (y,x,z)$.
$\tmy$ acts as $\tmy(x,y,z) = (x+\hf,-y+\hf,z)$ with the additional action by time reversal $T$.
We use a similar notation as well in two dimensions.
%%%%%%%%%%%%%%%%%%%%%%

%%%%%%%%%%%%%%%%%%%%%%
\section{Review of tight-binding theory}
\label{app:Tightbinding}
%%%%%%%%%%%%%%%%%%%%%%
In this section, we review the tight-binding theory and introduce the notations that are used repeatedly in symmetry analysis.
Let us consider a periodic lattice in $d$ dimensions with primitive lattice vectors $\bb a_i$ ($i=1,\dots,d$).
We denote lattice vectors as $\bR$ which have a form of $\sum_{i=1}^d n_i \bb a_i$ for $n_i \in \mathbb{Z}$.
A unit cell is labeled by each lattice vector $\bR$ and consists of $n_{\rm tot}$ sublattice sites.
The number of unit cells are $N_{\rm cell}=\sum_{\bR}$.
Sublattice sites within unit cell $\bR$ are located at $\bR+\bx_\alpha$ ($\alpha=1,\dots,n_{\rm tot}$).
We put atomic orbitals on these sublattice sites and denote them as $\ketr{\bR, \alpha}$.
$\ketr{\bR, \alpha}$ form the basis states for a tight-binding Hamiltonian and satisfy the orthogonality $\brkr{\bR, \alpha}{\bR', \beta} = \delta_{\bR,\bR'} \delta_{\alpha\beta}$.
A hopping process that hops from an orbital at $\bR+\bx_\alpha$ to an orbital at $\bR' + \bx_\beta$ has a strength $t(\bR,\alpha;\bR',\beta)$.
Then, a tight-binding Hamiltonian $H$ is defined as
\bg
H = \sum_{\bR,\bR'} \sum_{\alpha,\beta} \ketr{\bR,\alpha} \, t(\bR,\alpha;\bR',\beta) \, \brar{\bR', \beta}
\eg
in real-space basis.
We introduce the Fourier transform of $\ketr{\bR, \alpha}$, which forms the momentum-space basis,
\bg
\ketr{\bk, \alpha} = \frac{1}{\sqrt{N_{\rm cell}}} \sum_\bR e^{i \bk \cdot (\bR+\bx_\alpha)} \ketr{\bR, \alpha}.
\eg
Hence, the tight-binding Hamiltonian in momentum-space basis is given by
\bg
H(\bk)_{\alpha\beta} = \brar{\bk, \alpha} H \ketr{\bk, \beta} = \sum_{\bR} t(\bR,\alpha;\bb 0,\beta) e^{-i \bk \cdot (\bR+\bx_\alpha-\bx_\beta)}.
\eg
Now, we consider the energy eigenstates $\ket{u_n(\bk)}$ which satisfies $\sum_{\beta=1}^{n_{\rm tot}} H(\bk)_{\alpha \beta} \ket{u_n(\bk)}_{\beta} = E_n(\bk) \ket{u_n(\bk)}$, i.e. $H(\bk) \ket{u_n(\bk)} = E_n(\bk) \ket{u_n(\bk)}$ ($n=1,\dots,n_{\rm tot}$).
The eigenstates also satisfy the orthogonality and the completeness relation such that $\brk{u_n(\bk)}{u_m(\bk)} = \delta_{nm}$ and $\sum_{n=1}^{n_{\rm tot}} \kbr{u_n(\bk)}{u_n(\bk)} = \mathds{1}_{n_{\rm tot}}$.
Here, $\mathds{1}_{n_{\rm tot}}$ is the $n_{\rm tot} \times n_{\rm tot}$ identity matrix.
The energy eigenstates of $H$ also can be constructed by using $\ket{u_n(\bk)}$ and $\ketr{\bk,\alpha}$:
\ba
\ketr{\psi_n(\bk)} = \sum_{\alpha=1}^{n_{\rm tot}} \ket{u_n(\bk)}_\alpha \ketr{\bk,\alpha}, \quad
H \ketr{\psi_n(\bk)} = E_n(\bk) \ketr{\psi_n(\bk)}.
\label{eqapp:blochwf}
\ea
We call $\ket{\psi_n(\bk)}$ a Bloch wavefunction of $n$-th band.
The Bloch wave functions are orthogonal such that $\brkr{\psi_n(\bk)}{\psi_m(\bk')} = \delta_{\bk,\bk'} \delta_{nm}$, which is different to that of $\ket{u_n(\bk)}$ since $\brk{u_n(\bk)}{u_m(\bk')} \ne 0$ in general.
We use the same braket notation for $\ketr{\bR, \alpha}$, $\ketr{\bk,\alpha}$, $\ket{u_n(\bk)}$, and $\ketr{\psi_n(\bk)}$.
However, they can be distinguished without confusion by looking at the indices such as $\alpha$ and $n$.

$H(\bk)$ and $\ketr{\bk,\alpha}$ are not periodic in the Brillouin zone (BZ): $H(\bk+\bb G)_{\alpha\beta} = H(\bk)_{\alpha\beta} e^{-i \bb G (\bx_\alpha-\bx_\beta)}$ and $\ketr{\bk + \bb G, \alpha} = e^{i \bb G \cdot \bx_\alpha} \ketr{\bk,\alpha}$ for any reciprocal lattice vector $\bb G$.
By introducing the orbital embedding matrix $V(\bk)_{\alpha\beta} \coloneqq e^{-i \bk \cdot x_{\alpha}} \delta_{\alpha\beta}$, we represent the aperiodicity as
\bg
\ketr{\bk+\bb G, \alpha}
= [V(\bb G)^{-1}]_{\alpha\alpha} \ketr{\bk,\alpha}, \quad
H(\bk+\bb G) = V(\bb G) H(\bk) V(\bb G)^{-1}.
\label{eqapp:aperiodic}
\eg
Despite the aperiodicity, the eigenvalue spectra ${\rm spec}[H(\bk)]$ of $H(\bk)$ are periodic since $V(\bb G)$ is an unitary matrix.
Thus, $\{E(\bk+\bb G)\}=\{E(\bk)\}$ where $\{E(\bk)\} \coloneqq {\rm spec}[H(\bk)] = \{E_n(\bk)|n=1,\dots,n_{\rm tot}\}$. 
The aperiodicity in \eq{eqapp:aperiodic} implies that $V(\bb G) \ket{u_n(\bk)}$ is an energy eigenstate at $\bk+\bb G$ with energy eigenvalue $E_n(\bk)$.
Hence, we choose the periodic gauge,
\bg
\ket{u_n(\bk+\bb G)}_\alpha = V(\bb G)_{\alpha\beta} \ket{u_n(\bk)}_\beta = \ket{u_n(\bk)}_\beta V(\bb G)_{\beta\alpha},
\label{eqapp:pgauge}
\eg
which is always assumed throughout this work.
Note that $\ketr{\psi_n(\bk)}$ is periodic in the periodic gauge.

Now we review the action of symmetry operators on tight-binding Hamiltonian. 
We consider unitary symmetry $\sg$ and antiunitary symmetry $\sg'=\sg \mc{K}$, where $\mc{K}$ is complex-conjugation. 
Symmetries leave the Hamiltonian invariant and thus satisfy $H = \sg H \sg^{-1}$ and $H = \sg' H \sg'^{-1}$. 
Also, $\sg$ and $\sg'$ act on $\ketr{\bk,\alpha}$ as
\ba
\sg \ketr{\bk,\alpha} = \ketr{\sg\bk,\beta} U_\sg(\bk)_{\beta \alpha}, \quad
\sg' \ketr{\bk,\alpha} = \ketr{\sg'\bk,\beta} U_{\sg'}(\bk)_{\beta \alpha} \mc{K}
\eqqcolon \ketr{\sg'\bk,\beta} U^\mc{K}_{\sg'}(\bk)_{\beta \alpha},
\ea
for $n_{\rm tot} \times n_{\rm tot}$ unitary matrices $U_\sg(\bk)$ and $U_{\sg'}(\bk)$.
We call $U_\sg(\bk)_{\beta \alpha}$ and $U^\mc{K}_{\sg'}(\bk)_{\beta \alpha}$ symmetry operators.
Note that $\sg'\bk = - \sg \bk$ and $U^\mc{K}_{\sg'}(\bk) = U_{\sg'}(\bk) \mc{K}$ for antiunitary symmetry $\sg'=\sg \mc{K}$.
The tight-binding Hamiltonian in momentum-space basis $H(\bk)$ transforms as
\bg
H(\sg \bk)_{\alpha \beta}
= \brar{\sg \bk, \alpha} H \ketr{\sg \bk, \beta}
= [U_\sg(\bk) H(\bk) U_\sg(\bk)^{-1}]_{\alpha \beta}, \nn
H(\sg' \bk)_{\alpha \beta}
= \brar{\sg' \bk, \alpha} H \ketr{\sg' \bk, \beta}
= [ U_{\sg'}(\bk) \cm{H(\bk)} U_{\sg'}(\bk)^{-1} ]_{\alpha \beta}
= [ U^{\mc{K}}_{\sg'} (\bk) H(\bk) U^{\mc{K}}_{\sg'}(\bk)^{-1} ]_{\alpha \beta}.
\label{eqapp:symH}
\eg
When a set of bands is isolated from other bands by gap, we define them as the occupied bands.
Then, $U_\sg(\bk) \ketr{u_m(\bk)}$ belongs to the occupied-band subspace for an occupied band with $\ketr{u_n(\bk)}$.
The sewing matrices, $B_\sg(\bk)$ and $B^{(\mc{K})}_{\sg'}(\bk)$, are defined for the occupied-band subspace in terms of symmetry operators and occupied eigenstates:
\bg
B_\sg(\bk)_{nm}
= \bra{u_n(\sg\bk)} U_\sg(\bk) \ket{u_m(\bk)}, \nn
B^{\mc{K}}_{\sg'}(\bk)_{nm}
= B_{\sg'}(\bk)_{nm} \mc{K}
= \bra{u_n(\sg'\bk)} U^{\mc{K}}_{\sg'}(\bk) \ket{u_m(\bk)}
= \bra{u_n(\sg'\bk)} U_{\sg'}(\bk) \cm{\ket{u_m(\bk)}} \mc{K}.
\label{eqapp:sewB}
\eg
For $n_{\rm occ}$ number of occupied bands, $B_{\sg}(\bk)$ and $B_{\sg'}(\bk)$ are $n_{\rm occ} \times n_{\rm occ}$ unitary matrices.
We can also express the sewing matrices in terms of Bloch wave function in \eq{eqapp:blochwf} as
\bg
B_\sg(\bk)_{nm} = \brar{\psi_n(\sg \bk)} \sg \ketr{\psi_m(\bk)}, \quad
B^{\mc{K}}_{\sg'}(\bk)_{nm} = \brar{\psi_n(\sg' \bk)} \sg' \ketr{\psi_m(\bk)}.
\eg
Thus, in the periodic gauge, the sewing matrices are periodic in the BZ.

In many cases, we perform symmetry analysis at high-symmetry point $\bk^*_\sg$, which is left invariant under an unitary symmetry $\sg$, i.e. $\bk^*_\sg = \sg \bk^*_\sg + \bb G (\bk^*_\sg)$ for certain reciprocal lattice vector $\bb G(\bk^*_\sg)$. 
With the help of \eq{eqapp:aperiodic} and \eq{eqapp:symH}, we obtain
\bg
H(\bk^*_\sg)
= H(\sg \bk^*_\sg + \bb G (\bk^*_\sg))
%= V(\bb G (\bk^*_\sg)) H(\sg \bk^*_\sg) V(\bb G (\bk^*_\sg))^{-1}
= [V(\bb G (\bk^*_\sg)) U(\bk^*_\sg)] H(\bk^*_\sg) [V(\bb G (\bk^*_\sg)) U(\bk^*_\sg) ]^{-1},
\eg
or $[V(\bb G(\bk^*_\sg)) U_\sg(\bk^*_\sg), H(\bk^*_\sg)] = 0$ equivalently.
This implies that the energy eigenstates $\ket{u_n(\bk)}$ can be redefined such that they have well-defined eigenvalues for $V(\bb G(\bk^*_\sg)) U_\sg(\bk^*_\sg)$, which we simply refer to them as eigenvalues of $\sg$.
For $\sg$-invariant momentum, the relevant sewing matrix is given by
\bg
B_\sg(\bk^*_\sg)_{nm}
= \bra{u_n(\bk^*_\sg)} V(\bb G(\bk^*_\sg)) U_\sg(\bk^*_\sg) \ket{u_m(\bk^*_\sg)}.
\eg
Hence, the symmetry eigenvalues of occupied bands with respect to $\sg$ can be obtained by diagonalizing the sewing matrix $B_\sg(\bk^*_\sg)_{nm}$.
For antiunitary symmetry $\sg'$, we can show that, by applying similar procedure we have done for unitary symmetry, $[V(\bb G(\bk^*_{\sg'})) U^{\mc K}_{\sg'}(\bk^*_{\sg'}), H(\bk^*_{\sg'})] = 0$, $B^{\mc K}_{\sg'}(\bk^*_{\sg'})_{nm} = \bra{u_n(\bk^*_{\sg'})} V(\bb G(\bk^*_{\sg'})) U^{\mc K}_{\sg'}(\bk^*_{\sg'}) \ket{u_m(\bk^*_{\sg'})}$ hold.
%%%%%%%%%%%%%%%%%%%%%%

%%%%%%%%%%%%%%%%%%%%%%
\section{Fourfold Dirac fermions in magnetic wallpaper and space groups}
\label{app:MSGs}
%%%%%%%%%%%%%%%%%%%%%%
In the main text, we have shown that only three magnetic wallpaper groups (MWGs) can protect four-dimensional irreducible representation at the Brillouin zone (BZ) corner, $M=(\pi,\pi)$.
These three MWGs are Type-III $p4'g'm$, Type-IV $p'_cmm$, and Type-IV $p'_c4mm$.
When the band dispersion is linear near the fourfold-degenerate point, we call such fourfold irreducible representation a fourfold-degenerate Dirac fermion, or a Dirac fermion simply when the context is clear.
In nonmagnetic systems, only two wallpaper groups $p2gg1'$ and $p4gm1'$ can protect Dirac fermions~\cite{wieder2018wallpaper}.

Now, let us compare the Dirac fermions in nonmagnetic and magnetic wallpaper groups.
The main difference of them is the doubling theorem studied in Refs.~\cite{young2015dirac,wieder2018wallpaper}.
Wallpaper groups $p2gg1'$ and $p4gm1'$ in nonmagnetic systems have unitary nonsymmorphic symmetries.
Dirac fermions in $p2gg1'$ and $p4gm1'$ never appear as a unique nodal structure at Fermi level because the nonsymmorphic symmetries also protect another nodal point at the same filling~\cite{wieder2018wallpaper}.
However, three MWGs, $p4'g'm$, $p'_cmm$, and $p'_c4mm$, do not have unitary nonsymmorphic symmetries because magnetic orderings lower symmetry groups in general.
As a consequence a single Dirac fermion can appear as a unique nodal structure at Fermi level in magnetic MWGs and it evades the doubling theorem that holds in nonmagnetic wallpaper groups~\cite{young2017filling}.

In this work, we present the exhaustive list of MWGs protecting fourfold-degenerate Dirac fermions.
The symmetry algebra of Type-IV MWGs $p'_cmm$ and $p'_c4mm$ are discussed in \Rf{young2017filling}.
However, Type-III MWG $p4'g'm$ protects fourfold degeneracy in a different way.
Below, we prove the symmetry protection of fourfold degeneracies in MWGs $p'_cmm$, $p'_c4mm$, and  $p4'g'm$.
%%%%%%%%%%%%%%%%%%%%%%

%%%%%%%%%%%%%%%%%%%%%%
\tocless{\subsection{Type-IV MWGs $p'_cmm$ and $p'_c4mm$}
\label{appsub:fourfoldIV}}{}
%\subsection{Type IV MWGs $p'_cmm$ and $p'_c4mm$}
%\label{appsub:fourfoldIV}
%%%%%%%%%%%%%%%%%%%%%%
MWGs $p'_cmm$ and $p'_c4mm$ have symmetry elements $M_x=\{m_x|\bb 0\}$, $M_y=\{m_y|\bb 0\}$, and $\Tg = T\{E|\hf,\hf\}$.
Note that $p'_c4mm$ has an additional symmetry generator $C_{4z}=\{c_{4z}|\bb 0\}$.
Fourfold degeneracy in $p'_cmm$ and $p'_c4mm$ are protected at $M=(\pi,\pi)$ in the BZ, and this can be explained by considering $M_x$, $M_y$, and $\Tg$.
Let $\ketr{\psm}$ be an energy eigenstate (Bloch wavefunction) of arbitrary Hamiltonian $H$ that respects $p'_cmm$ or $p'_c4mm$.

Let us define a simultaneous eigenstate $\ketr{\psm}$ of $M_x$ and $H$ because $M_x$ is unitary and commutes with $H$.
$\ketr{\psm}$ has $M_x$ eigenvalue $\pm i$ such that $M_x \ketr{\psm} = \pm i \ketr{\psm}$.
Since $\Tg$ is antiunitary and $\Tg^2=\{\cm{E}|1,1\}$, $\ketr{\psm}$ and $\Tg \ketr{\psm}$ form a Kramers pair.
To show this, we use the identity $\brkr{\psi_1}{\psi_2} = \brkr{\sg' \psi_2}{\sg' \psi_1}$ which holds for any antiunitary operator $\sg'$.
\bg
\brkr{\psm}{\Tg \psm}
= \brkr{\Tg^2 \psm}{\Tg \psm}
= \brkr{\{\cm{E}|1,1\} \psm}{\Tg \psm}
= -\brkr{\psm}{\Tg \psm}
= 0.
\label{eqapp:antiKramers}
\eg
In the last equality, we use the identity $\{E|n_x,n_y\} \ketr{\psi(\bk)} = e^{-i \bk \cdot \bb n} \ketr{\psi(\bk)}$ that holds for $\bb n=(n_x,n_z) \in \mathbb{Z}^2$ because of translation symmetry.
Also note that $\cm{E} \ketr{\psi(\bk)} = -\ketr{\psi(\bk)}$.

From $M_x \Tg = \Tg M_x \{E|1,0\}$ and $M_x M_y = \cm{E} M_y M_x$, we obtain
\bg
M_x [ \Tg \ketr{\psm} ] = \pm i \Tg \ketr{\psm}, \quad
M_x [ M_y \ketr{\psm} ]= \mp i M_y \ketr{\psm},
\eg
respectively.
Hence, two eigenstates in a Kramers pair $\{\ketr{\psm}, \Tg \ketr{\psm}\}$ have $M_x$ eigenvalue $\pm i$ while those in a Kramers pair $\{M_y \ketr{\psm}, \Tg M_y \ketr{\psm}\}$ have $M_x$ eigenvalue $\mp i$.
Since $M_x$, $M_y$, and $\Tg$ leave $M$ invariant, two Kramers pairs have the same energy eigenvalue.

In summary, four energy eigenstates listed below form fourfold degeneracy.
\bg
\ketr{\psm}, \quad \Tg \ketr{\psm}, \quad
M_y \ketr{\psm}, \quad \Tg M_y \ketr{\psm}.
\label{eqapp:fourfold1}
\eg

Finally, let us comment on fourfold degeneracy in $p'_c4mm$.
MWG $p'_c4mm$ have two mirrors $\mxyb=\{m_{x\cm{y}}|\bb 0\}$, $\mxy=\{m_{xy}|\bb 0\}$, which are absent in $p'_cmm$.
Thus, we show the fourfold degeneracy at $M$ in $p'_c4mm$ by using the simultaneous eigenstate $\psm$ of $H$ and $\mxyb$.
In this representation, fourfold degeneracy is formed by two states with $\mxyb$ eigenstates $+i$ and two with $-i$.
This can be shown by considering $\mxyb$, $\mxy$, and the periodicity of bands along $\mxyb$-invariant line, i.e. $\bk = (k,k)$, which passes $M$.
Suppose that $\ketr{\varphi_\pm(k,k)}$ has $\mxyb$ eigenvalue $\pm i$ for all $k$.
Since $\mxyb$ and $\mxy$ anticommutes, $\mxy \ket{\varphi_\pm(k,k)}$ can be identified as $\ket{\varphi'_\mp (-k,-k)}$, an eigenstate at $(-k,-k)$ with $\mxyb$ eigenvalue $\mp i$.
This means that the number of bands at $(k,k)$ with $\mxyb$ eigenvalue $+i$ is equal to that of bands at $(-k,-k)$ with $\mxyb$ eigenvalue $-i$.
Hence, the fourfold degeneracy is formed by two bands with $\mxyb$ eigenvalue $+i$ and two with $-i$ because of the periodicity in the BZ.
In fact, one can find four states forming fourfold degeneracy explicitly.
Rather than showing the derivation, let us simply write down the four states.
$\ketr{\varphi_\pm(M)}$ and $(\Tg M_x+\Tg M_y)\ketr{\varphi_\pm(M)}$ have $\mxyb$ eigenvalue $\pm i$.
The other two states, $\mxy \ketr{\varphi_\pm(M)}$ and $(\Tg M_x-\Tg M_y)\ketr{\varphi_\pm(M)}$, have $\mxyb$ eigenvalue $\mp i$.
%%%%%%%%%%%%%%%%%%%%%%
\\

%%%%%%%%%%%%%%%%%%%%%%
\tocless{\subsection{Type-III MWG $p4'g'm$}
\label{appsub:fourfoldIII}}{}
%\subsection{Type III MWG $p4'g'm$}
%\label{appsub:fourfoldIII}
%%%%%%%%%%%%%%%%%%%%%%
MWG $p4'g'm$ has symmetry elements $C_{2z}=\{c_{2z}|\bb 0\}$, $\mxybo=\{m_{x\cm{y}}|\hf,\mhf\}$, $\mxyo=\{m_{xy}|\hf,\hf\}$, $\tmx= T\{m_x|\hf,\hf\}$, $\tmy= T\{m_y|\hf,\hf\}$, and $TC_{4z}=T\{c_{4z}|\bb 0\}$.
Note that $\mxybo$ and $\mxyo$ are off-centered mirrors and the tilde symbol distinguishes them from $\mxyb=\{m_{x\cm{y}}|\bb 0\}$ and $\mxy=\{m_{xy}|\bb 0\}$.
In $p4'g'm$, fourfold degeneracy is protected at $M$ point, as in $p'_cmm$ and $p'_c4mm$.
To explain this, we focus on $C_{2z}$, $\tmy$, and $\mxybo$, which leave $M$ invariant.
Since $C_{2z}$ is unitary, we consider a simultaneous eigenstate $\ketr{\psm}$ of $C_{2z}$ and $H$ such that $C_{2z} \ketr{\psm} = \pm i \ketr{\psm}$.
Four energy eigenstates $|\psm)$, $\tmy \ketr{\psm}$, $\mxybo \ketr{\psm}$, and $\tmy \mxybo \ketr{\psm}$ have the same energy because $M$ is left invariant under $\tmy$ and $\mxybo$.
Now, we show that these four eigenstates are independent to each other and form a fourfold degeneracy.

First, let us show that $\psm$ and $\tmy \psm$ is orthogonal, i.e. $\brkr{\psm}{\tmy \psm}=0$.
This can be proven in a way similar to \eq{eqapp:antiKramers} because $\tmy$ is antiunitary and $(\tmy)^2=\{E|1,0\}$.
Also, from $C_{2z} \tmy = \tmy C_{2z} \{\cm{E}|1,\mns1\}$, we obtain $
C_{2z} [\tmy \ketr{\psm}] = \pm i \tmy \ketr{\psm}$.
That is, $\tmy \ketr{\psm}$ has $C_{2z}$ eigenvalue $\pm i$.
Thus, two states in a Kramers pair $\{\ket{\psm}, \tmy \ket{\psm}\}$ formed by $\tmy$ have the same $C_{2z}$ eigenvalue $\pm i$.

We can find another Kramers pair $\{\mxybo \ketr{\psm}, \tmy \mxybo \ketr{\psm}\}$.
Two state in this Kramers pair have $C_{2z}$ eigenvalue $\mp i$.
Thus, two Kramers pairs $\{\ketr{\psm}, \tmy \ketr{\psm}\}$ and $\{\mxybo \ketr{\psm}, \tmy \mxybo \ketr{\psm}\}$ are independent to each other because they have opposite $C_{2z}$ eigenvalues, and they form a fourfold degeneracy.
We can prove this statement by using $C_{2z} \mxybo = \mxybo C_{2z} \{\cm{E}|\mns1,1\}$: $C_{2z} [ \mxybo \ketr{\psm} ] = \mp i \mxybo \ketr{\psm}$.

In summary, the four energy eigenstates listed below form fourfold degeneracy.
\bg
\ketr{\psm}, \quad \tmy \ketr{\psm}, \quad
\mxybo \ketr{\psm}, \quad \tmy \mxybo \ketr{\psm}.
\label{eq:fourfold1}
\eg

Above, we show the fourfold degeneracy at $M$ by using the simultaneous eigenstate $\psm$ of $H$ and $C_{2z}$.
As in \sn~\ref{appsub:fourfoldIV}, one represent fourfold degeneracy using a simultaneous eigenstate $\ketr{\varphi_\pm(M)}$ of $H$ and $\mxybo$.
Then, $\{\ketr{\varphi_\pm(M)}, (\tmx-\tmy)\ketr{\varphi_\pm(M)}\}$ have $\mxybo$ eigenvalue $\pm i$, and $\{\mxyo \ketr{\varphi_\pm(M)}, (\tmx+\tmy)\ketr{\varphi_\pm(M)}\}$ have $\mxybo$ eigenvalue $\mp i$.
%%%%%%%%%%%%%%%%%%%%%%
\\

%%%%%%%%%%%%%%%%%%%%%%
{
\renewcommand{\arraystretch}{1.2}
\begin{table*}[b]
\centering
\begin{minipage}{\textwidth}
\caption{
The magnetic space groups (MSGs) that have magnetic wallpaper groups (MWGs) $p4'g'm$, $p'_c4mm$, and $p'_cmm$ on the (001) surface.
The first column denotes the MWGs and their symmetry generators, where $E$, $\cm{E}$ and the lattice translations are omitted.
Here, the symmetry elements are given by $M_{x,y}=\{m_{x,y}|\bb 0\}$, $\mxyb=\{m_{x\cm{y}}|\bb 0\}$, $\mxybo=\{m_{x\cm{y}}|\hf,\mhf,0\}$,  $\Tg=T\{E|\hf,\hf,0\}$, $\tmy=T\{m_y|\hf,\hf,0\}$, and $t_z=\{E|0,0,1\}$.
In the second column, the MSG number is given along with the SG symbol in the BNS notation~\cite{bradley2010}.
Since these MSGs can be obtained by adding to the MWGs additional generators, along with a translation generator in the (001) direction, we give these generators in the third column.
Note that 16 MSGs related to MWGs $p4'g'm$ and $p_c'4mm$ can host TMDIs when mirror Chern number $\mchd$ is nonzero (See \sns~\ref{app:mCherns} and \ref{app:Wilsonloop} for the definition of mirror Chern number $\mchd$).
}
\label{Table:MSGs}
\end{minipage}
\begin{tabular*}{0.85\textwidth}{c| @{\extracolsep{\fill}} cc}
\hline \hline
MWG (Generators) & MSG & Additional symmetry generators
\\ \hline
\multirowcell{11}{$p4'g'm$ \\ $(\mxybo, TC_{4z}, \tmy)$}
& 100.173 $P4'b'm$ & $t_z$ \\
& 101.186 $P_{I}4_2cm$ & $t_z, T\{E|\hf,\hf,\hf\}$ \\
& 102.192 $P_{c}4_2nm$ & $t_z, T\{E|0,0,\hf\}$ \\
& 125.366 $P4'/nb'm$ & $t_z, \{i|\hf,\hf,0\}$ \\
& 125.368 $P4'/n'b'm$ & $t_z, T\{i|\hf,\hf,0\}$ \\
& 127.390 $P4'/mb'm$ & $t_z, \{i|\bb 0\}$ \\
& 127.392 $P4'/m'b'm$ & $t_z, T\{i|\bb 0\}$\\
& 132.458 $P_{I}4_2/mcm$ & $t_z, T\{E|\hf,\hf,\hf\}, \{i|\bb 0\}$ \\
& 134.480 $P_{c}4_2/nnm$ & $t_z, T\{E|0,0,\hf\}, \{i|\hf,\hf,0\}$ \\
& 136.504 $P_{c}4_2/mnm$ & $t_z, T\{E|0,0,\hf\}, \{i|\bb 0\}$ \\
& 138.530 $P_{I}4_2/ncm$ & $t_z, T\{E|\hf,\hf,\hf\}, \{i|\hf,\hf,0\}$
\\ \hline
\multirowcell{5}{$p'_c4mm$ \\ $(M_x, \mxyb, C_{4z}, \Tg)$}
& 99.169 $P_{C}4mm$ & $t_z$ \\
& 107.232 $I_{c}4mm$ & $t_z, \{E|\hf,\hf,\hf\}$ \\
& 123.349 $P_{C}4/mmm$ & $t_z, \{i|\bb 0\}$ \\
& 129.421 $P_{C}4/nmm$ & $t_z, \{i|\hf,\hf,0\}$ \\
& 139.540 $I_{c}4/mmm$ & $t_z, \{E|\hf,\hf,\hf \}, \{i|\bb 0\}$
\\ \hline
\multirowcell{6}{$p'_cmm$ \\ $(M_x, M_y, \Tg)$}
& 25.63 $P_{C}mm2$ & $t_z$ \\
& 38.194 $A_Bmm2$ & $t_z, \{E|0,\hf,\hf \}$ \\
& 44.233 $I_cmm2$ & $t_z, \{E|\hf,\hf,\hf \}$ \\
& 47.255 $P_{C}mmm$ & $t_z, \{i|\bb 0\}$ \\
& 51.303 $P_{C}mma$ & $t_z, \{i|\hf,0,0\}$ \\
& 59.415 $P_{C}mmn$ & $t_z, \{i|\hf,\hf,0\}$ \\
& 71.538 $I_{c}mmm$ & $t_z, \{E|\hf,\hf,\hf \}, \{i| \bb{0} \}$ \\
& 74.561 $I_cmma$ & $t_z, \{E|\hf,\hf,\hf \}, \{i|0, \hf, 0\}$ \\
& 105.217 $P_{C}4_2mc$ & $t_z,  \{c_{4z}|0,0,\hf\}$ \\
& 109.244 $I_c4_1md$ & $t_z, \{c_{4z}|0,\hf,\frac{1}{4} \}$ \\
& 115.289 $P_{C} \cm{4}m2$ & $t_z, \{i c_{4z}|\bb 0\}$ \\
& 119.320 $I_c \cm{4}m2$ & $t_z, \{E|\hf,\hf,\hf \}, \{ic_{4z}|\bb{0} \}$\\
& 131.445 $P_{C}4_2/mmc$ & $t_z, \{c_{4z}|0,0,\hf\}$, $\{i|\bb 0\}$ \\
& 137.517 $P_{C}4_2/nmc$ & $t_z, \{c_{4z}|0,0,\hf\}$, $\{i|\hf,\hf,0\}$ \\
& 141.560 $I_c4_1/amd$ & $t_z, \{i|0,\hf,0\}, \{c_{4z}|0,\hf,\frac{1}{4} \}$
\\ \hline \hline
\end{tabular*}
\end{table*}
}
%%%%%%%%%%%%%%%%%%%%%%

%%%%%%%%%%%%%%%%%%%%%%
\tocless{\subsection{Magnetic space groups with (001)-surface MWGs $p'_cmm$, $p'_c4mm$, and $p4'g'm$}
\label{app:MSG}}{}
%\subsection{Magnetic space groups generated by $p'_cmm$, $p'_c4mm$, and $p4'g'm$} %\label{app:MSG}
%%%%%%%%%%%%%%%%%%%%%%
TMDIs proposed in this work have fourfold-degenerate Dirac fermions on the (001) surface.
To host a fourfold-degenerate Dirac fermion on the (001) surface, the (001) surface MWG must be one of the following three, $p'_cmm$, $p'_c4mm$, or $p4'g'm$.
We confirmed that at least 31 magnetic space groups (MSGs) satisfy this criterion, as listed in \stable~\ref{Table:MSGs}.
Among these 31 MSGs, 16 MSGs that have $p'4'g'm$ or $p'_c4mm$ on the (001) surface can be a TMDI, as explained in the main text and \sn~\ref{app:Wilsonloop}.
In this section, we briefly introduce a method for finding MSGs that have the desired MWG on their surface.

Let's first explain how to find MSGs with (001)-surface MWG $p'_c4mm$. Since the generators of MWGs $p'_c4mm$ are $M_x$, $\mxyb$, $C_{4z}$ and $T_G=T\{E|\hf,\hf,0\}$, the MSGs must have these four symmetry elements as generators. 
Due to $C_{4z}$ and $T_G=T\{E|\hf,\hf,0\}$, a Type-IV TMDI belongs to a Type-IV MSG with either a tetragonal or cubic unit cell.
Further, we require candidate MSGs to have two mirror operators $M_x$ and $\mxyb$.
Actually, it's easy to know the information of symmetry operators in each MSG by checking the BNS (or OG) symbol. 
We found that tetragonal MSGs with the BNS notations such as $P_C x mm$ or $I_c x mm$ can satisfy all the above conditions.
Here, in BNS setting $P_C$ means primitive lattice with antiunitary translation $T\{E|\hf,\hf,0\}$ and $I_c$ means body-center lattice with antiunitary translation $T\{E|0,0,\hf\}$ and $T\{E|\hf,\hf,0\}$. 
For a tetragonal lattice, the primary axis is [001], the secondary axis is $\{[100],[010]\}$ and the tertiary axis is $\{[110],[1\bar10]\}$.
Therefore, $xmm$ suggests that both $M_x$ and $\mxyb$ exist, and $x$ is any symmetry compatible with a tetragonal unit cell.
By checking the BNS notations and the detailed surface symmetries, we totally found 5 MSGs can host $p'_c4mm$ as (001)-surface MWG as shown in \stable~\ref{Table:MSGs}.
On the other hand, these MSGs can be understood by vertically stacking 2D layers with MWG $p'_c4mm$ supplemented by additional compatible symmetries.
For instance, 99.169 $P_C4mm$ can be obtained by simply stacking MWG $p'_c4mm$ along $z$ direction. MSG 123.349 $P_C4/mmm$ can be obtained by stacking MWG $p'_c4mm$ while keeping inversion symmetry.
In order to deeply understand the relationship between MSGs and MWGs, we list the additional symmetry generators for each MSGs in \stable~\ref{Table:MSGs}.
A similar process can be used to find MSGs with MWG $p'_cmm$. 
However, $\mxyb$ and $C_{4z}$ are absent in this case. So both orthorhombic and tetragonal lattices can have $p'_cmm$ as (001)-surface MWG.
Finally, we found 15 MSGs can host $p'_cmm$ on (001) surface as shown in \stable~\ref{Table:MSGs}.

Next, we employ the same method to find MSGs with (001)-surface MWG $p4'g'm$.
The most difference is there are no antiunitary translation symmetries in Type-III MWG $p4'g'm$.
Hence, Type-III MWG $p4'g'm$ can be (001)-surface MWG of Type-III MSGs or Type-IV MSGs, since the antiunitary translation of Type-IV MSGs can be broken on the surface.
Specifically, since the generators of MWGs $p4'g'm$ are $\mxybo, TC_{4z}, \tmy$, the MSGs must contain these three generators. 
Similar to the case $p'_c4mm$, $TC_{4z}$ make MSGs to describe tetragonal or cubic lattice.
Furthermore, for Type-III MSGs, we found the MSGs with BNS notation $Pxb’m$ satisfy all the above symmetry conditions, where $x$ is any symmetry compatible with a tetragonal unit cell, and $b'm$ means $\tmy$ and $\mxybo$.
For Type-IV MSGs, we found the MSGs with BNS notation $P_c xnm$ and $P_I xcm$ satisfy all the above symmetry requirements.
Here, $P_c$ and $P_I$ represent primitive cell with antiunitary translation $T\{E|0,0,\hf\}$ and $T\{E|\hf,\hf,\hf\}$, respectively.
%
%And perpendicular to secondary axis $\{[100],[010]\}$, $n$ represents $n$ glide mirror whose glide is along the half of a diagonal of a face and $c$ represents $c$ glide mirror whose glide is along $c$ axis.
And $n$ and $c$ represents glide mirrors perpendicular to secondary axis $\{[100],[010]\}$,
while $m$ represents a diagonal mirror.
Finally, we found at least 11 MSGs satisfy one of these three kinds of BNS notation, Type-III $Pxb’m$, Type-IV $P_c xnm$ or $P_I xcm$, as shown in \stable~\ref{Table:MSGs}.
Note that the list of possible MSGs in \stable~\ref{Table:MSGs} is not guaranteed to be exhaustive.
Other MSGs not listed in \stable~\ref{Table:MSGs} may have MWGs $p'_cmm$, $p'_c4mm$, and  $p4'g'm$ on 2D surfaces.
%%%%%%%%%%%%%%%%%%%%%%
\\

%%%%%%%%%%%%%%%%%%%%%%
\section{Mirror Chern numbers in MSGs related to MWGs $p'_cmm$, $p'_c4mm$, and $p4'g'm$}
\label{app:mCherns}
%%%%%%%%%%%%%%%%%%%%%%
In this section, we study the mirror Chern numbers and their symmetry constraints in the MSGs related to MWGs $p'_c4mm$, $p4'g'm$, and $p'_cmm$.
Throughout this work, we denote a Chern number of occupied bands with mirror eigenvalue $\pm i$ in $\Sigma$-plane as $\mch{\pm}{\Sigma}$.
Before we study the mirror Chern numbers in the MSGs under consideration, let us give a brief introduction to the mirror Chern number with an example.
Suppose an insulator with $n_{\rm occ}$ number of occupied bands has mirror symmetry $M_y=\{m_y|\bb 0\}$ which flips the sign of $y$ coordinate, i.e. $m_y(x,y,z)=(x,-y,z)$.
Then, $k_y=k^*$ plane ($k^*=0$ or $\pi$) is left invariant under $M_y$, and the occupied bands have well-defined mirror eigenvalues $\pm i$ in $k_y=k^*$ plane.
Hence, we can divide the eigenstates of occupied bands in $k_y=k^*$ plane into two mirror sectors $\pm i$ according to their mirror eigenvalues.
The occupied bands with eigenstates $\ket{u^+_n(k_x,k^*,k_z)}$ for $n=1,\dots,n_+$ belong to the mirror sector $+i$.
Similarly, the mirror sector $-i$ consists of the occupied bands with eigenstates $\ket{u^-_m(k_x,k^*,k_z)}$ for $m=1,\dots,n_-$.
Note that $n_++n_-=n_{\rm occ}$.
For each mirror sector, we define the (abelian) Berry connection $A^\pm_{n,i}(\bk^*)$ and the (abelian) Berry curvature $F^\pm_{n,ij}(\bk^*)$:
\ba
A^\pm_{n,i}(\bk^*) = -i \langle u^\pm_n(\bk^*)| \der_i | u^\pm_n(\bk^*) \rangle, \quad
F^\pm_{n,ij}(\bk^*) = \der_i  A^\pm_{n,j}(\bk^*) - \der_j A^\pm_{n,i}(\bk^*)
\label{eq:Berrys}
\ea
where $\bb k^*=(k_x,k^*,k_z)$, $\der_i=\der/\der k_i$, and $i,j=x,y,z$.
Then, we can define Chern numbers for each mirror sector as
\ba
\mch{\pm}{k_y=k^*}=\frac{1}{2\pi} \int_{-\pi}^{\pi} dk_z \int_{-\pi}^{\pi} dk_x \, \sum_{n=1}^{n_\pm} F^\pm_{n,zx}(k_x,k^*,k_z).
\label{eq:Cy}
\ea
Similarly, we can define the mirror Chern numbers $\mch{\pm}{k_x=k^*}$ and $\mch{\pm}{k_z=k^*}$ if there are relevant mirror symmetries.

Depending on MSGs, mirror $M_z=\{m_z| \bb 0\}$ can exist as in the case of MGSs 47.255 $P_Cmmm$, 71.538 $I_cmmm$, 123.349 $P_C4/mmm$, 127.390 $P4'/mb'm$, 131.445 $P_C4_2/mmc$, 132.458 $P_I4_2/mcm$, 136.504 $P_c4_2/mnm$, and 139.540 $I_c4/mmm$ (see \stable~\ref{Table:MSGs}).
In this work, we mainly study surface states on (001) surface.
For this, we focus on the mirror symmetries whose normal vectors or the Miller indices are orthogonal to $\hat{z}$, such as $M_x$, $M_y$, and $\mxy$, primarily.
$M_z$ and the relevant mirror Chern numbers $\mc{C}^{k_z=0,\pi}$ are important only when we discuss higher-order topology and hinge modes of topological magnetic Dirac insulators in open boundary along $x$ and $y$ directions.
It is because, when $M_z$ exists and $\mc{C}^{k_z=0,\pi}$ are nonzero, (anti)chiral edge modes or Dirac cones appear on $(100)$ and $(010)$ surfaces.
In such case, we cannot observe hinge modes in open boundary along $x$ and $y$ directions because the surface gap is closed due to (anti)chiral edge modes or twofold-degenerate Dirac cones.
Unless otherwise noted, we assume that $\mc{C}^{k_z=0,\pi}=0$ and focus on the mirror Chern numbers defined in $k_i=k^*$ planes ($i=x,y,z$ and $k^*=0,\pi$) or $k_x=\pm k_y$ planes.
%%%%%%%%%%%%%%%%%%%%%%

%%%%%%%%%%%%%%%%%%%%%%
\tocless{\subsection{MSGs related to MWG $p'_cmm$}
\label{appsub:mC_pcmm}}{}
%\subsection{MSGs related to MWG \texorpdfstring{$p'_cmm$}{p'_cmm}} \label{appsub:mC_pcmm}
%%%%%%%%%%%%%%%%%%%%%%
Fifteen MSGs are relevant to MWG $p'_cmm$, as listed in \stable~\ref{Table:MSGs}.
These MSGs have two orthogonal mirrors $M_x=\{m_x|\bb 0\}$ and $M_y=\{m_y|\bb 0\}$.
The mirrors $M_x$ and $M_y$ define 8 mirror Chern numbers, $\mch{\pm}{k_i=0}$ and $\mch{\pm}{k_i=\pi}$ with $i=x,y$.
$M_x$ give constraints on the mirror Chern numbers with respect to $M_y$ and vice versa.

Let us first explain how $M_y$ imposes the constraints on $\mch{\pm}{k_x=k_*}$ ($k_*=0,\pi$).
Since $M_x$ and $M_y$ anticommutes such that $M_x M_y = \cm{E} M_y M_x$, $M_y$ maps an occupied band with $M_x$ eigenvalue $\pm i$ at $(k_*,k_y,k_z)$ to another occupied band with $M_x$ eigenvalue $\mp i$ at $(k_*,-k_y,k_z)$.
Therefore, $\ket{u^\pm_n(k_*,k_y,k_z)}$ and $\ket{u^\mp_m(k_*,-k_y,k_z)}$ are related by $M_y$.
To show this, we introduce the Bloch wavefunctions corresponding to $\ketr{\psi^\pm_n(k_*,k_y,k_z)}=\sum_{\alpha=1}^{n_{\rm tot}} \ket{u^\pm_n(k_*,k_y,k_z)}_\alpha \ketr{k_*,k_y,k_z, \alpha}$. [see \eq{eqapp:blochwf}.]
Note that $M_x \ketr{\psi^\pm_n(k_*,k_y,k_z)} = \pm i \ketr{\psi^\pm_n(k_*,k_y,k_z)}$ since it is already assumed that $\ket{u^\pm_n(k_*,k_y,k_z)}$ is a $M_x$ eigenstate with eigenvalue $\pm i$.

Now we apply the identity $M_x M_y=\cm{E} M_y M_x$ to $\ketr{\psi^\pm_n(k_*,k_y,k_z)}$:
\bg
M_x \left[ M_y \ketr{\psi^\pm_n(\bk^*)} \right]
= \cm{E} M_y M_x \ketr{\psi^\pm_n(\bk^*)}
= \mp i M_y \ketr{\psi^\pm_n(\bk^*)}
\eg
where $\bk^*=(k_*,k_y,k_z)$.
Thus, we can identify $M_y \ketr{\psi^\pm_n(\bk^*)}$ as $\ketr{\psi^\mp_m(M_y\bk^*)}$ whose $M_x$ eigenvalue is $\mp i$.
The same conclusion holds for the tight-binding eigenstates $\ket{u^\pm_n(\bk^*)}$ as well.
Therefore, we conclude that $n_+=n_-=n_{\rm occ}/2$ and $\sum_{n=1}^{n_+} \, F^+_{n,yz}(k_*,k_y,k_z)= - \sum_{n=1}^{n_-} \, F^-_{n,yz}(k_*,-k_y,k_z)$ since the Berry curvature $F_{yz}(\bk)$ is transformed to $-F_{yz}(M_y\bk)$ by $M_y$.
Hence, we obtain $\mch{+}{k_x=k_*}=-\mch{-}{k_x=k_*}$.
Note that for unitary symmetry $\sg$, which acts on momentum $\bk$ as $\bk \to O_{\sg} \bk$ with an orthogonal matrix $O_\sg$, $\sg$ transforms the Berry curvature as $F_{ij}(\bk)=F_{kl}(O_\sg \bk) [O_\sg]_{ki} [O_\sg]_{lj}$.
For antiunitary symmetry, similar equation holds with an additional minus sign.

Now, we consider the symmetry constraint by $\Tg=\{T|\hf,\hf,0\}$ and further show that $\mch{\pm}{k_x=\pi} = 0$.
The identity $M_x \Tg=\Tg M_x \{E|\bb a_1\}$ gives
\bg
M_x \left[ \Tg \ketr{\psi^\pm_n(\bk^*)} \right]
= \Tg M_x \{E|\bb a_1\} \ketr{\psi^\pm_n(\bk^*)}
= \Tg (\pm i e^{-i \bk^* \cdot \bb a_1}) \ketr{\psi^\pm_n(\bk^*)}
= \mp i e^{i k^*} \Tg \ketr{\psi^\pm_n(\bk^*)}
\label{eq:MxTrel}
\eg
Hence, $\Tg \ketr{\psi^\pm_n(\bk^*)}$ can be identified as another Bloch wavefunction at $\Tg\bk^*$, which is an eigenstate of $M_x$ with eigenvalue $\mp i e^{ik^*}$.
This implies $\sum_{n=1}^{n_\pm} \, F^\pm_{n,yz}(\bk^*) = - \sum_{n=1}^{n_\pm} \, F^\pm_{n,yz}(\Tg\bk^*)$ at $k^*=\pi$.
Thus, $\mch{\pm}{k_x=\pi} =0$.

Similarly, we can also derive $\mch{+}{k_y=k_*}=-\mch{-}{k_y=k_*}$ and $\mch{\pm}{k_y=\pi}=0$ by exchanging the roles of $M_x$ and $M_y$.
Hence, we have two independent mirror Chern numbers $\mc{C}_m^x \equiv \mch{+}{k_x=0} = -\mch{-}{k_x=0}$ and $\mc{C}_m^y \equiv \mch{+}{k_y=0} = - \mch{-}{k_y=0}$.
These two mirror Chern numbers are equal modulo 2, i.e. $\mc{C}_m^x = \mc{C}_m^y$ (mod 2), as shown in \sn~\ref{app:Wilsonloop}.
%%%%%%%%%%%%%%%%%%%%%%
\\

%%%%%%%%%%%%%%%%%%%%%%
\tocless{\subsection{MSGs related to MWG $p'_c4mm$}
\label{appsub:mC_pc4mm}}{}
%\subsection{MSGs related to MWG \texorpdfstring{$p'_c4mm$}{p'_c4mm}}
%\label{appsub:mC_pc4mm}
%%%%%%%%%%%%%%%%%%%%%%
A similar analysis can be applied to 5 MSGs that are relevant to MWG $p'_c4mm$.
They are MSGs 99.169 $P_C4mm$, 107.232 $I_c4mm$, 123.349 $P_C4/mmm$, 129.421 $P_C4/nmm$, and 139.540 $I_c4/mmm$, as listed in \stable~\ref{Table:MSGs}.
All these MSGs have $C_{4z}=\{c_{4z}|\bb 0\}$, $\Tg=\{T|\hf,\hf,0\}$, $M_x=\{m_x|\bb 0\}$, $M_y=\{m_y|\bb 0\}$, $\mxyb=\{m_{x\cm{y}}| \bb 0\}$, and $\mxy=\{m_{xy}|\bb 0\}$.
Hence, we can define 12 mirror Chern numbers with respect to $M_x$, $M_y$, $\mxyb$, and $\mxy$: $\mch{\pm}{k_i=0}$, $\mch{\pm}{k_i=\pi}$, $\mch{\pm}{k_x=k_y}$, $\mch{\pm}{k_x=-k_y}$ ($i=x,y$).
Note that the mirror Chern numbers $\mch{\pm}{k_x=k_y}$ and $\mch{\pm}{k_x=-k_y}$ are defined as
\bg
\mch{\pm}{k_x=sk_y}=\frac{1}{2\pi} \int_{-\sqrt{2}\pi}^{\sqrt{2}\pi} dk_s \int_{-\pi}^{\pi} dk_z \, \sum_{n=1}^{n_\pm} F^\pm_{n,zs}(\bk),
\eg
where $k_\pm=\frac{1}{\sqrt{2}}(k_x \pm k_y)$ and $F^\pm_{n,zs}=\der_z \langle u^\pm_n|\der_s|u^\pm_n \rangle-\der_s \langle u^\pm_n|\der_z|u^\pm_n \rangle$ with $s=\pm$ and $\der_s=\der/\der k_s$.
Here, we omit the arguments in $F^\pm_{n,sz}$ and $\ket{u^\pm_n}$ for simplicity.
Note that $\sqrt{2} F^\pm_{n,z+}(\bk)=F^\pm_{n,zx}(\bk)-F^\pm_{n,yz}(\bk)$ and $\sqrt{2}F^\pm_{n,z-}=F^\pm_{zx}(\bk)+F^\pm_{yz}(\bk)$.

Since $p'_c4mm$ have all the symmetry elements in $p'_cmm$, we can use the results obtained in \sn~\ref{appsub:mC_pcmm}.
Hence, $\mch{+}{k_i=0}=-\mch{-}{k_i=0}$ and $\mch{\pm}{k_i=\pi}=0$.
Now, let us consider symmetry constraints by additional symmetry elements such as $C_{4z}$, $\mxyb$, and $\mxy$.
First, let us consider $\mxyb$ and $\mxy$.
Since they anticommute to each other, we obtain $\mch{+}{k_x=k_y} = -\mch{-}{k_x=k_y}$ and $\mch{+}{k_x=-k_y} = -\mch{-}{k_x=-k_y}$.
Second, $C_{4z}$ imposes $\mch{\pm}{k_x=0} = \mch{\pm}{k_y=0}$ and $\mch{\pm}{k_x=k_y}=-\mch{\pm}{k_x=-k_y}$.
It is  because $M_{xy}C_{4z} = C_{4z} \mxyb$ and $M_yC_{4z}  = C_{4z} M_x$.
Hence, we have two independent mirror Chern numbers $\mchd \equiv \mch{+}{k_x=k_y}=-\mch{-}{k_x=k_y}$ and $\mc{C}_m^y \equiv \mch{+}{k_y=0}=\mch{-}{k_y=0}$.
These two mirror Chern numbers are equal modulo 2, i.e. $\mchd = \mc{C}_m^y$ (mod 2).
We prove this relation using the Wilson loop method in \sn~\ref{app:Wilsonloop}.
%%%%%%%%%%%%%%%%%%%%%%
\\

%%%%%%%%%%%%%%%%%%%%%%
\tocless{\subsection{MSGs related to MWG $p4'g'm$}
\label{appsub:mC_p4gm}}{}
%\subsection{MSGs related to MWG \texorpdfstring{$p4'g'm$}{p4'g'm}} 
%\label{appsub:mC_p4gm}
%%%%%%%%%%%%%%%%%%%%%%
Eleven MSGs are relevant to MWG $p4'g'm$, as listed in \stable~\ref{Table:MSGs}.
These MSGs have antiunitary fourfold rotation $TC_{4z}=T\{c_{4z}|\bb 0\}$ and two off-centered mirrors $\mxybo=\{m_{x\cm{y}}|\hf,\mhf,0\}$ and $\mxyo=\{m_{xy}|\hf,\hf,0\}$.
Hence, we can define four mirror Chern numbers $\mch{\pm}{k_x=k_y}$ and $\mch{\pm}{k_x=-k_y}$.

Now, let us show that only one mirror Chern number is independent.
Since $\mxybo$ and $\mxyo$ anticommute, we obtain $\mch{+}{k_x=k_y}=-\mch{-}{k_x=k_y}$ and $\mch{+}{k_x=-k_y}=-\mch{-}{k_x=-k_y}$, as in \sn~\ref{appsub:mC_pc4mm}.
Then, we consider the symmetry constraint by $TC_{4z}$.
The identity $\mxyo TC_{4z} = TC_{4z} \mxybo$ leads to $\sum_{n=1}^{n_+} \, F^+_{n,z+}(k_x,k_x,k_z)=\sum_{n=1}^{n_-} \, F^-_{n,z-}(k_x,-k_x,-k_z)$, which implies $\mch{+}{k_x=k_y}=\mch{-}{k_x=-k_y}$.
In summary, all four mirror Chern numbers are equal to $\mchd \equiv \mch{+}{k_x=k_y}=-\mch{-}{k_x=k_y}$ up to sign.
%%%%%%%%%%%%%%%%%%%%%%
\\

%%%%%%%%%%%%%%%%%%%%%%
\section{Wilson loop spectra in MSGs related to MWGs, $p'_cmm$, $p'_c4mm$, and $p4'g'm$}
\label{app:Wilsonloop}
%%%%%%%%%%%%%%%%%%%%%%
%
In this section, we study the Wilson loop spectra for 31 MSGs whose (001) surfaces have one of MWGs $p'_cmm$, $p'_c4mm$, and $p4'g'm$.
We follow the notation and results in Ref.~\cite{alexandradinata2016topological} closely.
First, we introduce the discretized definition of Wilson loop $W_z(\bkp)$~\cite{yu2011equivalent},
\ba
[W_z(\bkp)]_{nm}
=& \sum_{n_1,\dots,n_{N-1}=1}^{n_{\rm occ}} \brk{u_n(\bkp,\pi)}{u_{n_{N-1}}(\bkp,k_{N-1})} \brk{u_{n_{N-1}}(\bkp,k_{N-1})}{u_{n_{N-2}}(\bkp,k_{N-2})} \bra{u_{n_{N-2}}(\bkp,k_{N-2})} \nn
& \cdots \ket{u_{n_2}(\bkp,k_2)} \brk{u_{n_2}(\bkp,k_2)}{u_{n_1}(\bkp,k_1)} \brk{u_{n_1}(\bkp,k_1)}{u_m(\bkp,-\pi)} \nn
=& \bra{u_n(\bkp,\pi)} \prod_{k_z}^{\pi \leftarrow -\pi} P_{\rm occ}(\bkp,k_z) \ket{u_n(\bkp,-\pi)}, 
\label{eqapp:discreteWL}
\ea
where $n,m=1,\dots,n_{\rm occ}$, $\bkp = (k_x,k_y)$, and $P_{\rm occ}(\bk) = \sum_{n=1}^{n_{\rm occ}} \kbr{u_n(\bk)}{u_n(\bk)}$ is a projection operator into occupied-band subspace.
Here, we discretize $k_z$ such that $k_i = -\pi+ 2\pi i/N$ for $i =1,\dots,N-1$.
Wilson loop in Eq.~\eqref{eqapp:discreteWL} is periodic in the BZ, i.e. $W_z(\bkp + \bb g_1) = W_z(\bkp + \bb g_2) = W_z(\bkp)$ for $\bb g_1=(2\pi,0)$ and $\bb g_2=(0,2\pi)$, because of the periodic gauge.
In $N \to \infty$ limit, $W_z(\bkp)$ is unitary and thus its eigenvalues are unimodular.
We label them as $\{e^{i \theta_a(\bkp)}\}$ with $a = 1, \dots, n_{{\rm occ}}$, i.e. 
\bg
W_z(\bkp) \ket{\theta_a(\bkp)} = e^{i \theta_a(\bkp)} \ket{\theta_a(\bkp)}.
\eg
Each phase $\theta_a(\bkp) \in (-\pi,\pi]$ of Wilson loop eigenvalues is continuous in $\bkp$, and $\{\theta_a(\bkp)\}$ defines the Wilson loop spectrum or the Wilson bands.

Now we consider unitary symmetry $\sg: \bb r \to \sg \bb r = O_\sg \bb r + \bb \delta_\sg$, $\bk \to \sg \bk = O_\sg \bk$.
For $\sg$, one can obtain symmetry transformation of the Wilson loop~\cite{alexandradinata2014wilson,alexandradinata2016topological,benalcazar2017electric,wieder2018wallpaper,hwang2019fragile},
\bg
W^\sg_z(\bkp) = e^{2\pi i(O_\sg^{-1} \bb \delta_\sg)_z} B_\sg(\bkp, -\pi) W_z(\bkp) B_\sg(\bkp, -\pi)^{-1},
\label{eqapp:unitaryTrWL}
\eg
where we use the definition and the periodicity of sewing matrix (see Eq.~\eqref{eqapp:sewB}).
The symmetry-transformed Wilson loop $W^\sg_z(\bkp)$ is defined along a loop $O_\sg(\bkp, -\pi) \to O_\sg(\bkp, \pi)$ whereas the Wilson loop $W_z(\bkp)$ is defined along a loop $(\bkp, -\pi) \to (\bkp, \pi)$.

For antiunitary symmetry $\sg': \bb r \to \sg' \bb r = O_{\sg'} \bb r + \bb \delta_{\sg'}$, $\bk \to \sg' \bk = O_{\sg'} \bk$, we can also obtain symmetry transformation of the Wilson loop similarly,
\bg
W^{\sg'}_z(\bkp) = e^{2\pi i(O_{\sg'}^{-1} \bb \delta_{\sg'})_z} B^{\mc{K}}_{\sg'}(\bkp, -\pi) W_z(\bkp) B^{\mc{K}}_{\sg'}(\bkp, -\pi)^{-1},
\label{eqapp:antiunitaryTrWL}
\eg
where $W^{\sg'}_z(\bkp)$ is defined along a loop $O_{\sg'}(\bkp, -\pi) \to O_{\sg'}(\bkp, \pi)$.
Most of the symmetries discussed in this work have $[O_{\sg}^{-1} \bb \delta_{\sg}]_z=0$ or $[O_{\sg'}^{-1} \bb \delta_{\sg'}]_z=0$.
For such symmetries, the phase factors in \eqs{eqapp:unitaryTrWL} and \eqref{eqapp:antiunitaryTrWL} are simply $1$.

From now on, we discuss the relation between the winding numbers in the Wilson loop spectrum and Chern number.
For this purpose, we rewrite the Wilson loop  in \eq{eqapp:discreteWL} in continuous form:
\bg
W_z(\bkp) = P \left[ \exp -i \int_{-\pi}^\pi dk_z \, \mc{A}_z(\bk) \right], \quad
\mc{A}_{nm,z}(\bk) = -i \bra{u_n(\bk)} \der_z \ket{u_m(\bk)}.
\label{eqapp:contWL}
\eg
The symbol $P$ is the path ordering and $\mc{A}_z(\bk)$ is the nonabelian Berry connection.
From \eq{eqapp:contWL}, we can relate the sum of Wilson loop eigenvalues and the Berry connection:
\bg
\sum_{a=1}^{n_{\rm occ}} \theta_a(\bkp)
= -i {\rm Tr} \log W_z(\bkp)
= - \int_{-\pi}^\pi dk_z {\rm Tr_{occ}} \mc{A}_z(\bkp,k_z)
= -\int_{-\pi}^\pi dk_z \sum_{n=1}^{n_{\rm occ}} A_{n,z}(\bkp,k_z).
\label{eqapp:relWLeigenConnection}
\eg
Chern number can be directly related to the Wilson-loop winding number~\cite{alexandradinata2014wilson}.
For example, let us consider the Chern number in $k_y=k^*$ plane ($k^* \in \{0,\pi\}$).
One can always choose a continuous and periodic gauge in the $k_z$ direction for the eigenstate~\cite{soluyanov2012smooth}.
In such gauge, the Chern number $\mc{C}^{k_y=k^*}$ is expressed as
\bg
\mc{C}^{k_y=k^*}
= - \frac{1}{2\pi} \int_{-\pi}^\pi dk_z \int_{-\pi}^\pi dk_x \sum_{n=1}^{n_{\rm occ}} \der_x A_{n,z}(k_x,k^*,k_z)
= \frac{1}{2\pi} \int_{-\pi}^\pi dk_x \sum_{a=1}^{n_{\rm occ}} \der_x \theta_a(k_x,k^*),
\label{eqapp:ChernWinding1}
\eg
where \eq{eqapp:relWLeigenConnection} is used in the last equality.
The last expression in Eq.~\eqref{eqapp:ChernWinding1} is equal to the winding number of $\theta_a(k_x,k^*)$ along $k_x$ direction.
We define the winding number $\Delta N_{(-\pi,k^*)\to(\pi,k^*)}$ as follows.
First, set a reference horizontal line, i.e. constant $\theta(k_x,k^*)$, in the Wilson loop spectrum.
Then, count $N^{>0}_{(-\pi,k^*)\to(\pi,k^*)}$ which is the number of Wilson bands that cross the reference line with positive slope, i.e. $\der \theta(k_x,k^*)/\der k_x>0$.
Similarly, count the number of Wilson bands with negative slope, $N^{<0}_{(-\pi,k^*)\to(\pi,k^*)}$.
Then, the winding number is defined by $\Delta N_{(-\pi,k^*)\to(\pi,k^*)}=N^{>0}_{(-\pi,k^*)\to(\pi,k^*)} - N^{<0}_{(-\pi,k^*)\to(\pi,k^*)}$.
Hence, we conclude that
\bg
\mc{C}^{k_y=k^*}
= \Delta N_{(-\pi,k^*)\to(\pi,k^*)}
= N^{>0}_{(-\pi,k^*)\to(\pi,k^*)} - N^{<0}_{(-\pi,k^*)\to(\pi,k^*)}
\label{eqapp:ChernWinding2}
\eg
For the other mirror Chern numbers, we can derive similar relations by further considering the mirror eigenvalues of Wilson bands~\cite{wang2016hourglass,alexandradinata2016topological}, as discussed in the following sections.
%%%%%%%%%%%%%%%%%%%%%%

%%%%%%%%%%%%%%%%%%%%%%
\tocless{\subsection{Wilson loop spectra in MSGs related to MWG $p'_cmm$}
\label{appsub:WLpcmm}}{}
%\subsection{Wilson loop spectra in MSGs related to MWG $p'_cmm$}
%\label{appsub:WLpcmm}
%%%%%%%%%%%%%%%%%%%%%%
In this section, we discuss band degeneracies and winding structures in the Wilson loop spectra in MSGs related to MWG $p'_cmm$.
Representative Wilson loop spectra are shown in Supplementary Figure (SFig)~\ref{sfig1}.
Wilson bands, i.e. continuous sets of Wilson loop eigenvalues, with opposite mirror eigenvalues are twofold degenerate at $\bG$, $\bX$, and $\bY$.
Along $\bX$-$\bM$ line and $\bY$-$\bM$, two Wilson bands with opposite mirror eigenvalues are degenerate.
At $M$, fourfold degeneracy is formed by two Wilson bands in mirror sector $+i$ and two in mirror sector $-i$.
Along $\bG$-$\bX$ ($\bY$-$\bG$) line, the winding of Wilson bands determines the mirror Chern number $\mc{C}_m^y$ ($\mc{C}_m^x$) defined in \sn~\ref{app:mCherns}.
Note that we label the high-symmetry points in the (001) surface BZ: $\bG=(0,0)$, $\bX=(\pi,0)$, $\bY=(0,\pi)$, and $\bM=(\pi,\pi)$.
Below, we provide the proofs of all these properties.
%%%%%%%%%%%%%%%%%%%%%%
\\

%%%%%%%%%%%%%%%%%%%%%%
\tocless{\subsubsection{Twofold degeneracy at $\bG$, $\bX$, and $\bY$}}{}
%%%%%%%%%%%%%%%%%%%%%%
The twofold degeneracy of Wilson bands is protected by two orthogonal mirros $M_x=\{m_x|\bb 0\}$ and $M_y=\{m_y|\bb 0\}$.
First, let us consider the symmetry transformations of Wilson loop under $M_x$ and $M_y$ at $\bkp^* \in \{\bG,\bX,\bY,\bM\}$, which can be obtained by using Eq.~\eqref{eqapp:unitaryTrWL}:
\bg
W_z(\bkp^*) = B_{M_{x,y}}(\bkp^*,-\pi) W_z(\bkp^*) B_{M_{x,y}}(\bkp^*,-\pi)^{-1}.
\label{eq:app_WL_pcmm_1}
\eg
Since $B_{M_{x,y}}(\bkp^*,-\pi)$ commutes with $W_z(\bkp^*)$, we consider a simultaneous eigenstates$\ket{\theta_\pm(\bkp^*)}$ of $B_{M_x}(\bkp^*,-\pi)$ and $W_z(\bkp^*)$ such that $B_{M_x}(\bkp^*,-\pi) \ket{\theta_\pm(\bkp^*)} = \pm i \ket{\theta_\pm(\bkp^*)}$ and $W_z(\bkp^*) \ket{\theta_\pm(\bkp^*)} = e^{i \theta(\bkp^*)} \ket{\theta_\pm(\bkp^*)}$.

Now, let us use the fact that $M_{x}$ and $M_{y}$ anticommutes to each other, i.e. $M_x M_y = \cm{E} M_y M_x$.
The sewing matrices $B_\sg(\bk)$ and $B^{\mc{K}}_{\sg'}(\bk)$ respect the group multiplication faithfully, $B_{M_x}(\bkp^*,-\pi) B_{M_y}(\bkp^*,-\pi) = - B_{M_y}(\bkp^*,-\pi) B_{M_x}(\bkp^*,-\pi)$.
Thus, we can deduce that $B_{M_y}(\bkp^*,-\pi) \ket{\theta_\pm(\bkp^*)}$ has $M_x$ eigenvalue $\mp i$:
\bg
B_{M_x}(\bkp^*,-\pi) B_{M_y}(\bkp^*,-\pi) \ket{\theta_\pm(\bkp^*)}
= \mp i B_{M_y}(\bkp^*,-\pi) \ket{\theta_\pm(\bkp^*)}.
\eg
Also, \eq{eq:app_WL_pcmm_1} implies that $B_{M_y}(\bkp^*,-\pi) \ket{\theta_\pm(\bkp^*)}$ has $W_z(\bkp^*)$ eigenvalue $e^{i\theta(\bkp^*)}$.
Hence, the Wilson bands with opposite $M_x$ eigenvalues form twofold degeneracy at $\bG$, $\bX$, and $\bY$.
(Note that $\bM$ exhibits fourfold degeneracy.)
If we set a simultaneous eigenstates of $W_z(\bkp^*)$ and $B_{M_y}(\bkp^*)$ (rather than $B_{M_x}(\bkp^*))$, we can conclude that the twofold degeneracies are formed by Wilson bands with opposite $M_y$ eigenvalues.
%%%%%%%%%%%%%%%%%%%%%%
\\

%%%%%%%%%%%%%%%%%%%%%%
\tocless{\subsubsection{Twofold degeneracy in $\bX$-$\bM$ and $\bY$-$\bM$ lines}}{}
%%%%%%%%%%%%%%%%%%%%%%
Let us explain the twofold degeneracy in $\bX$-$\bM$ line first.
The $\bX$-$\bM$ line is invariant under  $M_x$ and $\tmy=T\{m_y|\hf,\hf,0\}$.
Note that antiunitary symmetry $\tmy$ is a combination of $M_y$ and $\Tg=T\{E|\hf,\hf,0\}$, i.e. $\tmy=\Tg M_y$.
We denote generic momenta in the $\bX$-$\bM$ line as $\bkp^*=(\pi,k_y)$.
Symmetry transformation of the Wilson loop by $M_x$ leads to
\bg
[B_{M_x}(\bkp^*,-\pi), W_z(\bkp^*)]=0.
\eg
Thus, we consider a simultaneous eigenstate $\ket{\theta_\pm(\bkp^*)}$ of them such that $B_{M_x}(\bkp^*,-\pi) \ket{\theta_\pm(\bkp^*)} = \pm i \ket{\theta_\pm(\bkp^*)}$ and $W_z(\bkp^*) \ket{\theta_\pm(\bkp^*)} = e^{i\theta(\bkp^*)} \ket{\theta_\pm(\bkp^*)}$.
Now, let us consider the symmetry transformation of Wilson loop under $\tmy$,
\bg
B^{\mc{K}}_{\tmy}(\bkp^*,-\pi) W_z(\bkp^*) B^{\mc{K}}_{\tmy}(\bkp^*,-\pi)^{-1} = W_z(\bkp^*)^\dg,
\label{eq:app_WL_pcmm_2}
\eg
according to Eq.~\eqref{eqapp:antiunitaryTrWL}.
In Eq.~\eqref{eq:app_WL_pcmm_2}, the Wilson loop $W_z(\bkp^*)$ is mapped to its hermitian conjugate $W_z(\bkp^*)^\dg$ by $\tmy$ because $\tmy$ reverses the sign of $k_z$.
Conjugating \eq{eq:app_WL_pcmm_2}, we obtain $W_z(\bkp^*) B^{\mc{K}}_{\tmy}(\bkp^*,-\pi) = B^{\mc{K}}_{\tmy}(\bkp^*,-\pi) W_z(\bkp^*)^\dg$.
Using this relation we can show that $B^{\mc{K}}_{\tmy}(\bkp^*,-\pi) \ket{\theta(\bkp^*)}$ has $W_z(\bkp^*)$ eigenvalue $e^{i\theta(\bkp^*)}$:
\bg
W_z(\bkp^*) \left[ B^{\mc{K}}_{\tmy}(\bkp^*,-\pi) \ket{\theta(\bkp^*)} \right] = B^{\mc{K}}_{\tmy}(\bkp^*,-\pi) W_z(\bkp^*)^\dg \ket{\theta(\bkp^*)} = e^{i\theta(\bkp^*)} \left[ B^{\mc{K}}_{\tmy}(\bkp^*,-\pi) \ket{\theta(\bkp^*)} \right].
\eg
In the second equality, $W_z(\bkp^*) ^\dg \ket{\theta_\pm(\bkp^*)} = e^{-i\theta(\bkp^*)} \ket{\theta_\pm(\bkp^*)}$ is used.

Furthermore, we can show that $\ket{\theta(\bkp^*)}$ and $B^{\mc{K}}_{\tmx}(\bkp^*,-\pi) \ket{\theta(\bkp^*)}$ form a Kramers pair at every $\bkp^*$.
To show this, we first note that $(\tmy)^2 = \{E|\bb a_1\}$ and the corresponding relation for the sewing matrices, $[B^{\mc{K}}_{\tmy}(\bkp^*,-\pi)]^2=e^{-i (\bkp^*,-\pi) \cdot \bb a_1} \mathds{1}=-\mathds{1}$.
Since $B^{\mc{K}}_{\tmy}(\bkp^*,-\pi)$ is antiunitary and its squared is $-\mathds{1}$, one can prove that
\bg
\bra{\theta_\pm(\bkp^*)} B^{\mc{K}}_{\tmy}(\bkp^*,-\pi) \ket{\theta_\pm(\bkp^*)} = 0,
\label{eq:app_WL_pcmm_3}
\eg
in a way similar to \eq{eqapp:antiKramers}.
Thus, $\ket{\theta(\bkp^*)}$ and $B^{\mc{K}}_{\tmy}(\bk^*) \ket{\theta(\bkp^*)}$ are independent to each other.

Now, let us show that the two states forming the Kramers pair $\{\ket{\theta(\bkp^*)}, B^{\mc{K}}_{\tmy}(\bk^*,-\pi) \ket{\theta(\bkp^*)}\}$ have opposite $M_x$ eigenvalues.
We start the derivation from noting that $M_x \tmy = \tmy M_x \{\cm{E}|\bb a_1\}$ which implies $B_{M_x}(\bkp^*,-\pi) B^{\mc{K}}_{\tmy}(\bkp^*,-\pi) = B^{\mc{K}}_{\tmy}(\bkp^*,-\pi) B_{M_x}(\bkp^*,-\pi)$.
Using this relation, we can show that
\bg
B_{M_x}(\bkp^*,-\pi) B^{\mc{K}}_{\tmy}(\bkp^*,-\pi) \ket{\theta(\bkp^*)}
= \mp i B^{\mc{K}}_{\tmy}(\bkp^*,-\pi) \ket{\theta(\bkp^*)}.
\eg
Thus, the Wilson bands are twofold degenerate along $\bX$-$\bM$ line.

The twofold degeneracy along $\bY$-$\bM$ line can be explained similarly by simply exchanging the roles of $M_x$ and $M_y$.
%%%%%%%%%%%%%%%%%%%%%%
\\

%%%%%%%%%%%%%%%%%%%%%%
\tocless{\subsubsection{Fourfold degeneracy at $\bM$}}{}
%%%%%%%%%%%%%%%%%%%%%%
Wilson bands form fourfold degeneracy at $\bM$, which corresponds to fourfold-degenerate Dirac fermion on (001) surface.
To prove this, let us consider $C_{2z}$, $M_x$, $M_y$, and $\Tg=T\{E|\hf,\hf,0\}$.
These symmetries lead to
\bg
[W_z(\bM), B_{C_{2z}}(\bM,-\pi)] = 0, \quad
[W_z(\bM), B_{M_{x,y}}(\bM,-\pi)] = 0, \quad
W_z(\bM) B^{\mc{K}}_{\Tg}(\bM,-\pi) = B^{\mc{K}}_{\Tg}(\bM,-\pi) W_z(\bM)^\dg.
\label{eq:app_WL_pcmm_4}
\eg
First, let us consider a simultaneous eigenstate of $W_z(\bM)$ and $B_{M_x}(\bM,-\pi)$ such that $W_z(\bM) \ket{\theta_\pm(\bM)} = e^{i\theta(\bM)} \ket{\theta_\pm(\bM)}$ and $B_{M_x}(\bM,-\pi) \ket{\theta_\pm(\bM)} = \pm i\ket{\theta_\pm(\bM)}$.
The sewing matrix $B^{\mc{K}}_{\Tg}(\bM,-\pi)$ is antiunitary and its squared is $-\mathds{1}$ because $(\Tg)^2=\{\cm{E}|1,1,0\}$.
Thus, $\ket{\theta_\pm(\bM)}$ and $B^{\mc{K}}_{\Tg}(\bM,-\pi) \ket{\theta_\pm(\bM)}$ form a Kramers pair, as can be shown in a way similar with \eq{eq:app_WL_pcmm_3}.
The two states in this Kramers pair have the same $M_x$ eigenvalue, since $M_x \Tg = \Tg M_x \{E|\bb a_1\}$ and thus $B_{M_x}(\bM,-\pi) B^{\mc{K}}_{\Tg}(\bM,-\pi) = -  B^{\mc{K}}_{\Tg}(\bM,-\pi) B_{M_x}(\bM,-\pi)$:
\bg
B_{M_x}(\bM,-\pi) B^{\mc{K}}_{\Tg}(\bM,-\pi) \ket{\theta_\pm(\bM)}
= \pm i B^{\mc{K}}_{\Tg}(\bM,-\pi) \ket{\theta_\pm(\bM)}.
\eg

We can find another Kramers pair $\{B_{M_y}(\bM,-\pi) \ket{\theta_\pm(\bM)}, B^{\mc{K}}_{\Tg}(\bM,-\pi) B_{M_y}(\bM,-\pi) \ket{\theta_\pm(\bk)}\}$ which have $W_z(\bM)$ eigenvalue $e^{i\theta(\bM)}$ and $B_{M_x}(\bM,-\pi)$ eigenvalue $\mp i$.
It is because $M_x M_y = \cm{E} M_y M_x$ and $B_{M_x}(\bM,-\pi) B_{M_y}(\bM,-\pi) = - B_{M_y}(\bM,-\pi) B_{M_x}(\bM,-\pi)$.
Therefore, two Kramers pairs,
\bg
\{\ket{\theta_\pm(\bM)}, B^{\mc{K}}_{\Tg}(\bM,-\pi), \ket{\theta_\pm(\bk)}\}, \quad
\{B_{M_y}(\bM,-\pi) \ket{\theta_\pm(\bM)}, B^{\mc{K}}_{\Tg}(\bM,-\pi) B_{M_y}(\bM,-\pi) \ket{\theta_\pm(\bk)}\},
\eg
form fourfold degeneracy at $\bM$.
Each fourfold degeneracy is formed by two Wilson bands in mirror sector $+i$ and two in mirror sector $-i$.
%%%%%%%%%%%%%%%%%%%%%%
\\

%%%%%%%%%%%%%%%%%%%%%%
\begin{figure*}[t!]
\centering
\includegraphics[width=0.98\textwidth]{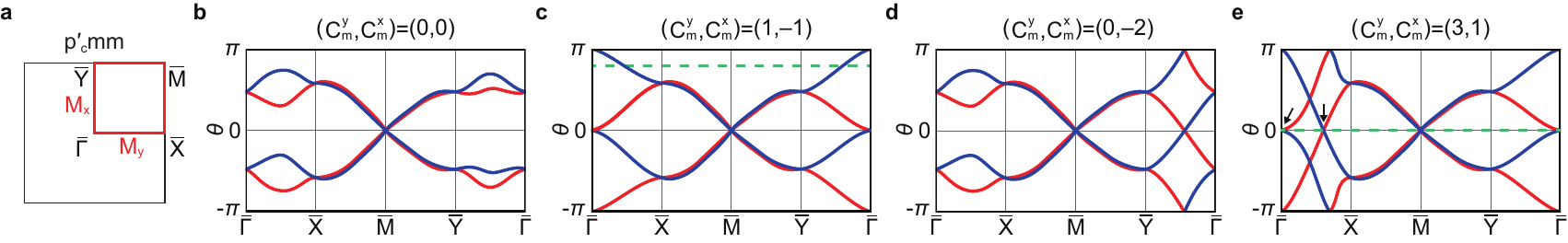}
\caption{
{\bf Wilson loop spectra in MSGs related to MWG $p'_cmm$.}
(a) (001) surface BZ.
(b)-(d) Wilson loop spectra $\{\theta(k_x,k_y)\}$ are characterized by mirror Chern numbers $\mc{C}_m^x=\mch{+}{k_x=0}$ and $\mc{C}_m^y=\mch{+}{k_y=0}$.
Note that $\mch{+}{k_i=0}=-\mch{-}{k_i=0}$ and $\mch{\pm}{k_i=\pi}=0$ for $i=x,y$.
Representative Wilson loop spectra for (b) $(\mc{C}_m^x,\mc{C}_m^y)=(0,0)$, (c) $(1,-1)$, (d) $(0,-2)$, and $(3,1)$.
Mirror sectors $+i$ and $-i$ are indicated with red and blue lines, respectively.
Here, the Wilson loop spectra are drawn symmetric such that $\{\theta(k_x,k_y)\} \leftrightarrow -\{\theta(k_x,k_y)\}$ for simplicity.
The connectivity of Wilson loop spectrum is interpreted as that of surface states.
The dispersion of fourfold-degenerate Dirac fermion is always trivial because of twofold degeneracy formed by opposite mirror eigenstates along $\bM$-$\bX$ and $\bM$-$\bY$ lines.
Black arrows in (e) denote chiral and antichiral surface states which manifest nonzero $\mc{C}_m^{x,y}$.
}
\label{sfig1}
\end{figure*}
%%%%%%%%%%%%%%%%%%%%%%

%%%%%%%%%%%%%%%%%%%%%%
\tocless{\subsubsection{Relation between the Wilson loop spectrum and mirror Chern numbers $\mc{C}_m^x$ and $\mc{C}_m^y$}}{}
%%%%%%%%%%%%%%%%%%%%%%
Analogous to \eqs{eqapp:ChernWinding1} and \eqref{eqapp:ChernWinding2}, we can show the relation between mirror Chern numbers and winding number in the Wilson loop spectrum.
For the mirror Chern number defined in $M_y=\{m_y|\bb 0\}$-invariant plane, i.e. $k_y=k^*$ plane ($k^*=0$ or $\pi$), is identical to the winding number in each mirror sector:
\bg
\mc{C}_\pm^{k_y=k^*}
%= \frac{1}{2\pi} \int_{-\pi}^\pi dk_z \int_{-\pi}^\pi dk_x \sum_{n=1}^{n_\pm} F^\pm_{n,zx}(k_x,k^*,k_z)
%= \frac{1}{2\pi} \int_{-\pi}^\pi dk_z \int_{-\pi}^\pi dk_x \sum_{n=1}^{n_\pm} \der_z A^\pm_{n,x}(k_x,k^*,k_z) - \der_x A^\pm_{n,z}(k_x,k^*,k_z) \nn
= - \frac{1}{2\pi} \int_{-\pi}^\pi dk_x \, \der_x \int_{-\pi}^\pi dk_z \sum_{n=1}^{n_\pm} A^\pm_{n,z}(k_x,k^*,k_z)
= \frac{1}{2\pi} \int_{-\pi}^\pi dk_x \sum_{a=1}^{n_\pm} \der_x \theta^\pm_a(k_x,k^*)
= \Delta N^\pm_{(-\pi,k^*)\to(\pi,k^*)}
\label{eq:app_mCy}
\eg
where the sign $\pm$ in lower/upper indices corresponds to $M_y$ mirror eigenvalue $\pm i$.
Here, as similar in \eq{eqapp:ChernWinding2}, we define the winding number $\Delta N^\pm_{(-\pi,k^*)\to(\pi,k^*)}$ as the difference between the number of Wilson bands that cross the reference horizontal line with positive and negative slopes, for a given mirror sector $\pm i$: $\Delta N^\pm_{(-\pi,k^*)\to(\pi,k^*)} = N^{\pm,>0}_{(-\pi,k^*)\to(\pi,k^*)} - N^{\pm,<0}_{(-\pi,k^*)\to(\pi,k^*)}$.
We can also obtain a similar relation for the mirror Chern number defined in $M_x=\{m_x|\bb 0\}$-invariant plane, i.e. $k_x=k^*$ plane,
\bg
\mc{C}_\pm^{k_x=k^*}
%= \frac{1}{2\pi} \int_{-\pi}^\pi dk_y \int_{-\pi}^\pi dk_z \sum_{n=1}^{n_\pm} F^\pm_{n,yz}(k^*,k_y,k_z)
%= \frac{1}{2\pi} \int_{-\pi}^\pi dk_y \int_{-\pi}^\pi dk_z \sum_{n=1}^{n_\pm} \der_y A^\pm_{n,z}(k^*,k_y,k_z) - \der_z A^\pm_{n,y}(k^*,k_y,k_z) \nn
= \frac{1}{2\pi} \int_{-\pi}^\pi dk_y \, \der_y \int_{-\pi}^\pi dk_z \sum_{n=1}^{n_\pm} A^\pm_{n,z}(k^*,k_y,k_z)
= - \frac{1}{2\pi} \int_{-\pi}^\pi dk_y \sum_{a=1}^{n_\pm} \der_y \theta^\pm_a(k^*,k_y) = - \Delta N^\pm_{(k^*,-\pi)\to(k^*,\pi)}.
\eg
The MSGs under consideration have two anticommuting mirrors $M_x$ and $M_y$.
Thus, a set of Wilson loop eigenvalues of $M_y$-invariant plane satisfies $\{\theta_+(k_x,k^*)\} = \{\theta_-(-k_x,k^*)\}$ because of $M_x$.
With the same logic, $\{\theta_+(k^*,k_y)\} = \{\theta_-(k^*,-k_y)\}$ holds.
Because of these spectral symmetries, $\Delta N^{\pm}_{(-\pi,k^*) \to (0,k^*)} = - \Delta N^{\mp}_{(0,k^*) \to (\pi,k^*)}$ and $\Delta N^{\pm}_{(k^*,-\pi) \to (k^*,0)} = - \Delta N^{\mp}_{(k^*,0) \to (k^*,\pi)}$.
Using these relations, we derive the mirror Chern numbers $\mc{C}_m^{x,y}$ and the number of intersections between the Wilson bands and the reference horizontal line in the Wilson loop spectrum.
First, by noting that $\mc{C}_m^x = \mch{+}{k_x=0} = -\mch{-}{k_x=0}$ and $N^{\pm}_{-\bY \to \bG}=-N^{\mp}_{\bG \to \bY}$,
we obtain
\bg
\mc{C}_m^x = - \Delta N^{+}_{\bG \to \bY} + \Delta N^{-}_{\bG \to \bY}.
\label{eqapp:Cmxrule}
\eg
Similarly, the mirror Chern number $\mc{C}_m^y$ is given by
\bg 
\mc{C}_m^y = \Delta N^{+}_{\bG \to \bX} - \Delta N^{-}_{\bG \to \bX}.
\label{eqapp:Cmyrule}
\eg
Therefore, the mirror Chern numbers $\mc{C}_m^{x,y}$ can be read off from the Wilson loop spectrum as follows.
First, divide Wilson bands $\theta^\pm(k_x,k_y)$ into mirror sectors $\pm i$ according to their mirror eigenvalues, and set a reference horizontal line in the Wilson loop spectrum.
Then, count the number of Wilson bands in each mirror sector $\pm i$ with positive slope ($N^{\pm,>0}_l$) and with negative slope ($N^{\pm,<0}_l$) along high-symmetry line $l:\bG \to \bX$ and $\bG \to \bY$.
Define $\Delta N^\pm_l = N^{\pm,>0}_l - N^{\pm,<0}_l$ and obtain the Chern numbers according to \eqs{eqapp:Cmxrule} and \eqref{eqapp:Cmyrule}.
We can apply this counting rule for the mirror Chern numbers to Wilson loop spectra in \sfig~\ref{sfig1}.
In \sfig~\ref{sfig1}(b), $\Delta N^{+}_{\bG \to \bY}=0$, $\Delta N^{-}_{\bG \to \bY}=-1$, $\Delta N^{+}_{\bG \to \bX}=0$, and $\Delta N^{-}_{\bG \to \bX}=-1$, which imply $\mc{C}_m^x=-1$ and $\mc{C}_m^y=1$.

We note that not all pairs $(\mc{C}_m^y,\mc{C}_m^x)$ are allowed for insulators.
Only the pairs satisfying $\mc{C}_m^y+\mc{C}_m^x = 0$ (mod 2) are allowed. 
Now, we show this in a similar method used in Ref.~\cite{wieder2018wallpaper}.
For insulators, the Berry flux or the integration of Berry curvature over any closed surface must vanish.
Thus, the Berry flux for a rectangular prism, which is formed by a rectangle $\bG$-$\bX$-$\bM$-$\bY$-$\bG$ and its translation along $z$ direction by unit translation, must be zero.
Otherwise, there are Weyl points.
Accordingly, the winding number along the closed loop $\bG \to \bX \to \bM \to \bY \to \bG$ is zero,
\ba
\sum_{s=\pm} \Delta N^s_{\bG \to \bX} + \Delta N^s_{\bX \to \bM} +\Delta N^s_{\bM \to \bY} +\Delta N^s_{\bY \to \bG} = 0.
\label{eqapp:Berryflux1}
\ea
As the Wilson bands along $\bX \to \bM$ and $\bM \to \bY$ are doubly degenerate with opposite mirror eigenvalues, $\Delta N^+_{\bX \to \bM}=\Delta N^-_{\bX \to \bM}$ and $\Delta N^+_{\bM \to \bY}=\Delta N^-_{\bM \to \bY}$ hold.
Also, from \eqs{eqapp:Cmxrule} and \eqref{eqapp:Cmyrule}, it can be shown that $\mc{C}_m^x = \Delta N^+_{\bG \to \bY} + \Delta N^-_{\bG \to \bY}$ (mod 2) and $\mc{C}_m^y = \Delta N^+_{\bG \to \bX} + \Delta N^-_{\bG \to \bX}$ (mod 2).
Thus, \eq{eqapp:Berryflux1} implies that $\mc{C}_m^x + \mc{C}_m^y = 0$ (mod 2).
%%%%%%%%%%%%%%%%%%%%%%
\\

%%%%%%%%%%%%%%%%%%%%%%
\tocless{\subsubsection{Absence of topological magnetic Dirac insulator in MSGs related to MWG $p'_cmm$}}{}
%%%%%%%%%%%%%%%%%%%%%%
We remark that there is no topological Dirac insulator phase in the MSGs related to MWG $p'_cmm$.
First, these MSGs have no diagonal mirror plane such as $\{m_{xy}|\bb 0\}$, the mirror Chern number $\mchd$ does not exist unlike other MSGs related to MWGs $p'_c4mm$ or $p4'g'm$.
Second, although the mirror Chern numbers $\mc{C}_m^{k_{x,y}=\pi}$ at $k_{x,y}=\pi$ planes can be defined with respect to for $M_{x,y}$, they are always zero as shown in \sn~\ref{appsub:mC_pcmm}.
Thus, the dispersion of fourfold-degenerate Dirac fermion (in each mirror sector) cannot be chiral along $M_{x,y}$-invariant lines, i.e. $\bM$-$\bX$ and $\bM$-$\bY$ lines.
In this sense, surface fourfold-degenerate Dirac fermion in the MSGs with (001)-surface MWG $p'_cmm$ is rather trivial.
However, topological phases with nonzero $\mc{C}_m^{x,y}$ exhibits usual chiral and antichiral surface states~\cite{hsieh2012topological} along $\bG$-$\bX$ and $\bG$-$\bY$ lines.
See \sfig~\ref{sfig1}(d) where these chiral and antichiral modes are indicated by black arrows.
Note that a chiral mode and a antichiral mode can form twofold-degenerate Dirac surface states if they have opposite mirror eigenvalues.
%
%%%%%%%%%%%%%%%%%%%%%%
\\

\begin{figure*}[b!]
\centering
\includegraphics[width=0.98\textwidth]{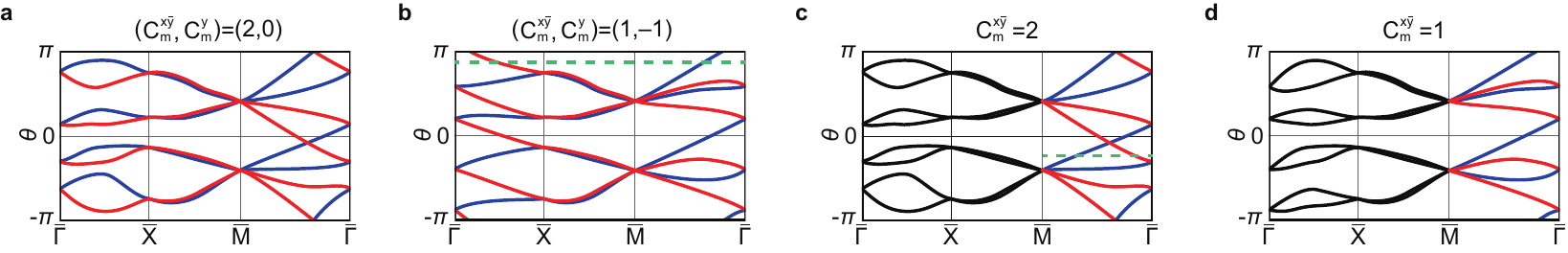}
\caption{
{\bf Wilson loop spectra in MSGs related to MWGs $p'_c4mm$ and $p4'g'm$ for many occupied bands.}
Wilson loop spectra with eight occupied bands ($n_{\rm occ}=8$).
Note that the cases with $n_{\rm occ}=4$ are illustrated in Fig.~2 in the main text.
(a)-(b) Wilson loop spectra in MSGs related to MWG $p'_c4mm$ with (a) $(\mc{C}_m^y,\mchd)=(0,2)$ and (b) $(\mc{C}_m^y,\mchd)=(-1,1)$.
(c)-(d) Wilson loop spectra in MSGs related to MWG $p4'g'm$ with (c) $\mchd=2$ and (d) $\mchd=1$.
Note the chiral dispersion of fourfold-degenerate Dirac fermion in each mirror sector along $\bM$-$\bG$-$(-\bM)$ line when $\mchd$ is nonzero.
The detailed discussion on chiral dispersion can be found in \sn~\ref{app:local_chiral}.
}
\label{sfig2}
\end{figure*}

%%%%%%%%%%%%%%%%%%%%%%
\tocless{\subsection{Wilson loop spectra in MSGs related to MWG $p'_c4mm$}
\label{appsub:WLpc4mm}}{}
%\subsection{Wilson loop spectra in MSGs related to MWG $p'_c4mm$}
%\label{appsub:WLpc4mm}
%%%%%%%%%%%%%%%%%%%%%%
In this section, we study the Wilson loop spectra in MSGs related to MWG $p'_c4mm$.
Representative Wilson loop spectra are shown in Figs.~2(g)-(j) in the main text and \sfigs~\ref{sfig2}(a)-(b).
Since $p'_c4mm$ has all the symmetry elements in $p'_cmm$, we use all the results obtained in \sn~\ref{appsub:WLpcmm}.
Compared to the Wilson loop spectra in the MSGs related to MWG $p'_cmm$, there are two additional properties in those in the MSGs related to MWG $p'_c4mm$.
First, the Wilson loop exhibit the same spectrum along $\bG$-$\bX$ line and $\bG$-$\bY$ line because of $C_{4z}=\{c_{4z}|\bb 0\}$.
The spectra along $\bX$-$\bM$ line and $\bY$-$\bM$ line are also same.
This can be deduced immediately by considering the symmetry transformation of $W_z(\bkp)$ with respect to $C_{4z}$ and thus we skip the derivation.
Second, the Wilson bands in $\bG$-$\bM$ line are divided into mirror sectors $\pm i$ with respect to $\mxyb=\{m_{x\cm{y}}|\bb 0\}$.
Also, the winding of Wilson bands in $\bG$-$\bM$ line determines $\mchd$.
Below, we explain this in detail.
%%%%%%%%%%%%%%%%%%%%%%
\\

%%%%%%%%%%%%%%%%%%%%%%
\tocless{\subsubsection{Mirror eigenvalues of fourfold degeneracy at $\bM$}}{}
%%%%%%%%%%%%%%%%%%%%%%
As shown in \sn~\ref{appsub:WLpcmm}, the Wilson bands at $M$ point are fourfold degenerate. 
The fourfold degeneracy is formed by two states with $\mxyb$ eigenvalue $+i$ and two with $-i$.
With two anticommuting mirrors $\mxy, \mxyb$ and periodicity of Wilson bands, we can show this.
Since $B_{\mxyb}(\bM,-\pi)$ commutes with $W_z(\bM)$, let $\ket{\theta_\pm(\bM)}$ be a simultaneous eigenstates of $B_{\mxyb}(\bM,-\pi)$ and $W_z(\bM)$ such that $B_{\mxyb}(\bM,-\pi) \ket{\theta_\pm(\bM)} = \pm i \ket{\theta_\pm(\bM)}$ and $W_z(\bM) \ket{\theta_\pm(\bM)} = e^{i\theta(\bM)} \ket{\theta_\pm(\bM)}$.
Due to anticommutation relation between two mirrors, $B_{\mxy}(\bM,-\pi) \ket{\theta_\pm(\bM)}$ can be identified as an $W_z(\bM)$ eigenstate with $B_{\mxyb}(\bM,-\pi)$ eigenvalue $\mp i$. 
This means that the number of bands at $\bM$ with $B_{\mxyb}(\bM,-\pi)$ eigenvalue $+i$ is equal to that of bands with $B_{\mxyb}(\bM,-\pi)$ eigenvalue $-i$.
Thus, two of each fourfold degeneracy belong to one mirror sector and the other two belong to the opposite mirror sector because of the periodicity of Wilson bands.
%%%%%%%%%%%%%%%%%%%%%%
\\

%%%%%%%%%%%%%%%%%%%%%%
\tocless{\subsubsection{Relation between the Wilson loop spectrum and mirror Chern numbers $(\mc{C}_m^y,\mchd)$}}{}
%%%%%%%%%%%%%%%%%%%%%%
We have already explained how to read off $\mc{C}_m^y$ from the Wilson loop spectrum in \sn~\ref{appsub:WLpcmm} [see Eq.~\eqref{eq:app_mCy}].
We can also relate mirror Chern number $\mchd$ to the winding number in the Wilson loop spectrum in a similar way to Eq.~\eqref{eq:app_mCy}:
\bg
\mch{\pm}{k_x=k_y} = \frac{1}{2\pi} \int_{-\sqrt{2}\pi}^{\sqrt{2}\pi} dk_+ \int_{-\pi}^\pi dk_z \sum_{n=1}^{n_\pm} F^\pm_{n,z+}(k_x,k_x,k_z) 
%= - \frac{1}{2\pi} \int_{-\sqrt{2}\pi}^{\sqrt{2}\pi} dk_+ \int_{-\pi}^\pi dk_z \sum_{n=1}^{n_\pm} \der_+ A^\pm_{n,z}(k_x,k^*,k_z)
= \frac{1}{2\pi} \int_{-\sqrt{2}\pi}^{\sqrt{2}\pi} dk_+ \sum_{a=1}^{n_\pm} \der_+ \theta^\pm_a(k_x,k_x) =  \Delta N^\pm_{(-\pi,-\pi) \to (\pi,\pi)}.
\eg
The MSGs under consideration have two anticommuting mirrors $\mxy$ and $\mxyb$.
Thus, a set of Wilson loop eigenvalues of $M_y$-invariant plane satisfies $\{\theta_+(k_x,k_x)\} = \{\theta_-(-k_x,-k_x)\}$ because of $\mxy$.
This symmetry constraint leads to $\Delta N^{\pm}_{-\bM \to \bG} = - \Delta N^{\mp}_{\bG \to \bM}$.
Since $\mch{\pm}{k_x=k_y}=\Delta N^\pm_{-\bM \to \bG}+\Delta N^\pm_{\bG \to \bM}$ and $\mchd=\pm \mch{\pm}{k_x=k_y}$, we obtain
\bg
\mchd = \Delta N^{+}_{\bG \to \bM} - \Delta N^{-}_{\bG \to \bM}.
\label{eqapp:Cmxyrule1}
\eg
Hence, the mirror Chern numbers $\mchd$ can be read off from the Wilson loop spectrum in $\bG$-$\bM$ line.
For example, $\Delta N^{+}_{\bG \to \bM}=0$ and $\Delta N^{-}_{\bG \to \bM}=-1$ in \sfig~\ref{sfig2}(b).
Thus, $\mchd=1$.
Note that $\mchd$ can also be counted by setting a reference line.
That is, draw first a reference line along $\bG$-$\bM$ line.
If there are $n_{\rm int}$ Wilson bands $\{\theta_A(\bkp)|A=1,\dots,n_{\rm int}\}$ that intersect with the reference line, 
$\mchd$ can be expressed as
\ba
\mchd = \sum_{A=1}^{n_{\rm int}} m_A \nu_A
\ea
by defining a sign of slope along the $\bG \to \bM$ direction, $\nu_A=\textrm{sign} \, \der_{\bkp} \theta_A(\bkp)$, and mirror eigenvalue $i m_A$ of each intersecting Wilson band.
Note that we can define the corresponding expression for $\mc{C}^{xy}_m$ in a similar way.
The only difference is that now a reference line lies along $\bG$-$\bM'$ line, where $\bM'=(\pi,-\pi)$, and the sign of slope for intersecting Wilson bands is defined along the $\bG$-$\bM'$ direction.

Finally, we note that only the pairs satisfying $\mc{C}_m^y+\mchd = 0$ (mod 2) are allowed for insulators. 
In insulators, the Berry flux for a triangular prism, which is formed by a rectangle $\bG$-$\bX$-$\bM$-$\bG$ and its translation along the $z$ direction by unit translation, must be zero:
\ba
\sum_{s=\pm} \Delta N^s_{\bG \to \bX} + \Delta N^s_{\bX \to \bM} +\Delta N^s_{\bM \to \bG} = 0.
\label{eqapp:Berryflux2}
\ea
In a similar way to the calculation below Eq.~\eqref{eqapp:Berryflux1}, one can show that $\mc{C}_m^y+\mchd = 0$ (mod 2).
%%%%%%%%%%%%%%%%%%%%%%
\\

%%%%%%%%%%%%%%%%%%%%%%
%%%%%%%%%%%%%%%%%%%%%%

%%%%%%%%%%%%%%%%%%%%%%
\tocless{\subsection{Wilson loop spectra in MSGs related to MWG $p4'g'm$}
\label{appsub:WLp4gm}}{}
%\subsection{Wilson loop spectra in MSGs related to MWG $p4'g'm$}
%\label{appsub:WLp4gm}
%%%%%%%%%%%%%%%%%%%%%%
First, let us summarize the symmetry analysis of Wilson loop spectra in MSGs related to MWG $p4'g'm$.
Wilson bands are twofold degenerate along $\bX$-$\bM$ line.
At $\bM$, fourfold degeneracy is formed by two Wilson bands in mirror sector $+i$ and two in mirror sector $-i$.
At $\bG$, two Wilson bands with opposite mirror eigenvalues form twofold degeneracy.
Along $\bM$-$\bG$ line, the winding of Wilson bands corresponds to the mirror Chern number $\mchd$ defined in \sn~\ref{app:mCherns}.
Representative Wilson loop spectra are shown in Fig.~2(b)-(e) in the main text and \sfig~\ref{sfig2}(c)-(d).
Below, we prove these properties.
%%%%%%%%%%%%%%%%%%%%%%
\\

%%%%%%%%%%%%%%%%%%%%%%
\tocless{\subsubsection{Twofold degeneracy at $\bG$}}{}
%%%%%%%%%%%%%%%%%%%%%%
We consider two orthogonal mirrors $\mxybo=\{m_{x\cm{y}}|\hf,\mhf,0\}$ and $\mxyo=\{m_{xy}|\hf,\hf,0\}$.
These two symmetries give the symmetry constraints on the Wilson loop $W_z(\bG)$:
\bg
W_z(\bG) = B_{\mxyo}(\bG,-\pi) W_z(\bG) B_{\mxyo}(\bG,-\pi)^{-1}, \quad
W_z(\bG) = B_{\mxybo}(\bG,-\pi) W_z(\bG) B_{\mxybo}(\bG,-\pi)^{-1}.
\label{eq:app_WL_p4gm_1}
\eg
This means that $B_{\mxybo}(\bG,-\pi)$ commutes with $W_z(\bG)$ and thus we can find simultaneous eigenstates of $B_{\mxybo}(\bG,-\pi)$ 
and $W_z(\bG)$ such that $B_{\mxybo}(\bG,-\pi) \ket{\theta_\pm(\bG)} = \pm i \ket{\theta_\pm(\bG)}$ and $W_z(\bG) \ket{\theta_\pm(\bG)} = e^{i \theta(\bG)} \ket{\theta_\pm(\bG)}$.
Since $\mxybo \mxyo = \cm{E} \mxyo \mxybo$ implies $B_{\mxybo}(\bG,-\pi) B_{\mxyo}(\bG,-\pi) = - B_{\mxyo}(\bG,-\pi) B_{\mxybo}(\bG,-\pi)$, we obtain
\bg
B_{\mxybo}(\bG,-\pi) \left[ B_{\mxyo}(\bG,-\pi) \ket{\theta_\pm(\bG)} \right]
= \mp i B_{\mxyo}(\bG,-\pi) \ket{\theta_\pm(\bG)}.
\eg
Also, \eq{eq:app_WL_p4gm_1} implies that $B_{\mxyo}(\bG,-\pi) \ket{\theta_\pm(\bG)}$ is also an eigenstate of $W_z(\bG)$ with eigenvalue $e^{i\theta(\bG)}$.
Hence, two Wilson bands with opposite $\mxyb$ eigenvalues form twofold degeneracy at $\bG$.
Note that the same algebra holds at $\bM$.
%%%%%%%%%%%%%%%%%%%%%%
\\

%%%%%%%%%%%%%%%%%%%%%%
\tocless{\subsubsection{Twofold degeneracy in $\bX$-$\bM$ line}}{}
%%%%%%%%%%%%%%%%%%%%%%
The $\bX$-$\bM$ line is invariant under  $\tmy=T\{m_y|\hf,\hf,0\}$.
Let us denote generic momenta in the $\bX$-$\bM$ line as $\bkp^*=(\pi,k_y)$.
Suppose that $\ket{\theta(\bkp^*)}$ is a $W_z(\bkp^*)$ eigenstate with eigenvalue $e^{i\theta(\bkp^*)}$.
Then, symmetry constraint on $W_z(\bkp^*)$ by $\tmx$, $B^{\mc{K}}_{\tmy}(\bkp^*,-\pi) W_z(\bkp^*) B^{\mc{K}}_{\tmy}(\bkp^*,-\pi)^{-1} = W_z(\bkp^*)^\dg$, implies that $B^{\mc{K}}_{\tmy}(\bkp^*,-\pi) \ket{\theta(\bkp^*)}$ is also an eigenstate of $W_z(\bkp^*)$ with eigenvalue $e^{i\theta(\bkp^*)}$.
Also, $\ket{\theta(\bkp^*)}$ and $B^{\mc{K}}_{\tmy}(\bkp^*,-\pi) \ket{\theta(\bkp^*)}$ form a Kramers pair because $(\tmy)^2 = \{E|\bb a_1\}$ and thus $[B^{\mc{K}}_{\tmy}(\bkp^*,-\pi)]^2=-\mathds{1}$.
Thus, the Wilson bands are twofold degenerate along $\bX$-$\bM$ line.
%%%%%%%%%%%%%%%%%%%%%%
\\

%%%%%%%%%%%%%%%%%%%%%%
\tocless{\subsubsection{Fourfold degeneracy at $\bM$}}{}
%%%%%%%%%%%%%%%%%%%%%%
Wilson bands form fourfold degeneracy at $\bM$.
To show this, let us consider $\mxybo$, $\tmx$, and $C_{2z}=\{c_{2z}|\bb 0\}$.
These symmetries leads to
\bg
[W_z(\bM), B_{C_{2z}}(\bM,-\pi)] = 0, \quad
[W_z(\bM), B_{\mxybo}(\bM,-\pi)] = 0, \quad
W_z(\bM) B^{\mc{K}}_{\tmx}(\bM,-\pi) = B^{\mc{K}}_{\tmx}(\bM) W_z(\bM)^\dg.
\eg
First, let us consider a simultaneous eigenstate of $W_z(\bM)$ and $B_{C_{2z}}(\bM,-\pi)$ such that $W_z(\bM) \ket{\theta_\pm(\bM)} = e^{i\theta(\bM)} \ket{\theta_\pm(\bM)}$ and $B_{C_{2z}}(\bM,-\pi) \ket{\theta_\pm(\bM)} = \pm i\ket{\theta_\pm(\bM)}$.
The symmetry constraint by $\tmx$ implies that $\ket{\theta_\pm(\bM)}$ and $B^{\mc{K}}_{\tmx}(\bM,-\pi) \ket{\theta_\pm(\bM)}$ form a Kramers pair.
The two states in a Kramers pair have the same $C_{2z}$ eigenvalue.
This can be shown by noticing that $C_{2z} \tmx = \tmx C_{2z} \{\cm{E}|\mns1,1,0\}$ which implies $B_{C_{2z}}(\bM,-\pi) B^{\mc{K}}_{\tmx}(\bM,-\pi) = -e^{- i M \cdot (-1,1,0)} B^{\mc{K}}_{\tmx}(\bM,-\pi) B_{C_{2z}}(\bM)$.
Thus, $B_{C_{2z}}(\bM,-\pi) B^{\mc{K}}_{\tmx}(\bM,-\pi) \ket{\theta_\pm(\bM)} = \pm i B^{\mc{K}}_{\tmx}(\bM,-\pi) \ket{\theta_\pm(\bM)}$.

We can find another Kramers pair $\{B_{\mxyb}(\bM,-\pi) \ket{\theta_\pm(\bM)}, B^{\mc{K}}_{\tmx}(\bM,-\pi) B_{\mxyb}(\bM,-\pi) \ket{\theta_\pm(\bM)}\}$ whose $W_z(\bM)$ eigenvalue is $e^{i\theta(\bM)}$.
This can be shown by using
\bg
C_{2z} \mxybo = \cm{E} \mxybo C_{2z} \{E|\mns1,1,0\}, \quad
B_{C_{2z}}(\bM,-\pi) B_{\mxybo}(\bM,-\pi) = - e^{- i M \cdot (-1,1,0)} B_{\mxybo}(\bM,-\pi) B_{C_{2z}}(\bM,-\pi).
\eg
The last equation implies $B_{C_{2z}}(\bM,-\pi) \left[ B_{\mxybo}(\bM,-\pi) \ket{\theta_\pm(\bM)} \right] = \mp i B_{\mxybo}(\bM,-\pi) \ket{\theta_\pm(\bM)}$.
Since two Kramers pairs $\{\ket{\theta_\pm(\bM)}, B^{\mc{K}}_{\tmx}(\bM,-\pi)\}$ and $\{B_{\mxybo}(\bM,-\pi) \ket{\theta_\pm(\bM)}, B^{\mc{K}}_{\tmx}(\bM,-\pi) B_{\mxybo}(\bM,-\pi) \ket{\theta_\pm(\bM)}\}$ are independent because their $C_{2z}$ eigenvalues are different.
Hence, the Wilson bands are fourfold degenerate at $\bM$.

Now, we further show that each fourfold degeneracy is formed by two Wilson bands with $\mxybo$ eigenvalue $+i$ and two with $-i$.
With two anticommuting mirrors $\mxyo, \mxybo$ and periodicity of Wilson bands, we can show this.
Since $B_{M'_{x\cm y}}(\bM,-\pi)$ commutes with $W_z(\bM)$, let $\ket{\theta_\pm(\bM)}$ be a simultaneous eigenstates of $B_{\mxybo}(\bM,-\pi)$ and $W_z(\bM)$ such that $B_{\mxybo}(\bM,-\pi) \ket{\theta_\pm(\bM)} = \pm i \ket{\theta_\pm(\bM)}$ and $W_z(\bM) \ket{\theta_\pm(\bM)} = e^{i\theta(\bM)} \ket{\theta_\pm(\bM)}$.
Due to anticommutation relation between two mirrors, $B_{\mxyo}(\bM,-\pi) \ket{\theta_\pm(\bM)}$ can be identified as an $W_z(\bM)$ eigenstate with $B_{\mxybo}(\bM,-\pi)$ eigenvalue $\mp i$. 
This means that the number of bands at $\bM$ with $B_{\mxybo}(\bM,-\pi)$ eigenvalue $+i$ is equal to that of bands with $B_{\mxybo}$ eigenvalue $-i$.
Thus, two of each fourfold degeneracy belong to one mirror sector and the other two belong to the opposite mirror sector because of the periodicity of Wilson bands.
%%%%%%%%%%%%%%%%%%%%%%
\\

%%%%%%%%%%%%%%%%%%%%%%
\tocless{\subsubsection{Relation between the Wilson loop spectrum and mirror Chern number $\mchd$}}{}
%%%%%%%%%%%%%%%%%%%%%%
The mirror Chern numbers $\mchd$ can be read off from the Wilson loop spectrum in $\bG$-$\bM$ line, as in the MSGs related to MWG $p'_c4mm$ (see \sn~\ref{appsub:WLpc4mm}).
For example, $\Delta N^{+}_{\bG \to \bM}=1$, $\Delta N^{-}_{\bG \to \bM}=-1$, and thus $\mchd=2$ in \sfig~\ref{sfig2}(c).
%%%%%%%%%%%%%%%%%%%%%%
\\

%%%%%%%%%%%%%%%%%%%%%%
\begin{figure}[b!]
\includegraphics[width=0.95\textwidth]{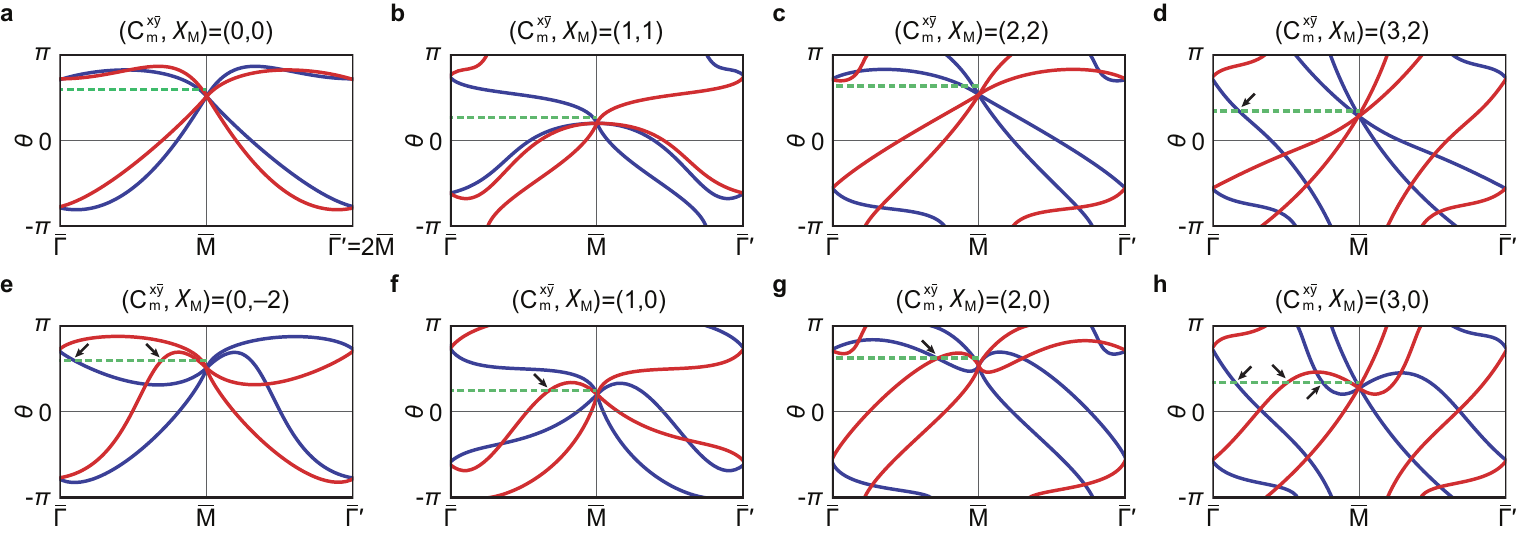}
\caption{
{\bf Relation between the mirror Chern number $\mchd$ and local chirality $\chi_M$ of fourfold-degenerate Dirac fermion at $\bM$.}
The corresponding $(\mchd, \chi_M, \delta N_{\bG \to \bM}^+, \delta N_{\bG \to \bM}^-)$ are (a) $(0,0,0,0)$, (b) $(1,1,0,0)$, (c) $(2,2,0,0)$, (d) $(3,2,0,-1)$, (e) $(0,-2,1,-1)$, (f) $(1,0,1,0)$ (g) $(2,0,1,-1)$, and (h) $(3,0,1,-2)$.
(a)-(c) When $|\mchd| \le 2$, a fourfold degeneracy at $M$ can appear without any additional node that crosses the green reference line or the Fermi level.
Then, $\mchd = \chi_M$.
(d)-(h) When $\mchd \ne \chi_M$, \eq{eq:app_localchiral} indicate how many additional surface states (black arrows) cross the green reference line or the Fermi level.
Comparing (a)-(c) with (e)-(g), we see that a mismatch between $\chi_M$ and $\mchd$ indicates additional nodes.
(d,h) Additional nodes must appear when $\mchd \neq \chi_M$.
The relation between $\mchd$, $\chi_M$, and the number of additional nodes ($\delta N_{\bG \to \bM}^\pm$) hold generally for multiband cases.
}
\label{sfig:local_chiral}
\end{figure}
%%%%%%%%%%%%%%%%%%%%%%

%%%%%%%%%%%%%%%%%%%%%%
\section{Relation between chiral dispersion of surface fourfold-degenerate Dirac fermion and mirror Chern number}
\label{app:local_chiral}
%%%%%%%%%%%%%%%%%%%%%%
In \sn~\ref{app:Wilsonloop}, we explain how the mirror Chern numbers can be read off from the Wilson loop spectrum.
[See Eqs.~\eqref{eqapp:Cmxrule}, \eqref{eqapp:Cmyrule}, and \eqref{eqapp:Cmxyrule1}].
For example, mirror Chern number $\mchd$ is expressed as $\mchd=\Delta N_{\bG \to \bM}^+ - \Delta N_{\bG \to \bM}^-$.
Here, we present a general formula that relates $\mchd$ to (i) the chiral dispersion of fourfold-degenerate Dirac fermion at $\bM$ and (ii) the number of additional nodes appearing at the Fermi level in the $\bM$-$\bG$ line.
To discuss the mirror-resolved chirality of Dirac fermion at $\bM$, we define the local chirality $\chi_M$,
\ba
\label{eq:app_chi_M}
\chi_M = \Delta N^+_{\bM-\cm{\delta} \to \bM} - \Delta N^-_{\bM-\cm{\delta} \to \bM}, \quad
|\chi_M| \le 2.
\ea
To define this quantity, we first set a reference line slightly above the Wilson loop eigenvalue of fourfold-degenerate point at $\bM$ (Green dashed lines in \sfig~\ref{sfig:local_chiral}).
If there are many ($n_{\rm occ}/4$ for $n_{\rm occ}$ occupied bands) Dirac points, we focus on one of them.
In \eq{eq:app_chi_M}, $\cm{\delta}$ is a point along $\bM$-$\bG$ line, and we set small $\cm{\delta}$, say $\cm{\delta}=0.05(\pi,\pi)$, to count the mirror-resolved chirality of Dirac point only near $\bM$.
Note that $|\chi_M| \le 2$ because a Dirac point is formed by two bands with mirror eigenvalue $+i$ and two with $-i$.

Now, we recast the formula for $\mchd$ to separate the contributions from $\bM$ and the additional nodes:
\ba
\label{eq:app_localchiral}
\mchd - \chi_M = \delta N_{\bG \to \bM}^+ - \delta N_{\bG \to \bM}^-.
\ea
Here, $\delta N_{\bG \to \bM}^\pm = \Delta N_{\bG \to \bM-\cm{\delta}}^\pm$ is defined in a similar way as $\Delta N_{\bG \to \bM}^\pm$, but neglects the contribution at $M$.
Note that $\delta N_{\bG \to \bM}^\pm \in \mathbb{Z}$ can be negative.  
Some representative examples are listed in \sfig~\ref{sfig:local_chiral}.
See also \sfig~\ref{sfig:DFT-Cm=1-Wlson}, where we discuss the local chirality $\chi_M$ of Dirac fermion in candidate material Nd$_4$Te$_8$Cl$_4$O$_{20}$.

In \sfigs~\ref{sfig:local_chiral}(a)-(c), the local chirality $\chi_M$ is equal to the mirror Chern number $\mchd$.
In this case, $\delta N_{\bG \to \bM}^+ - \delta N_{\bG \to \bM}^-=0$ according to \eq{eq:app_localchiral}.
On the other hand, in \sfigs~\ref{sfig:local_chiral}(d)-(h), $\chi_M \ne \mchd$ and $\delta N_{\bG \to \bM}^+ - \delta N_{\bG \to \bM}^- = \mchd - \chi_M \ne0$.
This means that $\delta N_{\bG \to \bM}^\pm$ cannot be zero at the same time.
For any $\mchd$, the minimal number of additional nodes along $\bM$-$\bG$ line is given by ${\rm min} \, N_{\rm add} = |\mchd - \chi_M|$ because $N_{\rm add}  \ge |\delta N_{\bG \to \bM}^+ + \delta N_{\bG \to \bM}^-| \ge |\delta N_{\bG \to \bM}^+ - \delta N_{\bG \to \bM}^-|$.
%
%%%%%%%%%%%%%%%%%%%%%%
\\

%%%%%%%%%%%%%%%%%%%%%%
\section{Higher-order bulk-boundary correspondence of topological magnetic Dirac insulators}
\label{app:higher-order}
%%%%%%%%%%%%%%%%%%%%%%
In this section, we discuss the higher-order bulk-boundary correspondence of topological magnetic Dirac insulators (TMDIs.)
The bulk topology of TMDIs is characterized by the mirror Chern numbers $\mchd$ and $\mc{C}_m^{xy}$ for diagonal mirror planes.
We can consider $\mxyb=\{m_{x\cm{y}}|\bb 0\}$ and $\mxy=\{m_{xy}|\bb 0\}$ by redefining the origin, regardless of the type of TMDI.
Note that $\mchd$ is identical to $\mc{C}_m^{xy}$ because of $TC_{4z}$ or $C_{4z}$, and now we focus on only $\mchd$.
When open boundaries along $x$ and $y$ directions preserve the diagonal mirror planes, hinge modes are protected by mirror when $\mchd$ is nonzero~\cite{langbehn2017reflection,schindler2018higher1}.

\begin{figure*}[t!]
\centering
\includegraphics[width=0.9\textwidth]{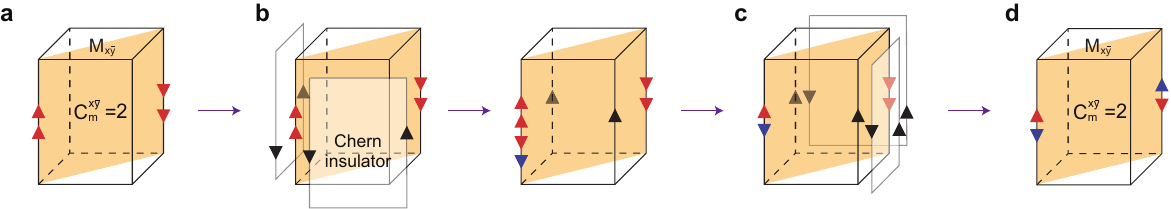}
\caption{
{\bf Mirror-protected hinge modes in 3D mirror-symmetric TCIs.}
Mirror-symmetric TCIs with $\mchd=2$ in open boundaries along $x$ and $y$ directions.
Diagonal mirror plane $\mxyb$ is represented by the orange plane.
(a) At mirror-invariant hinges (the lower left and upper right hinges), two chiral/antichiral hinge modes with the same mirror eigenvalue $+i$ (red arrows).
(b) Mirror-preserving perturbation that closes and reopens surface gap.
This can be described by a surface decoration of 2D surfaces 2D Chern insulators in a mirror-symmetric way.
Black arrows are chiral edge modes of Chern insulators.
Two copropagating and mirror-symmetric chiral edge modes become two chiral hinge modes with opposite mirror eigenvalues.
(c)-(d) Decoration of the other mirror-invariant hinge.
The chiral hinge modes in (a) become helical hinge modes with opposite mirror eigenvalues.
}
\label{sfig:hinge1}
\end{figure*}

First, let us review the higher-order bulk-boundary correspondence of mirror-symmetric topological crystalline insulators (TCIs)~\cite{langbehn2017reflection,schindler2018higher1,geier2018second,trifunovic2019higher}.
In open boundaries along $x$ and $y$ directions, a finite-size sample of TCI has two mirror-invariant hinges.
When $\mchd=\pm1$, a chiral (propagating along the positive $z$ direction) hinge mode with mirror eigenvalue $\pm i$, and an antichiral (propagating along the negative $z$ direction) hinge mode with mirror eigenvalue $\pm i$ exist on opposite hinges.
Note that, in fact, the relation between the mirror Chern number and the mirror eigenvalues of hinge modes depends on the definition of $\mchd$, because one can add an additional sign in the definition.
The mirror TCIs with $\mchd=\pm1$ are building blocks for understanding the cases with a general value of $\mchd$ because mirror Chern numbers are additive.
For example, a mirror TCI with $\mchd=2$ can be understood as a superposition of two mirror TCIs with $\mchd=1$.
Thus, its hinge modes are copropagating chiral hinge modes with the same mirror eigenvalue, as illustrated in \sfig~\ref{sfig:hinge1}(a).
However, a mirror TCI with $\mchd=2$ can also exhibit a helical hinge mode, formed by a chiral and an antichiral modes with opposite mirror eigenvalues, at each mirror-invariant hinge, as shown in \sfig~\ref{sfig:hinge1}(d).
This is because a mirror-preserving perturbation can close and reopen a surface gap and change the pattern of hinge modes.
Note that such perturbation does not close the bulk gap, thus the bulk topology does not change.

Now, let us explain how such mirror-preserving and bulk-gap-preserving perturbation can deform the dispersion and the number of hinge modes.
Closing and reopening the surface gap can be described by attaching 2D Chern insulators on 2D surfaces of a 3D TCI in a mirror-symmetric way~\cite{khalaf2018higher,geier2018second,schindler2018higher1,wieder2018axion}.
Note that we consider Chern insulators for the decoration because time-reversal symmetry is broken in magnetic systems.
See \sfigs~\ref{sfig:hinge1}(b)-(c) where this mirror-symmetric decoration is described.
In this decoration, chiral edge modes of Chern insulators become chiral hinge modes of the 3D mirror TCI.
Two copropagating chiral modes [black arrow in \sfig~\ref{sfig:hinge1}(b)] can be hybridized into two chiral hinge modes with opposite mirror eigenvalues.
Hence, there are two chiral hinge modes with mirror eigenvalue $+i$ and two antichiral hinge modes with opposite mirror eigenvalues, at the lower left hinge.
Note that a pair of chiral and antichiral modes can be gapped if they have the same mirror eigenvalues.
Hence, only a pair of chiral and antichiral modes with opposite mirror eigenvalues remain as shown in \sfig~\ref{sfig:hinge1}(c).
Repeating this decoration, one can obtain a mirror TCI with $\mchd=2$ and helical hinge modes.

We generalize this observation to a mirror TCI with a generic value of $\mchd$.
For this, we define the number of hinge modes: $n_{v,\lambda}$ for the hinge mode with group velocity of the sign $v=\pm$ along the $z$ direction and mirror eigenvalue $\lambda=\pm i$.
(For simplicity, let us focus on the lower left hinge.)
The surface decoration with Chern insulators changes $n_{+,+i}$ and $n_{+,-i}$, or $n_{-,+i}$ and $n_{-,-i}$, by the same amount, i.e. $\Delta n_{+,+i} = \Delta n_{+,-i}$ and  $\Delta n_{-,+i} = \Delta n_{-,-i}$.
Hence, one can define a quantity $n_{+,+i}-n_{-,+i}+n_{-,-i}-n_{+,-i}$ that is left invariant under the decoration.
In fact, this quantity is equal to $\mchd$:
\ba
\mchd = n_{+,+i}-n_{-,+i}+n_{-,-i}-n_{+,-i}.
\ea
This can be easily inferred from the fact that our building blocks correspond to $\mchd=\pm 1$, $n_{+,\pm i}=1$, and $n_{-,\pm i}=n_{+,\mp i}=n_{-,\mp i}=0$.
Thus, a mirror TCI with $\mchd$ can be constructed by superposing these building-block TCIs $|\mchd|$ times.
This implies that the number of hinge modes are not invariant but can be arbitrarily large.
However, there must be at least $|\mchd|$ hinge modes protected by mirror because
\ba
\sum_{v,\lambda} n_{v,\lambda} \ge |n_{+,+i}-n_{-,+i}+n_{-,-i}-n_{+,-i}| = |\mchd|.
\ea

\begin{figure*}[t!]
\centering
\includegraphics[width=0.9\textwidth]{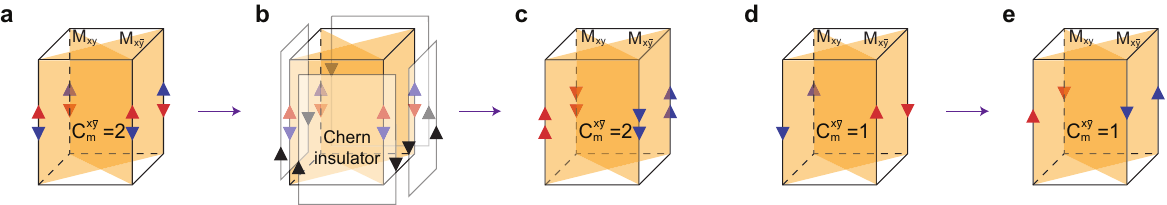}
\caption{
{\bf Mirror-protected hinge modes in 3D TMDIs.}
TMDIs and hinge modes in open boundaries along $x$ and $y$ directions.
Diagonal mirror planes $\mxyb$ and $\mxy$ are represented by the orange planes.
Note that $\mch{m}{xy}=-\mchd$.
(a)-(c) TMDI with $\mchd=2$.
(a) At each mirror-invariant hinge, a helical hinge mode is protected by $\mchd$ and $\mch{m}{xy}$.
For each diagonal mirror symmetry, the hinge mode dispersions correspond to \sfig~\ref{sfig:hinge1}(d) rather than \sfig~\ref{sfig:hinge1}(a).
This is because $C_{2z}=\mxyb \mxy$ flips the mirror eigenvalues of hinge modes according to $\{C_{2z},\mxyb\}=\{C_{2z},\mxy\}=0$.
(b) A mirror-preserving perturbation, which closes and reopens the surface gap, now should satisfy both $\mxyb$ and $\mxy$.
(c) The helical hinge modes in (a) become chiral hinge modes.
(d)-(e) Two possible realizations of hinge modes for TMDIs with $\mchd=1$.
}
\label{sfig:hinge2}
\end{figure*}

The bulk-boundary correspondence of mirror TCIs studied above can be directly applied to that of TMDIs by considering the additional mirror  plane $\mxy$.
We summarize possible hinge mode dispersions of TMDIs with $\mchd=1$ and $\mchd=2$ in \sfig~\ref{sfig:hinge2}.
The hinge modes of TMDIs with larger $\mchd$ can be obtained by superposing TMDIs with $\mchd=1$ and $\mchd=2$ as the building blocks.
%%%%%%%%%%%%%%%%%%%%%%

%%%%%%%%%%%%%%%%%%%%%%
%%%%%%%%%%%%%%%%%%%%%%

%%%%%%%%%%%%%%%%%%%%%%
%%%%%%%%%%%%%%%%%%%%%%

%%%%%%%%%%%%%%%%%%%%%%
\section{A scheme to search magnetic materials with Type-III and IV MSGs}
\label{app:mat_search}
%%%%%%%%%%%%%%%%%%%%%%
In this work, we study fourfold-degenerate Dirac fermions protected by three MWGs and propose 3D topological magnetic Dirac insulators (TMDIs) protected by 16 MSGs (see \stable~\ref{STable:materialCons}).
Now, we present a general scheme to search candidate materials for the TMDIs.
MSG can be derived from Fedorov space group (SG) by considering different magnetic orderings.
Therefore, we can obtain magnetic materials with the candidate MSGs for the TMDIs by inspecting how the magnetic moment sets on magnetic atoms in a material.
For materials whose structural information has not been studied experimentally, this method can be used to construct possible magnetic structures, which can then be further verified to be magnetic ground state or not in experiment.
%%%%%%%%%%%%%%%%%%%%%%
\\

%%%%%%%%%%%%%%%%%%%%%%
\tocless{\subsubsection{Review of definitions and notations of magnetic space groups}}{}
%%%%%%%%%%%%%%%%%%%%%%
First, let us revisit the definition of magnetic Shubnikov space groups (SSG) $\mbb M$.
There are totally 1651 SSGs which can be further divided into four types~\cite{bradley2010}.
Among them, 230 Type-I SSGs have the form: $\mbb M= \mbb G$ where $\mbb G$ is a Fedorov SG.
Type-I SSG has no antiunitary operators.
There are 230 Type-II SSGs of the form $\mbb M=\mbb G + \mbb G T$, which are paramagnetic SGs, where $T$ denotes time-reversal symmetry.
The nonmagnetic fourfold-degenerate Dirac fermions in Type-II SSGs have been investigated~\cite{wieder2018wallpaper}.
In general, Type-III and Type-IV (Shubnikov) MSGs $\mbb M$ can be expressed as $\mbb M = \mbb H + (\mbb G - \mbb H) T$.
For Type-III MSGs, $\mbb H$ is an equi-translation and index-2 subgroup of the Fedorov SG $\mbb G$, and the rest of point group symmetries are combined with time-reversal $T$ are anti-unitary operators.
So the ``magnetic unit cell" is the same as the unit cell of paramagnetic case.
For Type-IV MSGs, $\mbb H$ is an equi-class and index-2 MSG of the Fedorov SG $\mbb G$.
This means that $\mbb H$ has the same point group as $\mbb G$, but half the translations of Bravais lattice are combined with $T$ and become antiunitary magnetic translations. 
Thus, the unit cells of Type-IV MSGs are  larger than those of paramagnetic cases.

There are two different notations for denoting MSGs: Belov-Neronova-Smirnova (BNS)~\cite{belov1957Shubnikov} setting and Openchowsky-Guccione (OG) setting \cite{opechowski1965magnetism}.
For Type-I, II, and III MSGs, these BNS and OG settings are identical.
And these three kinds of MSGs are derived from Fedorov SG $\mbb G$.
However, for Type-IV MSGs, the BNS setting is derived from Fedorov SG $\mbb H$, and the OG setting is derived from Fedorov SG $\mbb G$.
Note that $\mbb H$ and $\mbb G$ are different in general because of changes of the Bravais lattice.
The OG notation are easily used to know the information of SG $\mbb G$.
Type-III and Type-IV magnetic materials can be  constructed based on their paramagnetic unit cell with SG $\mbb G^T=\mbb G+ \mbb G T$ or its supergroups, where $\mbb G^T$ is Type-II magnetic SSGs that describes nonmagnetic phase. 
Note that $\mbb G^T$ is supergroup of magnetic subgroup $\mbb M$.
In order to distinguish with Type-III and Type-IV magnetic SSGs $\mbb M$, we use $\mbb G^T$ denotes nonmagnetic SGs. 
In this work, we use the Hermann-Mauguin symbol instead of BNS symbol for denoting nonmagnetic space group, for simplicity.
For instance, the 123th nonmagnetic space group $\mbb G^T$ is expressed as SG 123 $P4/mmm$ in Hermann-Mauguin symbol and 123.340 in BNS symbol $P4/mmm1'$. 

Before we explain a scheme for constructing magnetic materials, let us introduce the propagation ``$\mbb k$-vector".
The ``$\mbb k$-vector" describes magnetic structure~\cite{perez2015symmetry}.
Magnetic structure of crystals can be described by the Fourier series as: $\mbb{m}_{j l}=\sum_{\mbb{k}} \mbb{S}_{\mbb{k} j} \exp \left(-2 \pi i \mbb{k} \cdot \mbb{R}_{l}\right)$, where $\mbb{m}_{j l}$ is the magnetic moment of atom $j$ within the unit cell located at $\mbb R_l$.
Lattice vectors $\mbb R_l$ that satisfy $e^{-2 \pi i \mbb{k} \cdot \mbb{R}_l}=1$ ($-1$) define a translation (antiunitary translation) symmetry.
If $n$ is the minimal integer such that $n \mbb k$ is a reciprocal lattice vector and $e^{ -2 \pi i \mbb{k} \cdot \mbb{R}_n}=1$, then the magnetic unit cell is $n$ times larger than the paramagnetic one.
If $n$ is an even integer and there exist another smaller integer $m$ such that $m \mbb k$ is a reciprocal lattice vector and $e^{-2 \pi i \mbb{k} \cdot \mbb{R}_m}=-1$, a given system is described by Type-IV MSGs.
If $n$ is an odd integer, the system is described by Type-I or Type-III MSGs.
Therefore, $\mbb k$-vector contains not only the information of MSGs also the size of magnetic unit cell.
Some magnetic systems can be described by a single $\mbb k$-vector, but some need multiple $\mbb k$-vectors which belong to the same $\mbb k$ star.
A set of distinct $\mbb k$-vectors is called a star, where distinct $\mbb k$-vectors are obtained under the point group symmetry operator $R$ if $R \mbb k \neq \mbb k + \bb G$.
%%%%%%%%%%%%%%%%%%%%%%

%%%%%%%%%%%%%%%%%%%%%%
\tocless{\subsubsection{The scheme to search candidate materials with Type-III and Type-IV MSGs}}{}
%%%%%%%%%%%%%%%%%%%%%%
Based on the definition of MSGs, the general process to obtain magnetic structure from paramagnetic structure can be summarized as following several steps:

{\textbf{1. Determine the space group $\mbb G$ for paramagnetic phase.}}
As we mentioned above, magnetic materials with MSG $\mbb M$ can be constructed from some paramagnetic materials with parent SG $\mbb {G}^T$ or its supergroups $\mathbb{G}$, where $\mathbb{G} \supset \mbb G^T$.
The group-supergroup relationship between $\mbb G^T$ and $\mathbb{G}$ can be obtained by group theory or using {\SUPERGROUPS}~\cite{bilbao_supgroup,supergroups} in the Bilbao Crystallographic Server~\cite{aroyo2006bilbao1,aroyo2006bilbao2}.
The coordinate systems for supergroup and its subgroup can be related by a matrix $(P|\mbb p)$.
The form of the matrix $(P|\mbb p)$ consists two parts: rotation part $P$ with respect to the unit cell $(\mbb a, \mbb b, \mbb c)$ and translation part for the origin $O$.
After transformation, $(\mbb a', \mbb b', \mbb c')=(\mbb a, \mbb b, \mbb c)P$ and $O'= O + p_1 \mbb a + p_2 \mbb b + p_3 \mbb c$.

\textbf{2. Obtain transformation matrix of basis vectors between nonmagnetic and magnetic unit cells.}
The transformation matrix gives the information how the lattice vectors transform from nonmagnetic unit cell to magnetic one. 
For Type-III MSG, a transformation matrix can be obtained by the group-supergroup relationship as we mentioned in the first step. 
In this case, a matrix $(P|\mbb p)$ relates the nonmagnetic group $\mbb G^T$ and its supergroup $\mathbb{G}$, and this matrix is the transformation matrix between magnetic and nonmagnetic unit cells.
It is because group $\mbb G^T$ is minimal parent index-2 supergroup of MSG $\mbb M$ with transformation matrix $(I| \bb 0)$, where rotation part is identity matrix $I$.
Thus, the transformation matrix between nonmagnetic group $\mbb G^T$ and supergroup $\mathbb{G}$ is same as that between MSG $\mbb M$ and supergroup $\mathbb{G}$, i.e. the transformation matrix between the magnetic and parent unit cells.
However, for Type-IV MSG, the determinant of rotation part in transformation matrix is greater than 1 due to the change of lattice vectors, even though $\mbb{G^T}$ is still minimal parent index-2 supergroup of MSG $\mbb{M}$.
In other words, we usually need to construct Type-IV magnetic phase with MSG $\mbb{M}$, based on the supercell of paramagnetic phase with SG $\mbb{G^T}$.
For this reason, we introduce $\mbb k$-vector to describe the Type-IV case.
Note that $\mbb k$-vectors give the information of supercell with paramagnetic SG $\mbb G^T$.
Therefore, for a paramagnetic SG $\mbb G^T$ and chosen $\mbb k$-vectors, one can obtain the possible subgroup MSG $\mbb M$ and the corresponding transformation matrix between magnetic unit cell and crystal unit cell with the help of program {\kSubgroupsmag} tool in Bilbao Crystallographic Server~\cite{bilbao_kvec}.

\textbf{3. Build up the magnetic unit cell and restrict spin configuration to magnetic atoms.} 
Once the transformation matrix between the paramagnetic unit cell and magnetic unit cell is obtained, we can build up the magnetic lattice according to the transformation matrix without considering magnetic ordering. 
Even if magnetic lattice are built up by following transformation matrix, different spin configurations give different magnetic structures. 
Therefore, it is important to restrict magnetic moments to the magnetic atoms that preserve the symmetry of the corresponding MSG. 
These information can be accessible from {\MWYCKPOS}  tool~\cite{gallego2012magnetic,bilbao_mwp} in the Bilbao Crystallographic Server.
The positions and magnetic moments of the magnetic atoms are crucial for constructing the corresponding magnetic groups.

Following the above three steps, we can construct a magnetic unit cell with target MSG, based on a paramagnetic unit cell which contains magnetic atoms located at proper Wyckoff positions (WPs).
Below, we explain how candidate magnetic materials for TMDIs with Type-III and IV MSGs can be found.
%%%%%%%%%%%%%%%%%%%%%%
\\

%%%%%%%%%%%%%%%%%%%%%%
\tocless{\subsubsection{Type-III magnetic materials}}{}
%%%%%%%%%%%%%%%%%%%%%%
As explained above, Type-III magnetic materials can be directly obtained by taking into account the magnetic ordering in paramagnetic materials with SG $\mbb G^T$.
For Type-III material, the magnetic unit cell is same as paramagnetic unit cell, thus the transformation matrix between them is identity.
Finally, the location and direction of the magnetic moments can be determined by the symmetry of corresponding MSG.
In this work, we are interested in 5 Type-III MSGs, 100.173 $P4'b'm$, 125.366 $P4'/nb'm$, 125.368 $P4'/n'b'm$, 127.390 $P4'/mb'm$, and 127.392 $P4'/m'b'm$.
Let us take MSG 100.173 $P4'b'm$ as a representative example.
Magnetic material with MSG $P4'b'm$ can be obtained from nonmagnetic material in $\mbb G^T=P4bm$ (SG 100) with magnetic atoms, for instance, located at WP $4c$.
Next, keep the crystal unit cell and consider magnetic moments on magnetic atoms.
The symmetries of MSG $P4'b'm$ constrains the components of magnetic moment at representative position $(x,x+\frac{1}{2},z)$ for WP $4c$ as  $( m_x,-m_x,0)$.
Finally, we can obtain a magnetic structure with MSG $P4'b'm$.

\begin{figure}[t!]
\centering
\includegraphics[width=0.9\textwidth]{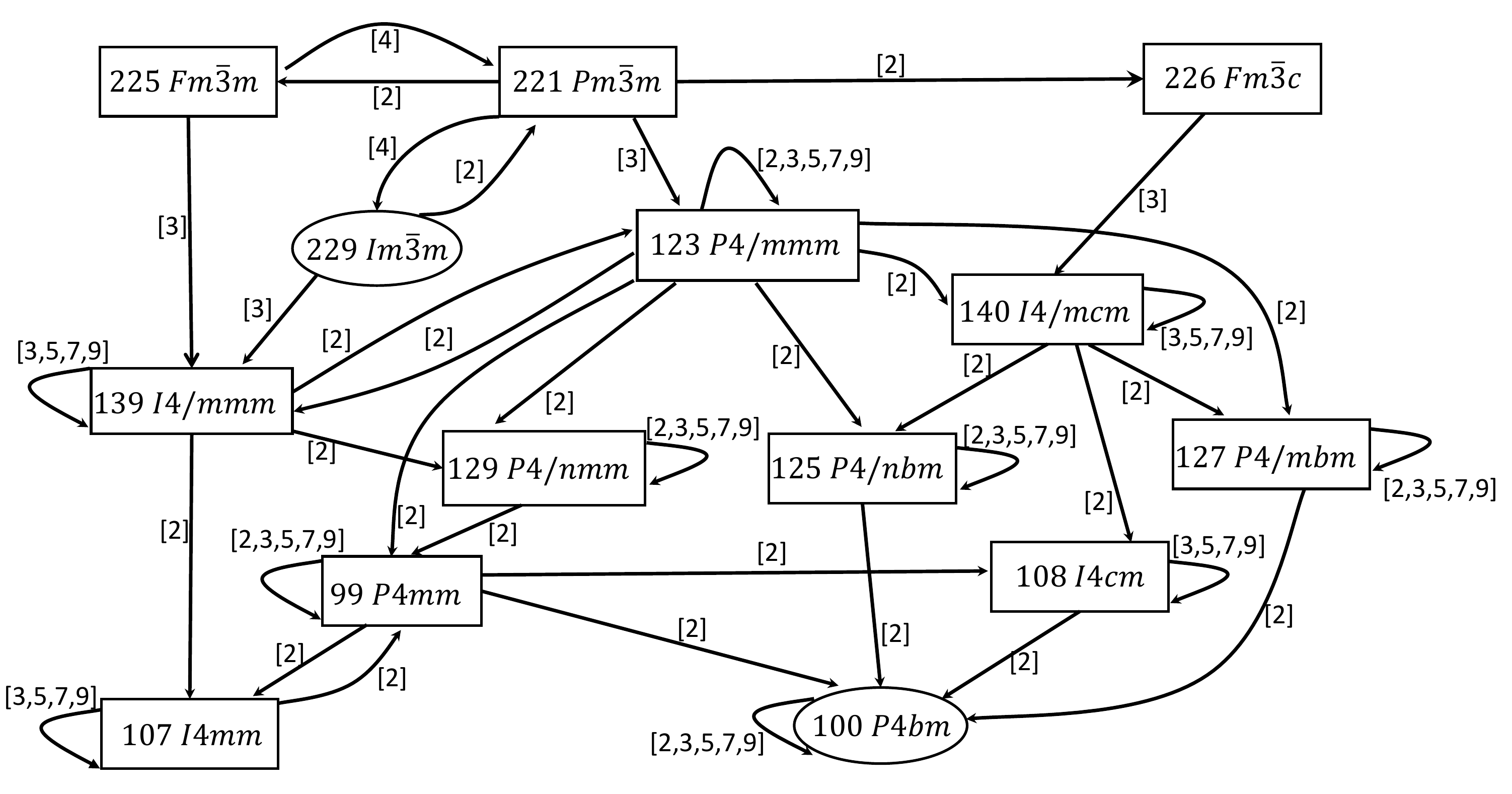}
\caption{
{\bf Lattice of maximal subgroups}
from $Im\bar3m$ (SG 229) to $P4bm$ (SG 100).
The numbers in square brackets are the index of a group-subgroup relation.
}
\label{sfig:maxgroups}
\end{figure}

It is also possible to obtain this MSG from the supergroups of $P4bm$ with appropriate atomic locations and magnetic moment.
The chains of maximal subgroups from $\mbb G^T=Im\bar3m$ (SG 229) to $P4bm$ (SG 100) are shown in \sfig~\ref{sfig:maxgroups}.
Therefore, we can also construct a magnetic structure with MSG $P4'b'm$ by starting from paramagnetic materials with supergroups of $P4bm$, such as $\mbb G^T$ 99 $P4mm$, 108 $I4cm$, 125 $P4/nbm$, and 127 $P4/mbm$ etc.
We take a supergroup $\mbb G^T=$ SG 125 as an example.
With the help of {\SUPERGROUPS}, the transformation matrix between SG 125 to 100 is $(P | \bb p)$,
\ba
(P | \bb p) =
\left(
\begin{array}{ccc|c}
1 & 0 & 0 & \frac{1}{4} \\
0 & 1 & 0 & \frac{1}{4} \\
0 & 0 & 1 & 0
\end{array}
\right).
\ea
This transformation matrix means that the original parent lattice characterized by lattice vectors $(\mbb a, \mbb b, \mbb c)$ and origin $O$ goes to magnetic unit cell with $(\mbb a', \mbb b', \mbb c')=(\mbb a, \mbb b, \mbb c)P=(\mbb a, \mbb b, \mbb c)$ and $O'= O + p_1 \mbb a + p_2 \mbb b + p_3 \mbb c= O+ (\frac{1}{4}, \frac{1}{4},0)$. 
After this transformation, WPs in SG 125 changes, for example, $4e/4f$ in SG 125 turns to $4c$ in SG 100. 
Hence, if there are magnetic atoms at WP $4e$ in SG 125, then these atoms go to WP $4c$ in magnetic unit cell with MSG $P4'b'm$. 
Finally, if we let magnetic moments on those atoms at WP $4c$ preserve the symmetries of MSG $P4'b'm$, then we can get a magnetic material with MSG $P4'b'm$.

In this work, we propose DyB$_4$ with Type-III MSGs 127.392 $P 4'/m'b'm$ as a TMDI candidate.
The corresponding magnetic structure, the $\Gamma_4$ phase, has been discussed in experiment~\cite{sim2016Spontaneous}.
For this case, our scheme is also useful to understand how magnetic structures with MSGs 127.392 $P4'/m'b'm$ comes from paramagnetic phases with $\mbb G^T=P4/mbm$ (SG 127).
Similarly, magnetic material with other Type-III MSGs can be constructed by the same procedure.
%%%%%%%%%%%%%%%%%%%%%%
\\

%%%%%%%%%%%%%%%%%%%%%%
%%%%%%%%%%%%%%%%%%%%%%

%%%%%%%%%%%%%%%%%%%%%%
\tocless{\subsubsection{Type-IV magnetic materials}}{}
%%%%%%%%%%%%%%%%%%%%%%
In the case of Type-IV MSG, the procedures involves more steps since the paramagnetic unit cell is usually changed.
Instead of getting the transformation matrix from group-supergroup relationship, we use $\mbb k$-vector to obtain it with the help of {\kSubgroupsmag} tool.
There are 5 Type-IV MSGs (99.169 $P_C4mm$, 107.232 $I_c 4mm$, 123.349 $P_C4/mmm$, 129.421 $P_C4/nmm$, and 139.540 $I_c4/mmm$) which can support Type-IV TMDIs, and 6 Type-IV MSGs (101.186 $P_{I}4_2cm$, 102.192 $P_{C}4_2nm$,132.458 $P_{I}4_2/mcm$,134.480 $P_c 4_2/nnm$, 136.504 $P_c4_2/mnm$ and 138.530 $P_I4_2/ncm$) that can support Type-III TMDIs as shown in \stable~\ref{Table:MSGs}. 
Among these, we take three of them, 99.169 $P_C4mm$, 123.349 $P_C4/mmm$, and 129.421 $P_C4/nmm$ as examples to illustrate our scheme.
In \stable~\ref{STable:materialCons}, in each row, we give the possible constructions from paramagnetic space groups for each MSG. For each paramagnetic SG, we list the information of allowed WPs [and the whole star(s)] of magnetic atoms in the parent paramagnetic SG and corresponding $\mbb k$-vectors which determine the supercell of the parent material.
If the $\mbb k$-vectors are symmetry related (belonging to the same star), we just list one of them.
Note that we only tabulate the cases where WPs have the multiplicity equal to or less than 4 and the components of $\mbb k$-vector equal to $\hf$, for simplicity of structure.
With this table, we can easily obtain the transformation matrix between magnetic unit cell and crystal unit cell, and further construct magnetic materials with the target MSG.
Among the 5 Type-IV MSGs, we take 129.421 $P_C4/nmm$ to illustrate our scheme.
It shows that MSG 129.421 $P_C4/nmm$ can be constructed from paramagnetic material with SG 123 $P4/mmm$, SG 139 $I4/mmm$ and SG 221 $Pm\bar3 m$.
For each case, the corresponding WPs of magnetic atoms and the $\mbb k$-vector are also given.
MSG 129.421 $P_C4/nmm$ in the BNS setting is identical to MSG 123.16.1014 $P_P4/m'mm$ in the OG setting.
The first number 129 (123) of the BNS (OG) symbol denotes a Fedorov SG $\mbb H =$ 129 $P4/nmm$ ($\mbb G =$ 123 $P4/mmm$).
Note that $\mbb H= P4/nmm$ is an equi-class subgroup of $\mbb G = P4/mmm$.
Thus, the simplest way is to consider a paramagnetic material with $\mbb G^T =$ 123 $P4/mmm$.
Now, we locate magnetic atoms, for example, at WP $1a$, and consider $\mbb k$-vector $(\frac{1}{2},0,0)$ [and the whole star(s)], in order to find a candidate material with MSG 129.421 $P_C4/nmm$.
With the help of program {\kSubgroupsmag} tool, one can obtain a transformation matrix, which shows that lattice vectors should be doubled along $\mbb a$ and $\mbb b$ directions.
(These directions are defined for the paramagnetic unit cell in SG 123.)
At last, WP $1a$ in SG 123 turns to $4f$ in MSG 129.421 $P_C4/nmm$.
Keeping magnetic moments on atoms at $4f$ preserving symmetries of MSG 129.421 $P_C4/nmm$, we can obtain a magnetic material with MSG 129.421 $P_C4/nmm$. 
At the end of this section, we give a real material to discuss this scheme in detail.

At the same time, materials with the Type-IV MSG $\mbb M$ can also be induced by other paramagnetic materials with SGs, which are supergroups of $\mbb G^T$. 
As we mentioned above, in case of MSG 129.421 $P_C4/nmm$ which is derived from Fedorov SG $\mbb G$ 123, materials with the supergroups of SG $\mbb G^T$ 123: $P4/mmm$ (123), $I4/mmm$ (139), $Pm\bar3m$ (221) can also be used to construct magnetic phase.
Similarly, we take 99.169 $P_C4mm$ as another example.
This group is derived from Fedorov space $\mbb G$ $P4mm$ (99).
The supergroups of paramagnetic SG $\mbb G^T$ $P4mm$ are $P4mm$ (99), $I4mm$ (107), $P4/mmm$ (123), $P4/nmm$ (129), $I4/mmm$ (139), $Pm\bar3m$ (221), $Fm\bar3m$ (225), and $Im\bar3m$ (229).
Then we can obtain 99.169 $P_C4mm$ based on such paramagnetic space groups with related magnetic atoms at proper WPs and considering reasonable $\mathbf{k}$-vectors.
Here we only consider magnetic atoms located at WPs with multiplicity up to 4 for the simplicity of the magnetic structure.
Therefore, even though 99.169 can be obtained from SG 221 with magnetic atoms, for instance, at $6e/f$, we discard this case.
It is the same reason that SGs 225 and 229 are absent in \stable~\ref{STable:materialCons}.
The possible parent paramagnetic SGs can be obtained according to the group-supergroup relationship.
The lattice of maximal subgroups from $Im\Bar{3}m$ (229) to $P4mm$ (100) contains all parent SGs that we are interested in,  as shown in \sfig~\ref{sfig:maxgroups}.
In next section, we give FeSe, FeTe, and AlGeMn as examples to show how to construct magnetic materials with MSG 99.169 $P_C4mm$ from paramagnetic phase with magnetic atoms at WP $2a$ in $\mbb G^T$ 129 and with $\mbb k$-vector $(\hf,\hf,0)$.
%%%%%%%%%%%%%%%%%%%%%%

%%%%%%%%%%%%%%%%%%%%%%
{
\renewcommand{\arraystretch}{1.2}
\begin{table}[t!]
\centering
\begin{minipage}{0.99\textwidth}
\caption{
{\bf Type-IV MSGs from paramagnetic space groups (PSGs).}
For each MSG, the possible constructions from parent PSGs $\mbb G^T = \mbb G + \mbb G T$ are given.
For each parent PSG $\mbb G^T$, the Wyckoff positions (WPs) of magnetic atoms and the corresponding propagation magnetic $\mathbf{k}$-vectors [and the whole star(s)] are listed.
We list one representative $\mbb k$-vector when $\mbb k$-vectors are symmetry related and belongs to the same star.
Note that `---' means a given MSG cannot be constructed from its parent paramagnetic SG or magnetic atoms are located at WPs with multiplicity more than 4.
For each MSG, both the BNS and OG (in parentheses) numbers are given.
}
\label{STable:materialCons}
\end{minipage}
%
%\resizebox{\textwidth}{30mm}{
\begin{tabular*}{0.99\textwidth}{c|c|c|c|c|c|c|c|c|c|c|c|c}
\hline \hline
%\diagbox{MSG}{WP,$\mathbf{k}$-vector}{$\mbb G$}
\multirow{2}{*}{MSG} & \multicolumn{12}{c}{PSG $\mbb G^T$ (WPs and $\mathbf{k}$-vector)} \\ \cline{2-13}
& \multicolumn{2}{c|}{99 $P4mm$} & \multicolumn{2}{c|}{107 $I4mm$} & \multicolumn{2}{c|}{123 $P4/mmm$} & \multicolumn{2}{c|}{129 $P4/nmm$} & \multicolumn{2}{c|}{139 $I4/mmm$} & \multicolumn{2}{c}{221 $Pm\bar3m$}
\\ \hline
\multirowcell{4}{99.169 $P_C4mm$ \\ (OG: 99.7.829)} & \multirowcell{3}{$1a/b$, \\ $2c$, \\ $4d/e/f$} & \multirowcell{3}{$(\frac{1}{2}, 0, 0)$} & \multirowcell{4}{$4b$} & \multirowcell{4}{$(\frac{1}{2}, 0, 0)$, \\ \\ $(\frac{1}{2},\frac{1}{2}, 0)$} & \multirowcell{3}{$2h/g$, \\ $4i/j/k/$ \\ $l/m/n/o$} & \multirowcell{3}{$(\frac{1}{2}, 0, 0)$} & \multirowcell{4}{$2a/b$, \\ \\ $4d/e/f$} & \multirowcell{4}{$(\frac{1}{2}, 0, 0)$, \\ \\ $(\frac{1}{2}, \frac{1}{2}, 0)$} & \multirowcell{4}{$4d$} & \multirowcell{4}{$(\frac{1}{2}, 0, 0)$, \\ \\ $(\frac{1}{2}, \frac{1}{2}, 0)$} & \multirowcell{4}{---} & \multirowcell{4}{---}
\\
& & & & & & & & & & & & \\
& & & & & & & & & & &  & \\
\cline{2-3} \cline{6-7}
& $2c$, $4d/e/f$ & $(\frac{1}{2}, \frac{1}{2}, 0)$ & & & $4i/l/m/n/o$ & $(\frac{1}{2}, \frac{1}{2}, 0)$ & & & & & &
\\ \hline
%%%%%%%%%%%%
\multirowcell{3}{123.349 $P_C4/mmm$ \\ (OG: 123.11.1009)} & \multirowcell{3}{---} & \multirowcell{3}{---} & \multirowcell{3}{---} & \multirowcell{3}{---} & \multirowcell{2}{$2h/g$, \\ $4i/l/m/n/o$} & \multirowcell{2}{$(\frac{1}{2}, 0, 0)$} & \multirowcell{3}{---} & \multirowcell{3}{---} & \multirowcell{3}{$4d$} & \multirowcell{3}{$(\frac{1}{2},0,0)$, \\ \\ $(\frac{1}{2},\frac{1}{2},0)$} & \multirowcell{3}{---} & \multirowcell{3}{---} \\
& & & & & & & & & & & & \\
\cline{6-7} 
& & & & & $4i/l/m/n/o$ & $(\frac{1}{2}, \frac{1}{2}, 0)$ & & & & & &
\\ \hline
%%%%%%%%%%%%
\multirowcell{5}{129.421 $P_C4/nmm$ \\ (OG: 123.16.1014)} & \multirowcell{5}{---} & \multirowcell{5}{---} & \multirowcell{5}{---} & \multirowcell{5}{---} & \multirowcell{3}{$1a/b/d$, $2h/g$, \\ $2f/e$,$4i/j/$ \\ $k/l/m/n/o$} & \multirowcell{3}{$(\frac{1}{2}, 0, 0)$} & \multirowcell{5}{---} & \multirowcell{5}{---} & \multirowcell{5}{$4c/d$} & \multirowcell{5}{$(\frac{1}{2}, 0, 0)$, \\ \\ $(\frac{1}{2}, \frac{1}{2}, 0)$} & \multirowcell{5}{$1a/b$, \\ \\ $3c/d$} & \multirowcell{5}{$(\frac{1}{2}, 0, 0)$} \\
& & & & & & & & & & & & \\
& & & & & & & & & & & & \\
\cline{6-7}
& & & & & \multirowcell{2}{$2f/e$, $4i/j/$ \\ $k/l/m/n/o$} & \multirowcell{2}{$(\frac{1}{2}, \frac{1}{2}, 0)$} & & & & & & \\
& & & & & & & & & & & &
\\ \hline \hline
\end{tabular*}
%}
%\end{sidewaystable}
\end{table}
}
%%%%%%%%%%%%%%%%%%%%%%

%%%%%%%%%%%%%%%%%%%%%%
Finally, let us take NdTe$_2$ClO$_5$ as a representative example to demonstrate how to use \stable~\ref{STable:materialCons} for searching Type-IV candidate magnetic materials.
NdTe$_2$ClO$_5$ with parent SG 123 $P4/mmm$ has magnetic atoms Nd located at $1a$ WP.
This information is contained in \stable~\ref{STable:materialCons}, which also indicates that a magnetic system with MSG 129.421 can be obtained from considering the magnetic propagation $\mathbf{k}$-vector are $(\frac{1}{2},0,0)$ and $(0,\frac{1}{2},0)$ (they belong to the same star) and magnetic atoms located at WPs $1a/b/d$, $2h/g$, $2/f/e$, and $4i/j/k/l/m/n/o$.
With the help of {\kSubgroupsmag} tool~\cite{bilbao_kvec}, we can get the transformation matrix $(P | \bb p)$,
\ba
(P | \bb p) =
\left(
\begin{array}{ccc|c}
2 & 0 & 0 & 0 \\
0 & 2 & 0 & 0 \\
0 & 0 & 1 & 0
\end{array}
\right).
\ea
This transformation matrix suggests that to obtain a magnetic unit cell with standard Bilbao convention, the parent unit cell with a basis vector $(\mbb a, \mbb b, \mbb c)$ and an origin $O$ should transform to $(\mbb a', \mbb b', \mbb c')=(\mbb a, \mbb b, \mbb c)P$ and $O'= O + p_1 \mbb a + p_2 \mbb b + p_3 \mbb c$, i.e. $(\mbb a', \mbb b', \mbb c')=(2\mbb a, 2\mbb b, \mbb c)$ and $O'=O$.
Hence, the unit cell is doubled along $\mbb a$ and $\mbb b$ directions.
This leads that WP $1a$ of Nd atoms in SG 123 corresponds to $4f$ in MSG 129.421.
Now, according to the symmetry of MSG 129.421, we can determine the direction of the magnetic moments.
In this example, WP $4f$ has four positions: $(\hf,0,0)$, $(\hf,\hf,0)$, $(0, \hf,0)$, $(0,0,0)$.
First, let us suppose that the magnetic moment on $(\hf,0,0)$
is $(m_x,m_y,m_z)$. 
Consider the generators of MSG 129.421: $\{ E | \mbb 0 \}$, $\{ C_{4z}| (\hf,0,0) \}$, $\{ C_{2x}| (\hf,0,0)\}$, $\{ C_{2y}| (0,\hf,0) \}$, inversion $\{ I | \mbb 0 \}$ and anti-unitary translation $\Tg = \{T |\hf,\hf, 0 \}$, where $\{ E | \mbb 0 \}$ is identity.
Under $\{ C_{4z}| (\hf,0,0) \}$, a position and magnetic moment transforms as $(\hf,0,0) \to (\hf,\hf,0)$ and  $(m_x,m_y,m_z) \to (-m_y,m_x,m_z)$.
For other symmetries, $\{ C_{2x}| (\hf,0,0)\}: (\hf,0,0) \to (0,0,0)$, $(m_x,m_y,m_z) \to (m_x,-m_y,-m_z)$, $\{ C_{2y}| (0,\hf,0) \}: (\hf,0,0) \to (\hf,\hf,0)$, $(m_x,m_y,m_z) \to (-m_x,m_y,-m_z)$, $\{ I | \mbb 0 \}: (\hf,0,0) \to (\hf,0,0)$, $(m_x,m_y,m_z) \to (m_x,m_y,m_z)$, and $\Tg = \{T |\hf,\hf, 0 \}: (\hf,0,0) \to (0,\hf,0)$, $(m_x,m_y,m_z) \to (-m_x,-m_y,-m_z)$.
Combining above transformations, we obtain $m_x = m_y$, $m_z = -m_z =0$.
In summary, we obtain the magnetic moment configurations on $4f$ are $(\hf,0,0 | m_x,m_x,0)$, $(\hf,\hf,0 | -m_x,m_x,0)$, $(0, \hf,0 | -m_x,-m_x,0)$, and $(0,0,0 | m_x,-m_x,0)$.
One can determine possible magnetic moments directly by using {\MWYCKPOS} tool~\cite{gallego2012magnetic,bilbao_mwp}.
In next section, we take N$_3$TaTh as an example to show MSG 129.421 $P_C4/nmm$ can also be constructed from paramagnetic SG $\mbb G^T$ 127 $P4/mbm$.
Based on this scheme, Type-IV magnetic materials can be studied. 
%
%%%%%%%%%%%%%%%%%%%%%%
\\

%%%%%%%%%%%%%%%%%%%%%%
\begin{figure}[b!]
\includegraphics[width=0.95\textwidth]{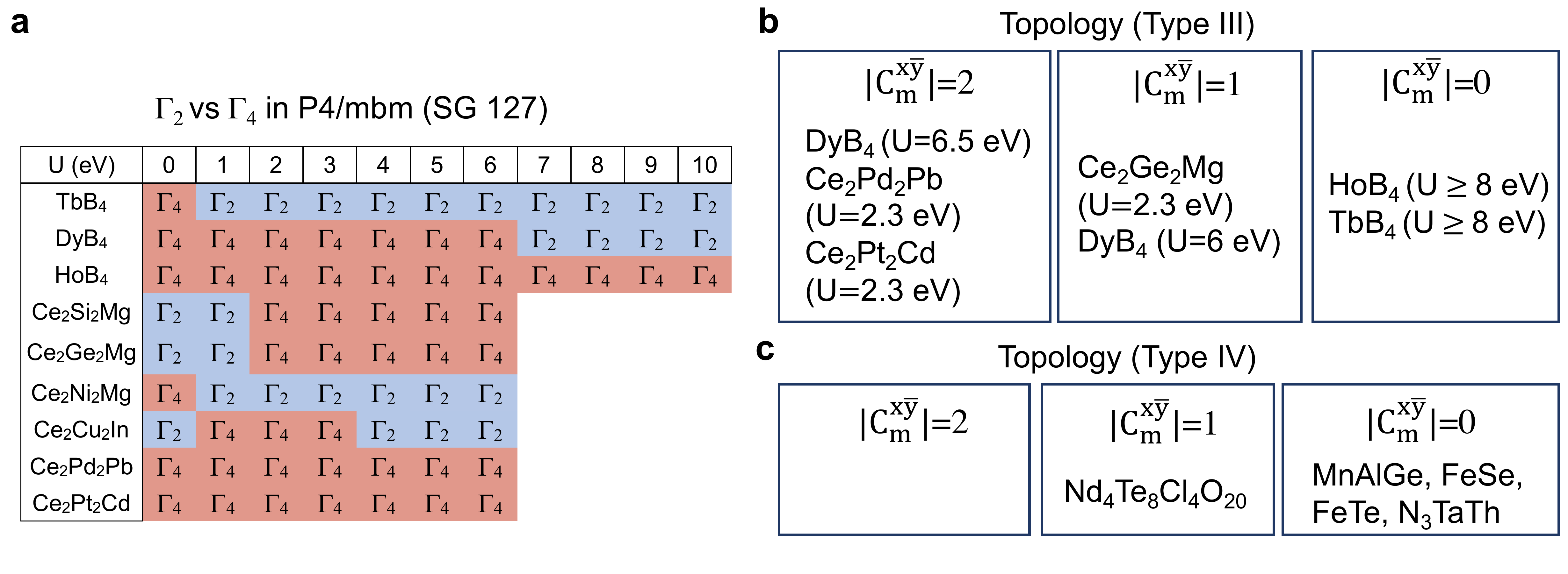}
\caption{
{\bf Material search of the magnetic wall-paper group.}
(a) Magnetic ground state using the DFT+$U$ method
($U$ is the on-site Coulomb interaction).
$\Gamma_4$ state can host magnetic Dirac insulating phases.
(b)-(c) Candidate materials for the (b) Type-III and (c) Type-IV TMDIs.
}
\label{sfig:candidates}
\end{figure}
%%%%%%%%%%%%%%%%%%%%%%

%%%%%%%%%%%%%%%%%%%%%%
\section{Details of ab initio calculations}
\label{app:DFT}
%%%%%%%%%%%%%%%%%%%%%%
For the computation of the band structure, we employ the Vienna ab initio simulation package (VASP)~\cite{kresse1996efficient} with the projector augmented-wave method (PAW)~\cite{kresse1999ultrasoft}.
The generalized gradient approximation (PBE-GGA) is employed for exchange-correlation potential~\cite{perdew1998perdew}.
We used the default VASP potentials and a 400 eV cutoff.
A $8 \times 8 \times 14$ Monkhorst-pack k-point mesh was used.
The spin-orbit coupling is taken into to account due to the presence of the heavy rare-earth atoms in the unit cell.
To capture correlation effect in localized $4f$ electron, on-site Coulomb interaction is taken into account with $U$ =6.5 eV and $J$ (Hund's coupling) = 1 eV in DyB$_4$ and $U$=6 eV and $J$=0 eV in NdTe$_2$ClO$_5$ [$J$=0 eV value is used for the other materials except $R$B$_4$ ($R$=Tb, Dy, Ho)].
The slab band structures and Wilson spectrum are obtained from the Wannier functions by using WANNIER90~\cite{mostofi2014updated} and WannierTools~\cite{wu2018wanniertools} packages.
We symmetrized the Wannier Hamiltonian by using WannSymm code~\cite{wannsymm} for each material to keep the corresponding MSG symmetries.
We employ the experimental crystal structure of DyB$_4$~\cite{sim2016Spontaneous} without structural relaxation. 
The information of crystal structure for candidate materials are summarized in \stable~\ref{tab:icsd}.
For the cases of TbB$_4$ and HoB$_4$, the pseudopotential of the corresponding rare-earth atom is replaced with the same the structure.
The crystal structures in Figs. 3 and 4 in the main text and \sfigs~\ref{sfig:Ce2Pd2Pb}-\ref{sfig:AlGeMn} are drawn using the VESTA package~\cite{Vesta}.
The details on how projectors are chosen to get a Wannier function tight-binding model for candidate materials are provided in \stable~\ref{tab:icsd}.
%

%%%%%%%%%%%%%%%%%%%%%%
\begin{figure}[t!]
\includegraphics[width=0.7\textwidth]{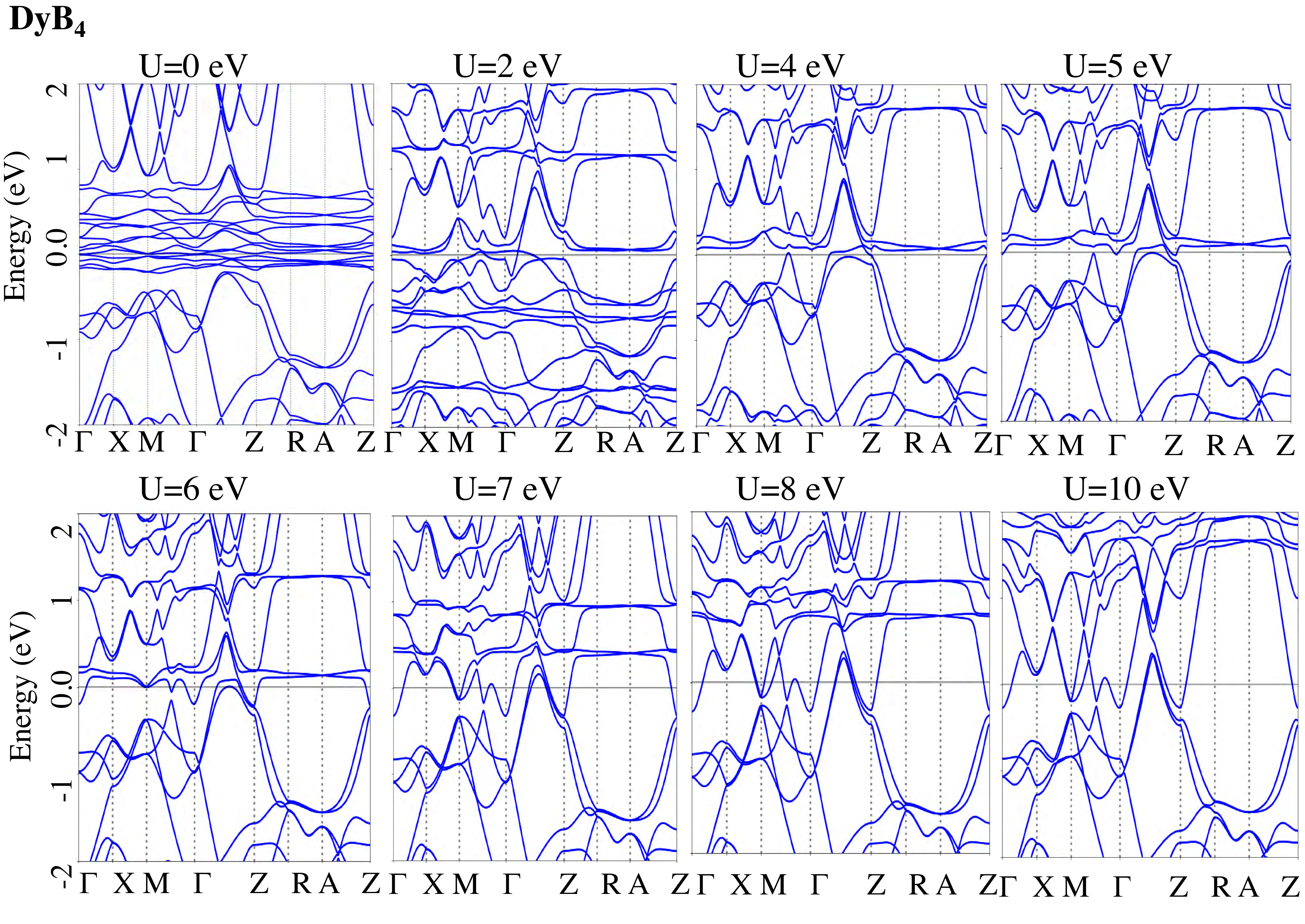}
\caption{
{\bf Band structures of DyB$_4$ as a function of $U$ from 0 to 10 eV.}
When $4f$ electrons give rise to a magnetic ground state, they are expected to be in a localized limit, which itinerant $4f$ electrons in the conventional first principles calculation could not describe.
For this reason, we employ dubbed DFT+ $U$ method.
The band structures of DyB$_4$ are shown for a range of $U$ from 0 eV to 10 eV.
The Hund's coupling $J$ is fixed to 1 eV through the calculations.
More than $U=$7 eV, the band structures near the Fermi level are not change a lot,
because localized $4f$ electrons are pushed away from the Fermi level at large $U$.
Note that $\Gamma=(0,0,0)$, $X=(\pi,0,0)$, $M=(\pi,\pi,0)$, $Z=(0,0,\pi)$, $R=(\pi,0,\pi)$, and $A=(\pi,\pi,\pi)$.
}
\label{sfig:DFT+U}
\end{figure}
%%%%%%%%%%%%%%%%%%%%%%

%%%%%%%%%%%%%%%%%%%%%%
\begin{figure}[t!]
\includegraphics[width=0.7\textwidth]{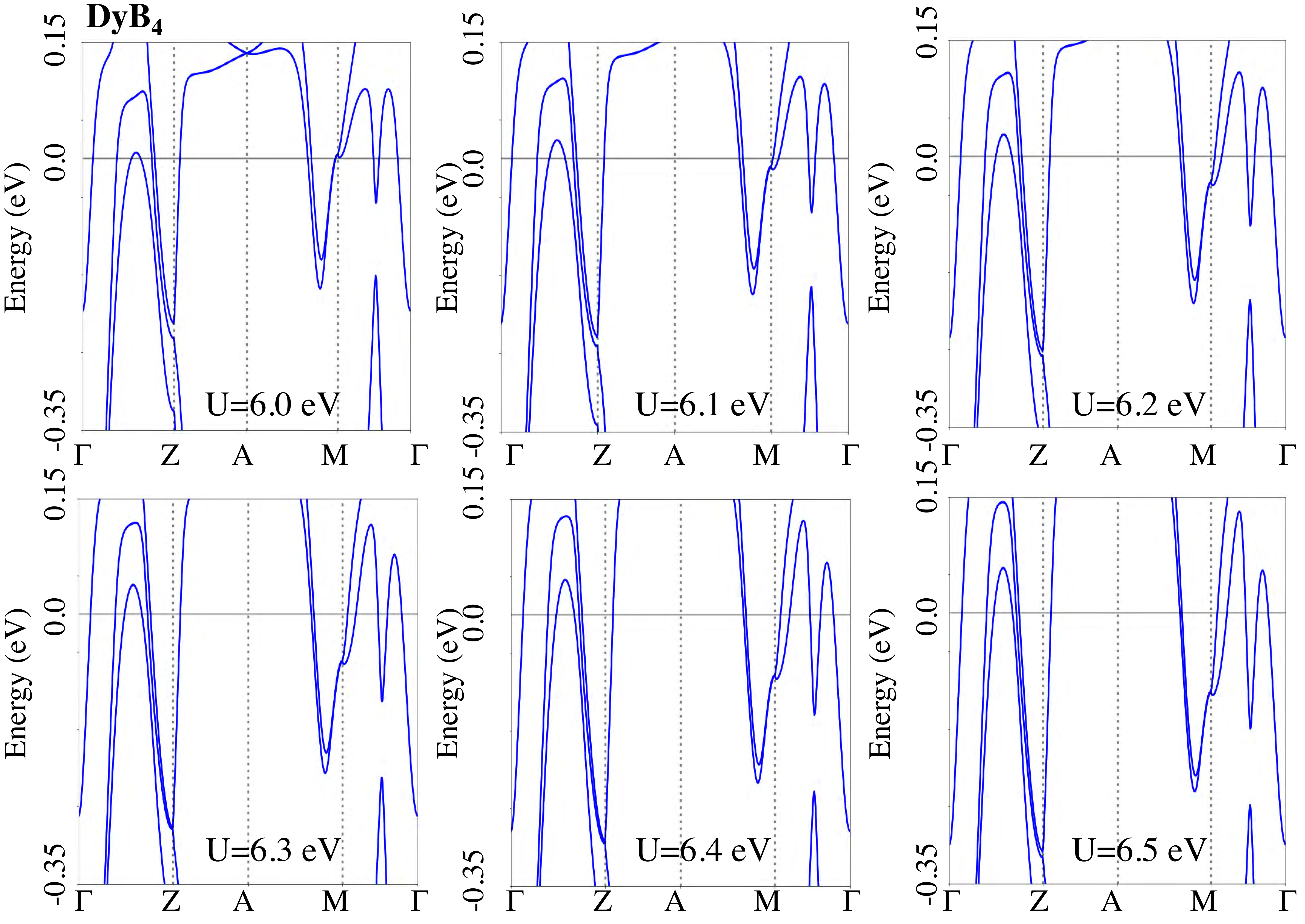}
\caption{
{\bf Band structures of DyB$_4$ as a function of $U$ from 6.0 to 6.5 eV along $\Gamma$-$Z$-$A$-$M$-$G$ line.}
Along $\Gamma$ to $Z$, the gap feature becomes narrower as $U$ is changed from 6.0 eV to 6.4 eV and is opened at $U$ = 6.5 eV.
Note that $\Gamma=(0,0,0)$, $Z=(0,0,\pi)$, $A=(\pi,\pi,\pi)$, and $M=(\pi,\pi,0)$.
}
\label{sfig:DFT+U:detail}
\end{figure}
%%%%%%%%%%%%%%%%%%%%%%

%%%%%%%%%%%%%%%%%%%%%%
\begin{figure}[t!]
\includegraphics[width=0.5\textwidth]{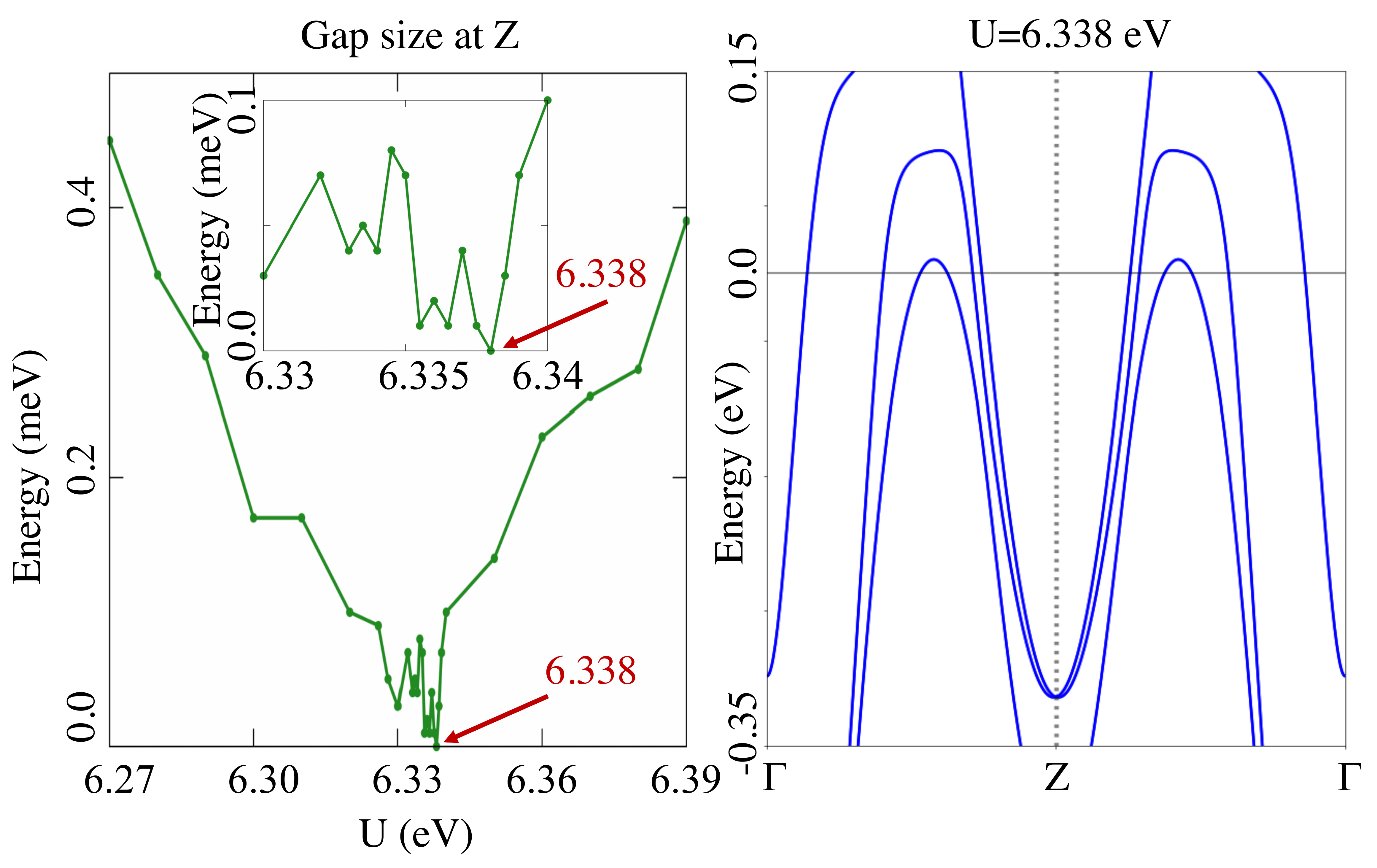}
\caption{
{\bf A gap closing at $Z$ in DyB$_4$.}
(a) The gap size at $Z$ as a function of $U$. The gap is closed at $U=$6.338 eV, where is pointed by red arrows.
Detailed features around $U=$6.338 eV are provided in the inset.
(b) The band structure around $Z$ at $U=$ 6.338 eV.
Energy gaps less than 0.0001 eV are regarded to be closed.
}
\label{sfig:gap_closing}
\end{figure}
%%%%%%%%%%%%%%%%%%%%%%

We study the competitions between the $\Gamma_2$ and $\Gamma_4$ states in rare-earth tetraborides and other compounds with SG 127 $P4/mbm$ based on the density functional theory (DFT) calculations.
The result is summarized in \sfig~\ref{sfig:candidates}.
Other rare-earth atoms near Dy among Lanthanoids are substituted by Dy to change the number of $4f$ electrons such as $R$B$_4$ with $R$=Tb and Ho.
Experimentally, they all exhibit antiferromagnetism at low temperature.
Specifically TbB$_4$~\cite{matsumura2007non} and DyB$_4$~\cite{koehler1982neutron} have been reported to show three-dimensional $\Gamma_2$ ground states, while
HoB$_4$ would be an incommensurate magnetic state or a frustrated spin structure.
Here, the crystal structures are fixed with the experimental structure of DyB$_4$ to understand how their electronic structures could affect their magnetic structure.
In weakly correlated regime (small $U$), $4f$ electrons would be like a normal metallic (itinerant) phase.
Large $U$ drives $4f$ electrons into a localized moment.
In all cases, the $\Gamma_4$ state is expected at small $U$, where itinerant $4f$ states reside near the Fermi level.
Bulk band structures of DyB$_4$ for various values of $U$ are summarized in \sfig~\ref{sfig:DFT+U}.
In HoB$_4$, the $\Gamma_4$ state has lower energy than the $\Gamma_2$ state does regardless of $U$.
HoB$_4$ might be an ideal candidate to retain a $\Gamma_4$ state.
On the other hand, TbB$_4$ has the stable $\Gamma_2$ states at large $U$.
Interestingly, DyB$_4$ shows the $\Gamma_4$ would be more stable than $\Gamma_2$ phases at $U$ $\leq$ 6.5 eV [see \sfig~\ref{sfig:candidates}(a)].
Around $U$ = 6 eV, the $\Gamma_4$ state has $\mchd=-1$, while at $U$=6.5 eV it becomes $\mchd=-2$.
Figure~\ref{sfig:DFT+U:detail} shows band structures on the $k_x=k_y$ plane along $\Gamma$-$Z$-$A$-$M$-$\Gamma$ between $U=$6.0 eV and $U=$6.5 eV.
The DFT+$U$ calculations as a function of $U$ indeed show a gap closing at $U =$6.338 eV, as summarized in \sfig~\ref{sfig:gap_closing}.
This gap closing induce a topological phase transition from a topological phase with $\mchd=-1$ at $U=$6.0 eV to one with $\mchd=-2$ at $U=$6.5 eV.
(The detailed analysis on this topological phase transition can be found in \sn~\ref{app:phasetransition}.)
%
%These results imply that one can manipulate magnetism in the localized DyB$_4$ close to the $\Gamma_4$ state by enhancing the kinetic energy (or reducing the strength of $U$).
%
We also examine Ce$_2$$A_2B$ series.
(see \sfig~\ref{sfig:candidates} for $A$ and $B$.)
Most of them exhibit $\Gamma_4$ ground state.
However, their gap features are not seen clearly and the bulk bands hide the surface states.
%

%%%%%%%%%%%%%%%%%%%%%%
\begin{figure}[t!]
\includegraphics[width=0.98\textwidth]{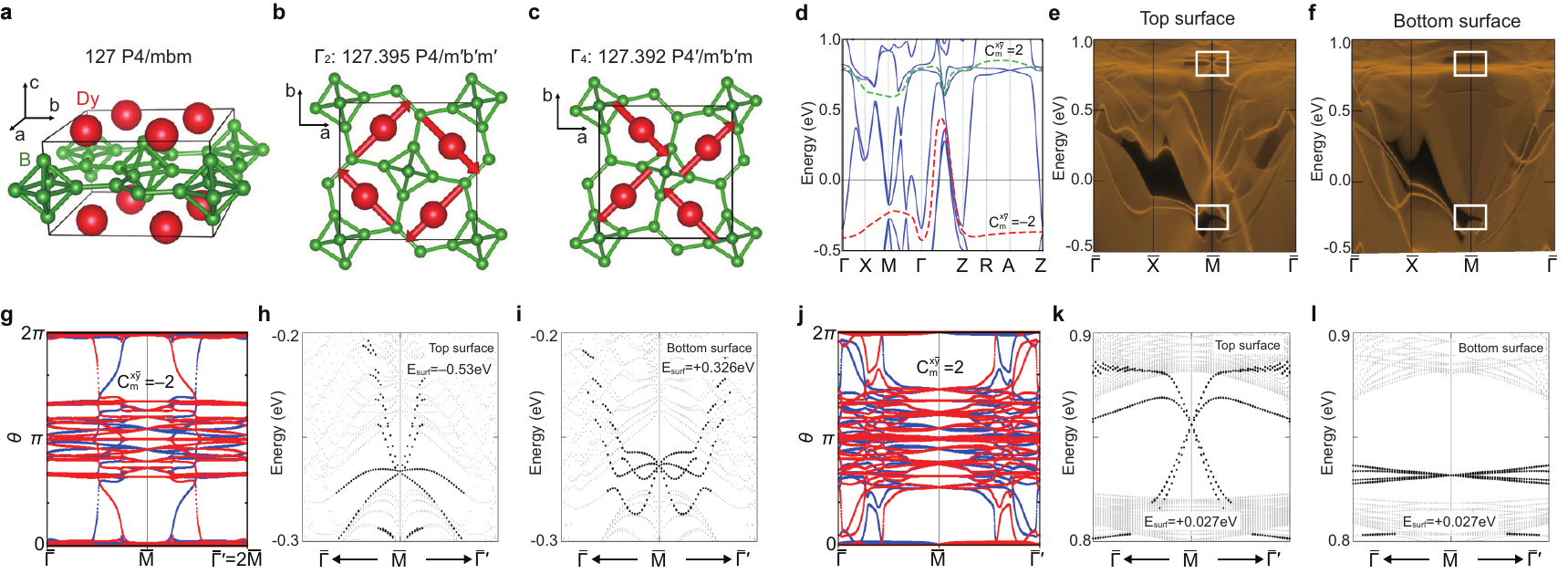}
\caption{
{\bf Type-III material with $\mchd=\pm 2$, DyB$_4$ at $U$ = 8 eV.}
(a) Tetragonal crystal structure of DyB$_4$.
(b)-(c) Magnetic spin structures in (b) $\Gamma_2$ and (c) $\Gamma_4$ states.
$\Gamma_4$ state corresponds Type-III MSG 127.392 $P4'/m'b'm$ with (001)-surface MWG $p4'g'm$.
(d) Bulk band structure from the DFT+$U$ calculations ($U$ = 8 eV and $J$ = 1 eV) for the $\Gamma_4$ state.
The red dashed and green dashed lines represent two different gap features examined in (g)-(i) and (j)-(l), respectively.
(e)-(f) (001)-surface spectra for (e) the top (B-terminated) and (f) the bottom (Dy-terminated) surfaces using the surface Green's function method.
(g) Wilson loop spectrum below the red dashed line along $\bG$-$\bM$-$\bGp$.
Winding structure exhibits $\mchd=-2$.
(h)-(i) (001)-surface band structures, which corresponds to the lower white rectangles in (e) and (f), from 60-layer slab calculation.
Bulk and surface bands are represented by gray and black respectively.
Surface Dirac fermions are moved inside the gap by applying surface potentials of -0.53 eV for the top and +0.326 eV for the bottom surfaces.
(j) Wilson loop spectrum below the green dashed line ($\mchd=+2$).
(k)-(l) (001)-surface band structures corresponding to the upper white rectangles in (e) and (f), with a surface potential of 0.027 eV.
}
\label{sfig:DyB4_U8}
\end{figure}
%%%%%%%%%%%%%%%%%%%%%%

Here, we present  the electronic topological properties of DyB$_4$ at $U$ = 8 eV, because this state would be appropriate to describe a very localized $4f$ states, which would be expected in magnetic DyB$_4$.
Our first-principle calculation of bulk electronic band structure of DyB$_4$ in the $\Gamma_4$ state exhibits two gap features around $-0.25$ eV and 0.85 eV [the red dashed line within $-0.5$ eV and 0.5 eV and the green dashed line within 0.5 eV and 1.0 eV in \sfig~\ref{sfig:DyB4_U8}(d)].
The bands below the first gap feature (red dashed line) have the mirror Chern number $\mchd=-2$, as identified by the Wilson loop spectrum in \sfig~\ref{sfig:DyB4_U8}(g).
Because of a very small gap at $\bM$, both the top (B-terminated) and bottom (Dy-terminated) surfaces do not exhibit any identifiable surface Dirac fermion [\sfigs~\ref{sfig:DyB4_U8}(e) and (f)].
To observe surface Dirac fermions hidden in bulk bands, we apply a surface potential to top and bottom layers, as shown in \sfigs~\ref{sfig:DyB4_U8}(h) and (i).
Such procedure can be realized though intrinsic surface deformation or extrinsic impurity atom, in experiments.
Similarly, the bands below the second gap feature (green dashed line) have the mirror Chern number $\mchd=2$.
The corresponding Wilson loop spectrum and the surface Dirac fermions are demonstrated in \sfigs~\ref{sfig:DyB4_U8}(j)-(l).

\begin{figure}[!ht]
\includegraphics[width=0.95\textwidth]{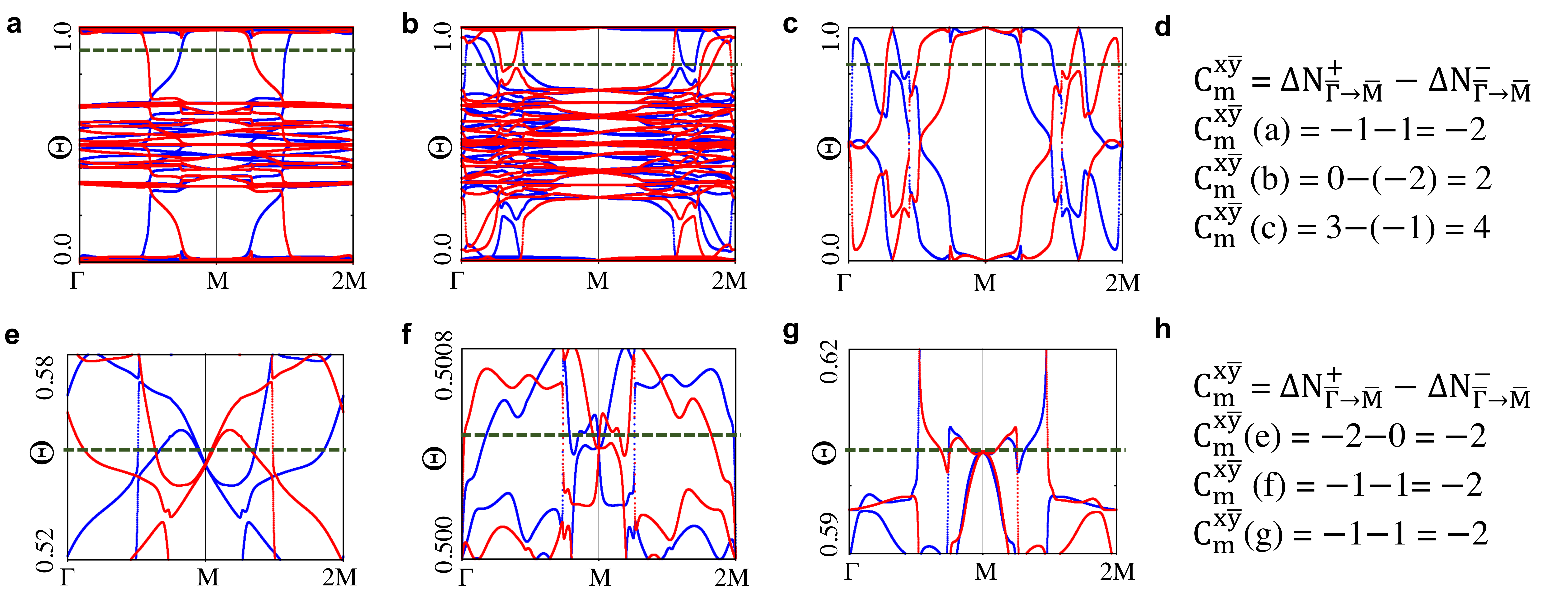}
\caption{
{\bf Detailed analysis on the Wilson spectra of DyB$_4$ at $U$ = 8 eV.} 
The Wilson spectrum along the diagonal path from $\Gamma$ to $2M=(2\pi,2\pi)$ for (a) the bands below the red line
(b) the bands below the green line
(c) the bands between the red and green lines in \sfig~\ref{sfig:DyB4_U8}.
Red (blue) lines represent positive (negative) mirror sectors.
(d,h) Demonstration of how to count $\mchd$  for (a)-(c) and (e)-(g).
Note that $\mchd$(a)+$\mchd$(c)=$\mchd$(b). 
(e-h) Different choices of reference lines in (a).
}
\label{sfig:DFT-Cm=2-Wlson}
\end{figure}

\begin{figure}[t!]
\includegraphics[width=0.8\textwidth]{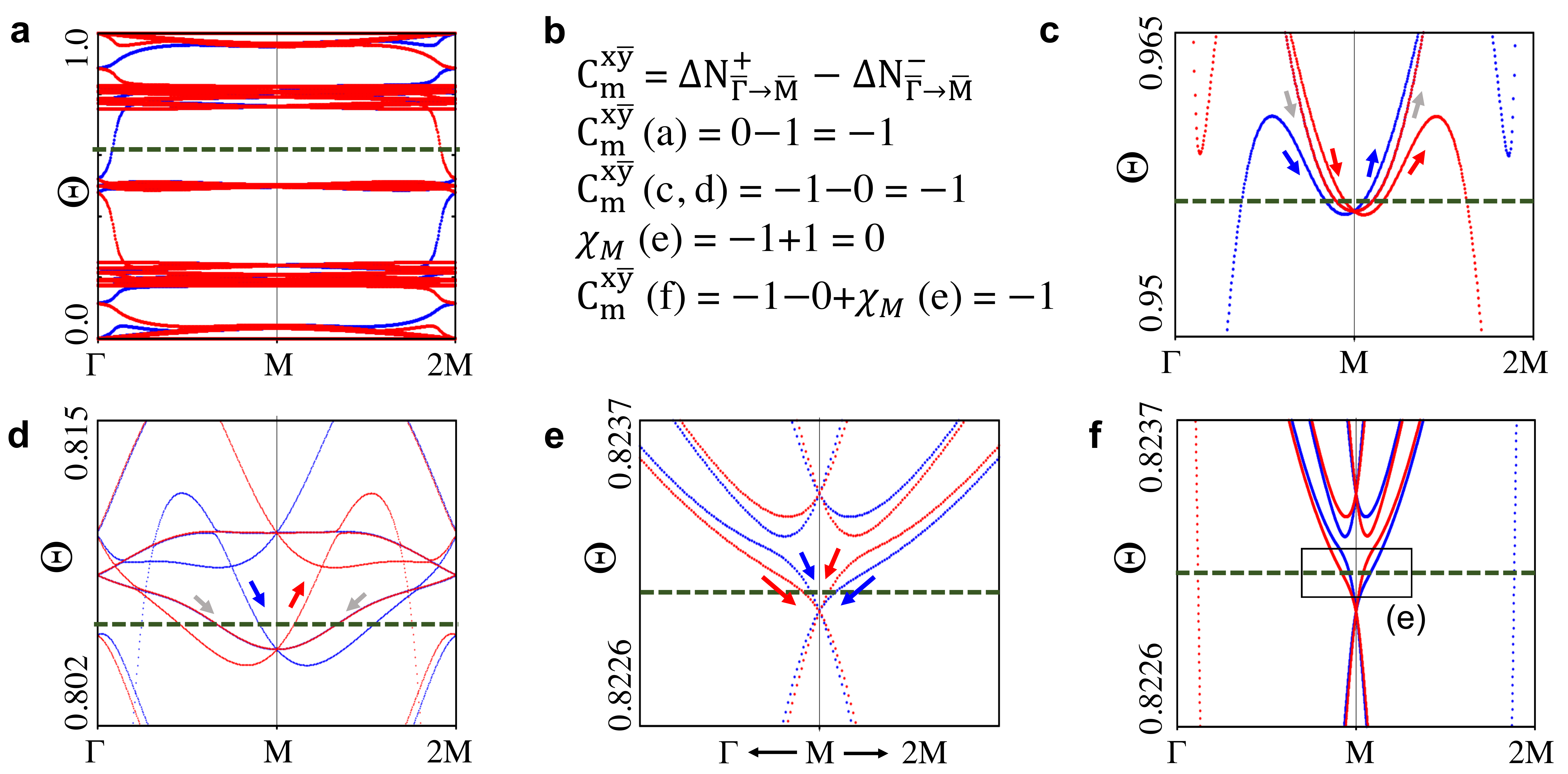}
\caption{
{\bf Detailed analysis on the Wilson spectra of $\mchd =-1$ material (Nd$_4$Te$_8$Cl$_4$O$_{20}$).} 
(a-b) We illustrate how to determine $\mchd$, similar to the case of DyB$_4$ in \sfig~\ref{sfig:DFT-Cm=2-Wlson}.
We also examine the local chirality $\chi_M$ at $M$ defined in Eq.~\eqref{eq:app_chi_M}.
(c-f) Detailed figures for narrow ranges of $\Theta$.
The red (blue) arrows indicate the slopes of Wilson bands with positive (negative) mirror eigenvalue.
The gray arrows are for the nearly twofold-degenerate Wilson bands with opposite mirror eigenvalues.
The bands with these arrows are counted for $\chi_M$.
Note that (c) $\chi_M=0$ (d) $\chi_M=1$ (e)-(f) $\chi_M=0$.
As explained in \sn~\ref{app:local_chiral}, there must be additional nodes crossing the reference line if $\mchd \ne \chi_M$.
}
\label{sfig:DFT-Cm=1-Wlson}
\end{figure}

\begin{figure}[!ht]
\includegraphics[width=0.95\textwidth]{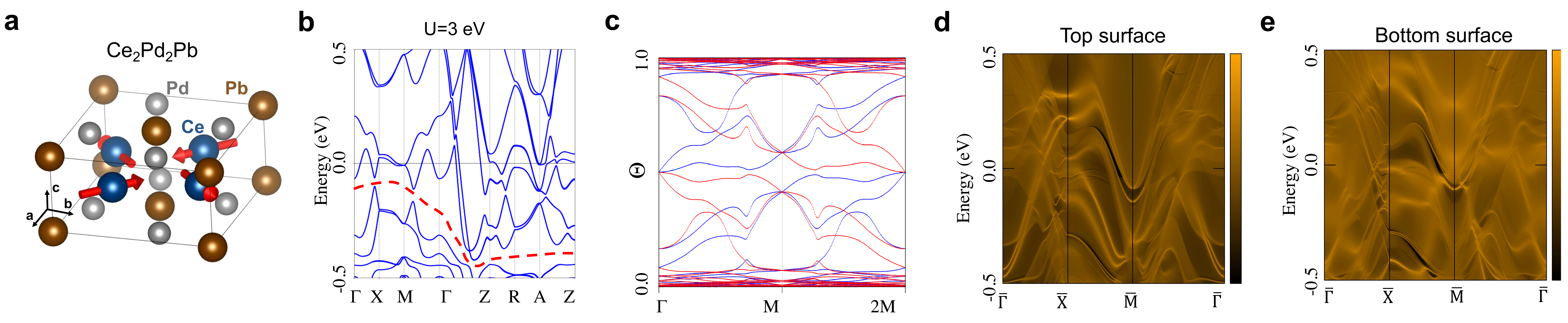}
\caption{
{\bf Computational results of $\mchd=-2$ material (Ce$_2$Pd$_2$Pb).} 
(a) Crystal structure in $\Gamma_4$ state.
(b) The bulk band structure of the DFT+$U$ calculations ($U$ = 3 eV) for the $\Gamma_4$ state.
The states below the red dashed line are considered to calculate the Wilson loop in (c).
(c) The $k_z$-directed Wilson loop spectrum with $\mchd=-2$.
(d)-(e) (001)-surface spectra for (d) the top surface and (e) the bottom surface using the surface Green's function method.
}
\label{sfig:Ce2Pd2Pb}
\end{figure}

\begin{figure}[!ht]
\includegraphics[width=0.95\textwidth]{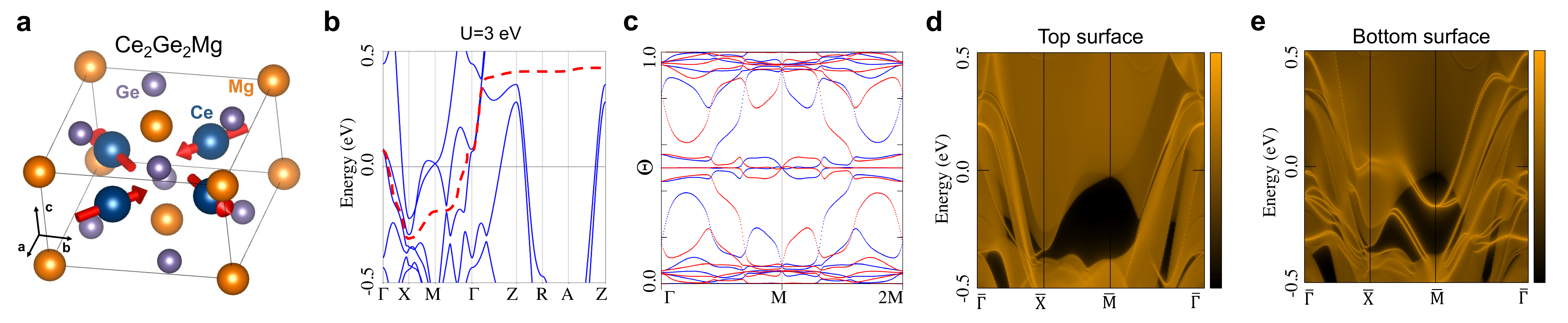}
\caption{
{\bf Computational results of $\mchd=-1$ material (Ce$_2$Ge$_2$Mg).} 
(a) Crystal structure in $\Gamma_4$ state.
(b) The bulk band structure of the DFT+$U$ calculations ($U$ = 3 eV) for the $\Gamma_4$ state.
The states below the red dashed line are considered to calculate the Wilson loop in (c).
(c) The $k_z$-directed Wilson spectrum with $\mchd=-1$.
(d)-(e) (001)-surface spectra for (d) the top surface and (e) the bottom surface  using the surface Green's function method.
}
\label{sfig:Ce2Ge2Mg}
\end{figure}

\begin{figure}[!ht]
\includegraphics[width=0.95\textwidth]{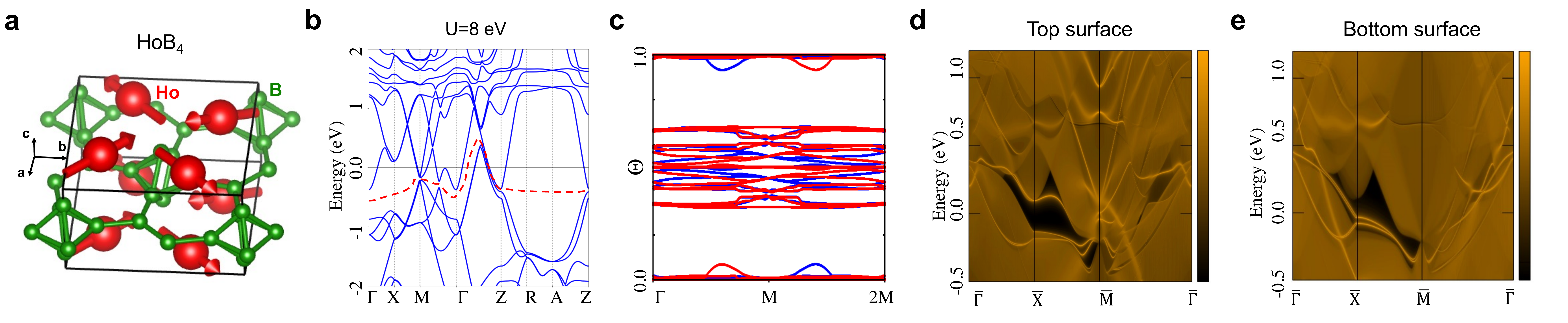}
\caption{
{\bf The computational results of HoB$_4$.}
(a) Crystal structure in $\Gamma_4$ state.
(b) The bulk band structure of the DFT+$U$ calculations ($U$ = 8 eV) for the $\Gamma_4$ state.
The states below the red dashed line are considered to calculate the Wilson loop in (c).
(c) The $k_z$-directed Wilson spectrum with $\mchd=0$.
(d)-(e) (001)-surface spectra for (d) the top surface and (e the bottom surface  using the surface Green's function method.
}
\label{sfig:HoB4}
\end{figure}

\begin{figure}[!ht]
\includegraphics[width=0.95\textwidth]{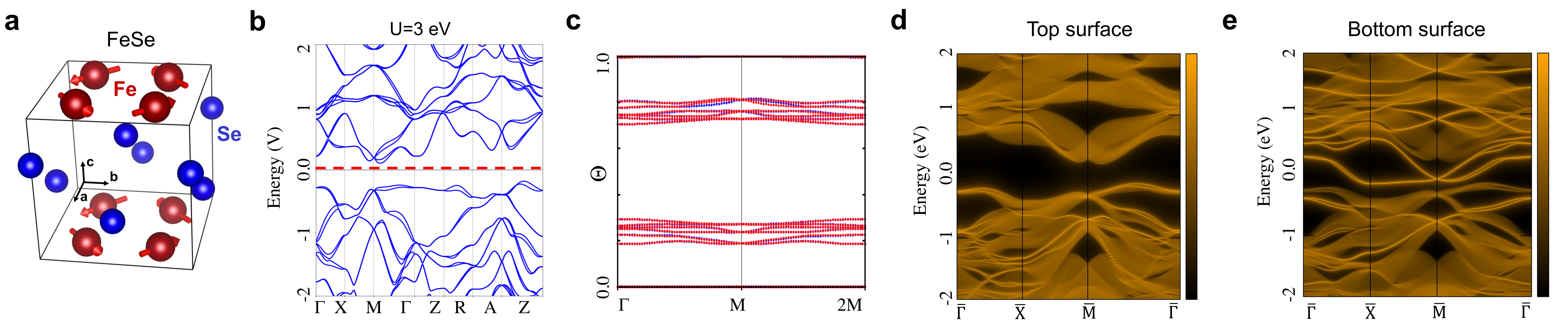}
\caption{
{\bf The computational results of FeSe.}
(a) Crystal structure with MSG 99.169 $P_C4mm$.
(b) The bulk band structure of the DFT+$U$ calculations ($U$ = 3 eV).
The states below the red dashed line are considered to calculate the Wilson loop in (c).
(c) The $k_z$-directed Wilson spectrum with $\mchd=0$.
(d)-(e) (001)-surface spectra for (d) the top surface and (e) the bottom surface  using the surface Green's function method.
}
\label{sfig:FeSe}
\end{figure}

\begin{figure}[!ht]
\includegraphics[width=0.95\textwidth]{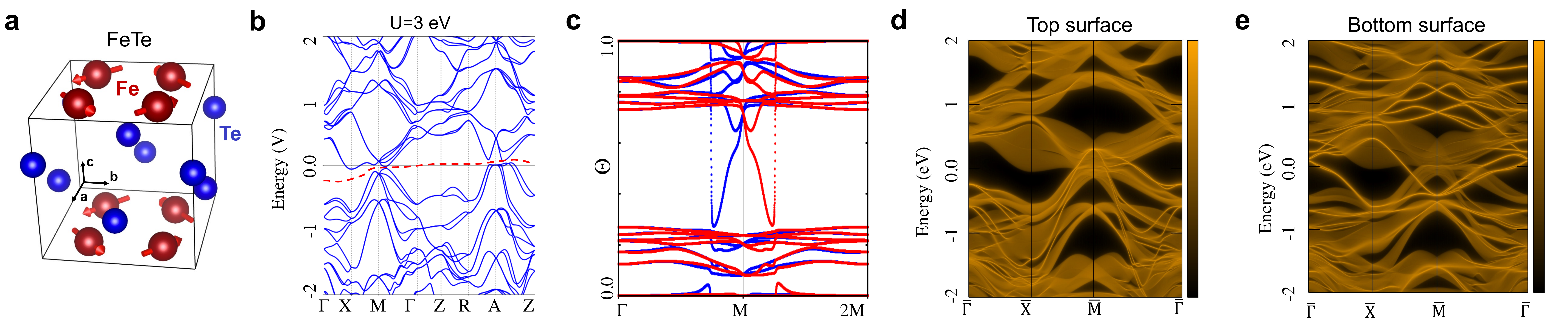}
\caption{
{\bf The computational results of FeTe.}
(a) Crystal structure with MSG 99.169 $P_C4mm$.
(b) The bulk band structure of the DFT+$U$ calculations ($U$ = 3 eV).
The states below the red dashed line are considered to calculate the Wilson loop in (c).
(c) The $k_z$-directed Wilson spectrum with $\mchd=0$.
(d)-(e) (001)-surface spectra for (d) the top surface and (e) the bottom surface  using the surface Green's function method.
}
\label{sfig:FeTe}
\end{figure}

\begin{figure}[!ht]
\includegraphics[width=0.95\textwidth]{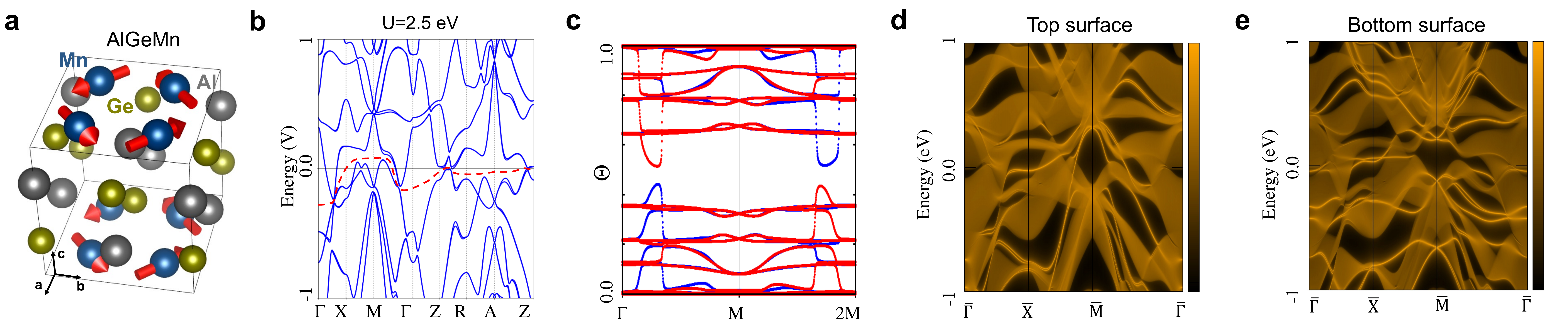}
\caption{
{\bf The computational results of AlGeMn.}
(a) Crystal structure with MSG 99.169 $P_C4mm$.
(b) The bulk band structure of the DFT+$U$ calculations ($U$ = 2.5 eV).
The states below the red dashed line are considered to calculate the Wilson loop in (c).
(c) The $k_z$-directed Wilson spectrum with $\mchd=0$.
(d)-(e) (001)-surface spectra for (d) the top surface and (e) the bottom surface  using the surface Green's function method.
}
\label{sfig:AlGeMn}
\end{figure}

The detailed analysis of Wilson spectrum for DyB$_4$ ($U$=8 eV, MSG 127.392 $P4'/m'b'm$) and Nd$_4$Te$_8$Cl$_4$O$_{20}$ ($U$=6 eV, MSG 129.421 $P_C 4/nmm$) are provided in \sfigs~\ref{sfig:DFT-Cm=2-Wlson} and \ref{sfig:DFT-Cm=1-Wlson}, respectively.
These figures illustrate how to evaluate $\mchd$ for the given Wilson spectrum and the dispersion of fourfold-degenerate Dirac fermions.
Specifically, \sfig~\ref{sfig:DFT-Cm=2-Wlson}(a) shows $\mchd=4$ for four bands between the red and green dashed lines in \sfig~\ref{sfig:DyB4_U8}(d).

\begin{figure}[!ht]
\includegraphics[width=0.95\textwidth]{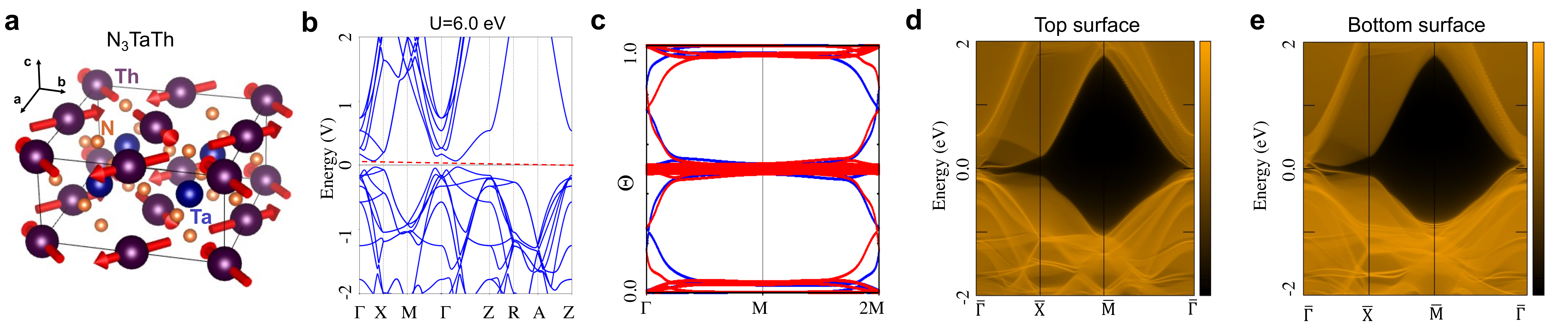}
\caption{
{\bf The computational results of N$_3$TaTh.}
(a) Crystal structure with MSG 129.421 $P_{C}4/nmm$.
(b) The bulk band structure of the DFT+$U$ calculations ($U$ = 6.0 eV).
The states below the red dashed line are considered to calculate the Wilson loop in (c).
(c) The $k_z$-directed Wilson spectrum with $\mchd=0$.
(d)-(e) (001)-surface spectra for (d) the top surface and (e) the bottom surface  using the surface Green's function method.
}
\label{sfig:N3TaTh}
\end{figure}

Finally, we present the DFT calculations for other materials, whose band topologies are examined by calculating the mirror Chern number $\mchd$ for diagonal mirror plane.
The electronic and topological properties of Ce$_2$Pd$_2$Pb, Ce$_2$Ge$_2$Mg, HoB$_4$, FeSe, FeTe, AlGeMn and N$_3$TaTh are shown in \sfigs~\ref{sfig:Ce2Pd2Pb}-\ref{sfig:N3TaTh}.
In particular, Ce$_2$Pd$_2$Pb and Ce$_2$Ge$_2$Mg are shown to be topological with $\mchd=-2$ and $-1$ respectively.
Although HoB$_4$, FeSe, FeTe, AlGeMn, and N$_3$TaTh correspond to $\mchd=0$, we can find a nonchiral surface Dirac cone in gap for HoB$_4$, FeSe, FeTe, and AlGeMn.
We take FeSe, FeTe, and AlGeMn as examples to show how to construct magnetic materials with MSG 99.169 $P_C4mm$ based on paramagnetic phase with magnetic atoms at WP $2a$ in SG 129 and $\mbb k$-vector $(\hf,\hf,0)$.
We show N$_3$TaTh as an example with MSG 129.421 $P_C 4/nmm$ constructed from paramagnetic phase with Th at WP $1a$ in SG 221 and $\mbb k$-vector $(\hf,0,0)$. 
(See also \sn~\ref{app:mat_search}.)
Members of Ce$_2A_2B$ family are examples possessing similar crystal structures with rare-earth tetraborides family with SG 127 $P4/mbm$.
Note that the $\Gamma_4$ state is considered when the paramagnetic unit cell of a material has SG 127 $P4/mbm$.
The information of crystal structure for candidate materials are summarized in \stable~\ref{tab:icsd}.
Their bulk topological invariants, i.e. $\mchd$, are summarized in \sfig~\ref{sfig:candidates}.
%%%%%%%%%%%%%%%%%%%%%%
\\

%%%%%%%%%%%%%%%%%%%%%%
{
\renewcommand{\arraystretch}{1.2}
\begin{table}[t]
\centering
\begin{minipage}{\textwidth}
\caption{
{\bf Tables of material structure information and projectors for a Wannier function tight-binding model.}
The  Inorganic Crystal Structure Database (ICSD) index~\cite{ICSD} for candidates materials are summarized.
We adopt the crystal structure~\cite{sim2016Spontaneous} of DyB$_4$ to calculations for other $R$B$_4$.
}
\label{tab:icsd}
\end{minipage}
\begin{tabular*}{0.98\textwidth}{@{\extracolsep{\fill}} c|c|c|c|c|c|c|c}
\hline \hline
Material & TbB$_4$ & DyB$_4$ & HoB$_4$ & NdTe$_2$ClO$_5$ & MnAlGe  & FeSe & FeTe
\\ \hline
ICSD \# & \multicolumn{3}{c|}{H.Sim et al, PRB (2016)~\cite{sim2016Spontaneous}} & 88673~\cite{ndteocl} & 607972~\cite{mnalge} & 169267~\cite{fese}   & 180602~\cite{fete}
\\ \hline
Projectors & \multicolumn{3}{c|}{$f$,$d$ for $R$, $p$ for B} & $f$ for Nd, $p$ for Te, Cl, O & $d$ for Mn, $p$ for Ge   & \multicolumn{2}{c}{$d$ for Fe, $p$ for Se, Te} \\ \hline \hline
\end{tabular*}
\\ \vspace{0.5em}
\begin{tabular*}{0.98\textwidth}{@{\extracolsep{\fill}} c|c|c|c|c|c|c|c}
\hline \hline
Material & Ce$_2$Pd$_2$Pb & Ce$_2$Pt$_2$Cd & Ce$_2$Si$_2$Mg & Ce$_2$Ge$_2$Mg & Ce$_2$Ni$_2$Mg & Ce$_2$Cu$_2$In & N$_3$TaTh
\\ \hline
ICSD \# & 99191~\cite{cepdpb} & 411009~\cite{ceptcd} & 83915~\cite{cesimg}  & 413849~\cite{cegemg} & 411012~\cite{ceptcd} & 620863~\cite{cecuin} & 247345~\cite{n3tath}
\\ \hline
Projectors & \multicolumn{6}{c|}{$f$,$d$ for Ce, $d$ for Pd, Pt, Ni, Cu, $p$ for Pb, Cd, Si, In} & $f$ for Th, $d$ for Ta, $p$ for N
\\ \hline \hline
\end{tabular*}
\end{table}
}
%%%%%%%%%%%%%%%%%%%%%%

%%%%%%%%%%%%%%%%%%%%%%
%%%%%%%%%%%%%%%%%%%%%%

%%%%%%%%%%%%%%%%%%%%%%
%%%%%%%%%%%%%%%%%%%%%%

%%%%%%%%%%%%%%%%%%%%%%
%%%%%%%%%%%%%%%%%%%%%%

%%%%%%%%%%%%%%%%%%%%%%
%%%%%%%%%%%%%%%%%%%%%%

%%%%%%%%%%%%%%%%%%%%%%
%%%%%%%%%%%%%%%%%%%%%%

%%%%%%%%%%%%%%%%%%%%%%
%%%%%%%%%%%%%%%%%%%%%%

%%%%%%%%%%%%%%%%%%%%%%
%%%%%%%%%%%%%%%%%%%%%%

%%%%%%%%%%%%%%%%%%%%%%
%%%%%%%%%%%%%%%%%%%%%%

%%%%%%%%%%%%%%%%%%%%%%
%%%%%%%%%%%%%%%%%%%%%%

%%%%%%%%%%%%%%%%%%%%%%
\section{Tight-binding models for MWG $p4'g'm$ and MSG 127.392 $P4'/m'b'm$}
\label{app:models}
%%%%%%%%%%%%%%%%%%%%%%
In this section, we provide the detailed description for the tight-binding models, $H_{p4'g'm}(\bk)$ and $\mc{H}_{P4'/m'b'm}(\bk)$ whose symmetry groups are MWG $p4'g'm$ and MSG 127.392 $P4'/m'b'm$, respectively.
%%%%%%%%%%%%%%%%%%%%%%

\begin{figure}[b!]
\centering
\includegraphics[width=0.6\textwidth]{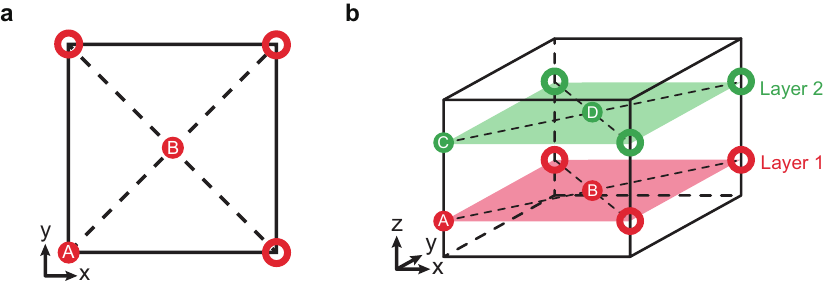}
\caption{
{\bf Description of tight-binding models.}
(a) Unit cell for the model $H_{p4'g'm}(\bk)$. $A=(0,0)$ and $B=(1/2,1/2)$.
$s$ orbitals with spin up and down are located at each sublattice sites.
(b) Unit cell for the model $\mc{H}_{P4'/m'b'm}(\bk)$. 
The unit cell is composed of two layers, Layer 1 and Layer 2, related by spacetime-inversion symmetry $I'=T\{i|\bb 0\}$.
Each layer have the same structure and symmetry to those of unit cell for $H_{p4'g'm}(\bk)$ shown in (a).
Sublattice sites are located at $A=(0,0,1/4)$, $B=(1/2,1/2,1/4)$, $C=(0,0,3/4)$, and $D=(1/2,1/2,3/4)$.
}
\label{sfig3}
\end{figure}

%%%%%%%%%%%%%%%%%%%%%%
\tocless{\subsection{Tight-binding models for MWG $p4'g'm$}
\label{appsub:TB_MWGIII}}{}
%%%%%%%%%%%%%%%%%%%%%%
MWG $p4'g'm$ has $\mxybo=\{m_{x\cm{y}}|\hf,\mhf\}$, $\mxyo=\{m_{xy}|\hf,\hf\}$, $TC_{4z}=T\{c_{4z}|\bb 0\}$, and $\tmy=T\{m_y|\hf,\hf\}$.
To construct a tight-binding model for MWG $p4'g'm$, we consider a unit cell composed of two sublattice sites $A=(0,0)$ and $B=(\frac{1}{2},\frac{1}{2})$.
We put $s$ orbitals with spin up and down on each sublattice sites~\sfig~\ref{sfig3}(a).

Then, symmetry operators , $U^{\mc{K}}_{\sg'}(\bk)$ for antiunitary symmetries $\sg' \in \{TC_{4z}, \tmy\}$ and $U_\sg(\bk)$ for unitary symmetries $\sg \in \{\mxybo, \mxyo\}$, are given by
\bg
U^{\mc{K}}_{TC_{4z}}(\bk) = -i \tau_0 \sg_y e^{i\frac{\pi}{4}\sg_z} \mc{K}, \quad
U^{\mc{K}}_{\tmy}(\bk) = -e^{\frac{i}{2} (k_x - k_y)} \tau_x \sg_0 \mc{K}, \nn
U_{\mxybo}(\bk) = -\frac{i}{\sqrt{2}} e^{\frac{i}{2}(k_x-k_y)} \tau_x (\sg_x-\sg_y), \quad
U_{\mxyo}(\bk) = -\frac{i}{\sqrt{2}} e^{\frac{i}{2}(k_x+k_y)} \tau_x (\sg_x+\sg_y).
\eg
Including all the symmetry allowed hopping interactions up to next nearest neighbor ones, we obtain the tight-binding Hamiltonian $H_{p4'g'm}(\bk)$:
\ba
H_{p4'g'm}(\bk) &= t_1 \cos \frac{k_x}{2} \cos \frac{k_y}{2} \tau_x + t_2 (\cos k_x + \cos k_y) + v_1 \cos \frac{k_x}{2} \cos \frac{k_y}{2} \tau_y \sg_z + v_2 \sin \frac{k_x}{2} \cos \frac{k_y}{2} \tau_x \sg_y \nn
&- v_2 \cos \frac{k_x}{2} \sin \frac{k_y}{2} \tau_x \sg_x + v_3 \sin k_x \tau_z \sg_x + v_3 \sin k_y \tau_z \sg_y + v_4 \sin k_x \sg_y - v_4 \sin k_y \sg_x \nn
&+ v_5 (\cos k_x - \cos k_y) \sg_z,
\label{eq:TB_MWGIII}
\ea
where the Pauli matrices $\tau$ and $\sg$ act on the sublattice sites and spin degrees of freedom, respectively.
Among the hopping parameters $t_{1,2}$ and $v_{1,\dots,5}$, $v_{1,\dots,5}$ denote the spin-orbit couplings.
Band structure in Fig. 1(d) of the main text is obtained by using the parameter values $t_1=1$, $t_2=0$, $v_1=0.7$, $v_2=0.4$, $v_3=-1.5$, $v_4=1$, $v_5=0.5$.
%%%%%%%%%%%%%%%%%%%%%%
\\

%%%%%%%%%%%%%%%%%%%%%%
%%%%%%%%%%%%%%%%%%%%%%

%%%%%%%%%%%%%%%%%%%%%%
\tocless{\subsection{Tight-binding models for MSG 127.392  $P4'/m'b'm$}
\label{appsub:TB_127.392}}{}
%%%%%%%%%%%%%%%%%%%%%%
Now, we discuss the tight-binding model $\mc{H}_{P4'/m'b'm}(\bk)$ for MSG 127.392 $P4'/m'b'm$.
Note that MSG $P4'/m'b'm$ is a MSG made by MWG $p4'g'm$ with additional symmetry generators.
Thus, $\mc{H}_{P4'/m'b'm}(\bk)$ is constructed based on the tight-binding model $H_{p4'g'm}(\bk)$.
We chose a unit cell with a layered structure such that each layer is symmetric under $p4'g'm$ and two layers in unit cell are related by spacetime-inversion symmetry $TI=T\{i|\bb 0\}$.
Hence, the tight-binding Hamiltonian $\mc{H}_{P4'/m'b'm}(\bk)$ are divided into two part that describe intralayer and interlayer hopping processes $\mc{H}_{P4'/m'b'm}(\bk)
=\mc{H}_{\rm intra}(\bk) + \mc{H}_{\rm inter}(\bk)$.
The intralayer Hamiltonian is given by
\ba
\mc{H}_{\rm intra}(\bk)
&= \cos \frac{k_x}{2} \cos \frac{k_y}{2} (t_1 \Gamma_{010} + v_1 \Gamma_{023}) + t_2 (\cos k_x + \cos k_y) \Gamma_{000} + v_2 (\sin \frac{k_x}{2} \cos \frac{k_y}{2} \Gamma_{312} - \cos \frac{k_x}{2} \sin \frac{k_y}{2} \Gamma_{311}) \nn
&+ v_3 (\sin k_x \Gamma_{331} + \sin k_y \Gamma_{332}) + v_4 (\sin k_x \Gamma_{302} - \sin k_y \Gamma_{301}) + v_5 (\cos k_x - \cos k_y) \Gamma_{303}.
\ea
The interlayer Hamiltonian is expressed as
\ba
\mc{H}_{\rm inter}(\bk)
&= \cos \frac{k_x}{2} \cos \frac{k_y}{2} \cos \frac{k_z}{2} (t_{l1} \Gamma_{110} + v_{l1} \Gamma_{123}) + \sin \frac{k_x}{2} \sin \frac{k_y}{2} \sin \frac{k_z}{2} (t_{l2} \Gamma_{110} + v_{l2} \Gamma_{123}) + t_{l5} \cos \frac{k_z}{2} \Gamma_{100} \nn
&+ \cos \frac{k_x}{2} \cos \frac{k_y}{2} \sin \frac{k_z}{2} (t_{l3}  \Gamma_{210} + v_{l3} \Gamma_{223}) + \sin \frac{k_x}{2} \sin \frac{k_y}{2} \cos \frac{k_z}{2} (t_{l4}  \Gamma_{210} + v_{l4} \Gamma_{223}) + t_{l6} \sin \frac{k_z}{2} \Gamma_{200} \nn
&+ v_{l5} \cos \frac{k_z}{2} (\sin \frac{k_x}{2} \cos \frac{k_y}{2} \Gamma_{221} + \cos \frac{k_x}{2} \sin \frac{k_y}{2} \Gamma_{222}) + v_{l6} \cos \frac{k_z}{2} (\sin \frac{k_x}{2} \cos \frac{k_y}{2} \Gamma_{122} + \cos \frac{k_x}{2} \sin \frac{k_y}{2} \Gamma_{121}) \nn
&+ v_{l7} \sin \frac{k_z}{2} (\sin \frac{k_x}{2} \cos \frac{k_y}{2} \Gamma_{121} + \cos \frac{k_x}{2} \sin \frac{k_y}{2} \Gamma_{122}) + v_{l8} \sin \frac{k_z}{2} (\sin \frac{k_x}{2} \cos \frac{k_y}{2} \Gamma_{222} + \cos \frac{k_x}{2} \sin \frac{k_y}{2} \Gamma_{221}).
\ea
Here, $\Gamma_{abc}=\mu_a \tau_b \sg_c$ with the Pauli matrices $\mu$, $\tau$, and $\sg$ acting on layer, sublattice, and spin degrees of freedom.
The intralayer and interlayer hopping parameters are denoted as $t_i$ and $t_{li}$, respectively, and $v_{i}$ and $v_{li}$ are intralayer and interlayer spin-orbit couplings, respectively.

The tight-binding Hamiltonian $\mc{H}_{P4'/m'b'm}(\bk)$ respects MSG 127.392 which have symmetry elements $TC_{4z}=T\{c_{4z}|\bb 0\}$, $\tmy=T\{m_y|\hf,\hf,0\}$, $TI=T\{i|\bb 0\}$, $\mxybo=\{m_{x\cm{y}}|\hf,-\hf,0\}$, and $\mxyo=\{m_{xy}|\hf,\hf,0\}$.
Symmetry operators are given by
\bg
U^{\mc{K}}_{TC_{4z}}(\bk)
= -i\mu_0 \tau_0 \sg_y e^{i\frac{\pi}{4}\sg_z} \mc{K}, \quad
U^{\mc{K}}_{\tmy}(\bk)
= -e^{\frac{i}{2} (k_x - k_y)} \mu_0 \tau_x \sg_0 \mc{K}, \quad
U^{\mc{K}}_{TI}(\bk)
= -i\mu_x \tau_0 \sg_y \mc{K}, \nn
U_{\mxybo}(\bk)
= -\frac{i}{\sqrt{2}} e^{\frac{i}{2}(k_x-k_y)} \mu_0 \tau_x (\sg_x-\sg_y), \quad
U_{\mxyo}(\bk)
= -\frac{i}{\sqrt{2}} e^{\frac{i}{2}(k_x+k_y)} \mu_0 \tau_x (\sg_x+\sg_y).
\eg

One can denote $\mc{H}_{\rm intra}(\bk)$ as $\mc{H}_{\textrm{Layer 1}}(\bk) +\mc{H}_{\textrm{Layer 2}}(\bk)$, where $\mc{H}_{\textrm{Layer 1,2}}(\bk)$ are the intralayer Hamiltonians for the Layers 1 and 2 illustrated in \sfig~\ref{sfig3}.
This can be done by noticing that the Layer 1 and Layer 2 are mapped to each other by $TI$ symmetry.
Hence $\mc{H}_{\textrm{Layer 2}}(\bk) = U^{\mc{K}}_{TI}(\bk) \mc{H}_{\textrm{Layer 1}}(\bk) U^{\mc{K}}_{TI}(\bk)^{-1}$.
Also, the intralayer Hamiltonian for the Layer 1 is identical to $H_{p4'g'm}(\bk)$ in \eq{eq:TB_MWGIII} in a sense that
\ba
\mc{H}_{\textrm{Layer 1}}(\bk) =& \bpm H_{p4'g'm}(\bk) & 0_{4\times4} \\ 0_{4\times4} & 0_{4\times4} \epm = H_{p4'g'm(\bk)} \otimes \bpm 1 & 0 \\ 0 & 0 \epm_\mu = H_{p4'g'm(\bk)} \otimes \frac{1}{2}(\mu_0+\mu_z).
\ea

Topological invariant in this system is mirror Chern number $\mchd$ for $\mxybo$, which is defined in the $k_x=k_y$ plane.
As defined in \sn~\ref{app:mCherns}, $\mchd$ is equal to $\mch{+}{k_x=k_y}=-\mch{-}{k_x=k_y}$ where the $\mch{\pm}{k_x=k_y}$ is the Chern number of occupied bands with $\mxybo$ eigenvalue $\pm i$ in $k_x=k_y$ plane.

\begin{figure}[t!]
\centering
\includegraphics[width=0.98\textwidth]{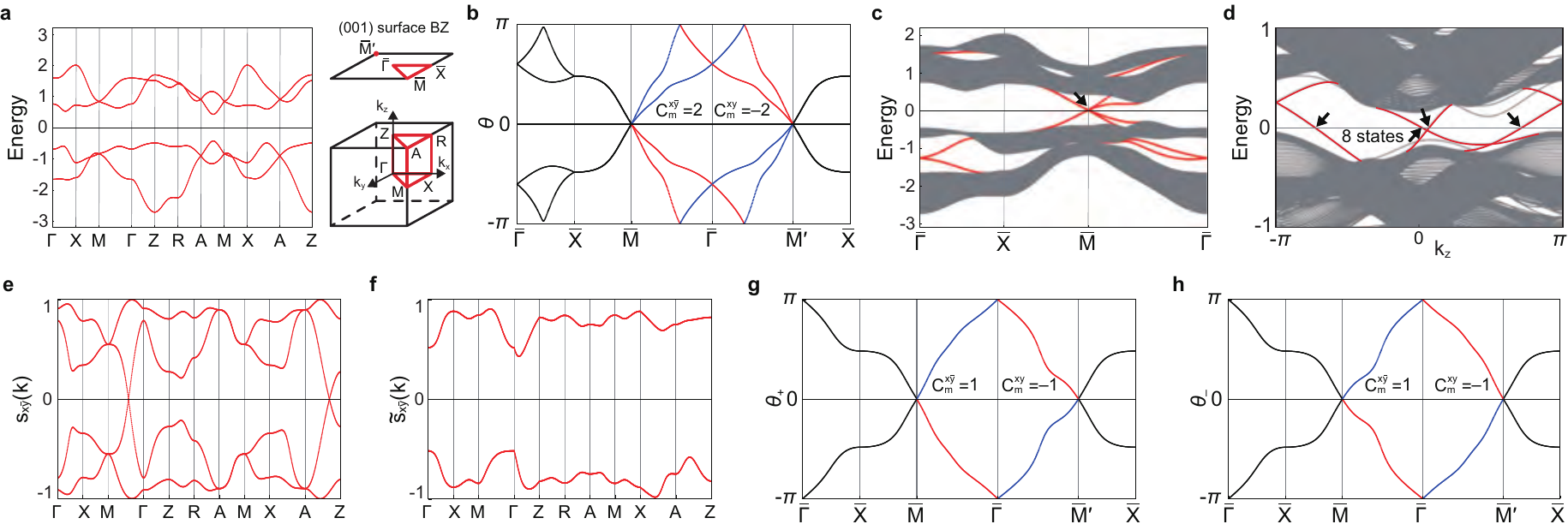}
\caption{
{\bf Description of the tight-binding model $\mc{H}_{P4'/m'b'm}(\bk)$ for Type-III TMDI with $\mchd=2$ and its pseudo-spin resolved topology.}
(a) Band structure of the tight-binding model $\mc{H}_{P4'/m'b'm}(\bk)$.
Four bands are occupied below $E=0$.
(b) Wilson loop spectrum for four occupied bands, which correspond to $\mchd=2$ and $\mchdp=-2$.
Red (blue) lines indicate the mirror eigenvalue $+i$ ($-i$).
(c) (001)-surface spectrum for slab geometry with $n_z = 60$ unit cells along the $z$-direction.
Surface-localized states are indicated by red lines.
A surface Dirac fermion appears near $E=0$ at $\bM$ (black arrow), consistent with the Wilson loop spectrum in (b).
(d) The band structure in open boundaries along $x$ and $y$ directions with $n_x=25$ and $n_y=25$ unit cells, respectively.
Hinge-localized states that interpolate the conduction and valence bands are indicated by red.
Each hinge mode has a mirror-symmetric partner at another hinge with the same dispersion.
Thus, the hinge modes are twofold-degenerate in surface band structure.
Eight hinge modes cross $E=0$ (black arrows).
Four of them have positive group velocity (chiral) while the other four have negative group velocity (antichiral).
The location and mirror eigenvalue of each hinge mode is illustrated in \sfig~\ref{sfig:hinge2}(c).
(e) Gapless spin spectrum for four occupied bands with $S_{x\cm{y}}$.
Note that $\{s_{x\cm{y},A}(\bk)|A=1, \dots, n_{\rm occ}\}= {\rm Spec}[P_{\rm occ}(\bk) S_{x\cm{y}} P_{\rm occ}(\bk)]$ where $P_{\rm occ}(\bk)$ is a projection into the occupied bands.
(f) Gapped pseudo-spin spectrum for four occupied bands with pseudo-spin operator $\td{S}_{x\cm{y}}.$
A pseudo-spin gap defines $\td{S}_{x\cm{y}}^{>0}$ and  $\td{S}_{x\cm{y}}^{<0}$ sectors according to the sign of $\td{s}_{x\cm{y},A}(\bk) = {\rm Spec}[P_{\rm occ}(\bk) \td{S}_{x\cm{y}} P_{\rm occ}(\bk)]$.
(g)-(h) Wilson loop spectrum for two occupied bands (g) in the $\td{S}_{x\cm{y}}^{>0}$ sector and (h) in the $\td{S}_{x\cm{y}}^{<0}$ sector.
Note that $\mchd=1$ and $\mchdp=-1$ in both $\td{S}_{x\cm{y}}$ sectors.
Also note that the Wilson loop spectrum for $\td{S}_{x\cm{y}}^{>0}$ and $\td{S}_{x\cm{y}}^{<0}$ sectors are related by $TC_{4z}$ because $\{\td{S}_{x\cm{y}},U^\mc{K}_{TC_{4z}}(\bk)\}=0$.
}
\label{sfig:TB}
\end{figure}

Now, let us study a phase with $\mchd=2$ as an example of Type-III magnetic Dirac insulator.
This phase is realized with a set of parameters $t_1=1$, $t_2=0$, $v_1=-0.1$, $v_2=0.7$, $v_3=0.1$, $v_4=0$, $v_5=0.5$, $t_{l1}=0.1$, $t_{l2}=-0.55$, $t_{l3}=0.7$, $t_{l4}=-0.1$, $t_{l5}=-0.45$, $t_{l6}=-0.6$, $v_{l1}=0.6$, $v_{l2}=-0.4$, $v_{l3}=-1.15$, $v_{l4}=0.7$, $v_{l5}=0.6$, $v_{l6}=-0.2$, $v_{l7}=0.6$, $v_{l8}=-0.4$.
The band structures in periodic and open boundary conditions, and the Wilson loop spectrum of $\mc{H}_{P4'/m'b'm}(\bk)$ are shown in \sfig~\ref{sfig:TB}.
The (001)-surface band structure in \sfig~\ref{sfig:TB}(c) exhibits a topologically protected Dirac fermion on the (001) surface as a unique nodal point at the Fermi level.
This is consistent with the Wilson loop spectrum in \sfig~\ref{sfig:TB}(b) showing $\mchd=2$.
In open boundaries along the $x$ and $y$ directions that preserve the diagonal mirror planes, this model exhibits mirror-protected hinge modes.
In \sfig~\ref{sfig:TB}(d), four chiral and four antichiral modes cross the Fermi level with the positive and negative velocity along the $z$ direction, respectively.
At each hinge, there are two copropagating (anti)chiral modes, as illustrated in \sfig~\ref{sfig:hinge2}(c) and also in the right panel of Fig.~1(e) in the main text.
Note that, in other MSGs, to observe mirror-protected hinge modes, additional topological invariants that protect metallic states on the (100) or (010) surfaces, such as $\mc{C}_\pm^{k_z=0,\pi}$,
must be zero.
%%%%%%%%%%%%%%%%%%%%%%
\\

%%%%%%%%%%%%%%%%%%%%%%
%%%%%%%%%%%%%%%%%%%%%%

%%%%%%%%%%%%%%%%%%%%%%
\section{Pseudo-spin resolved topology of 3D TMDI with even $\mchd$}
\label{app:pseudospin}
%%%%%%%%%%%%%%%%%%%%%%
In this section, we study the pseudo-spin resolved topology of the 3D TMDI, following Ref.~\cite{lin2022spin} closely.
Specifically, we focus on the tight-binding model $\mc{H}_{P4'/m'b'm}(\bk)$ describing a Type-III TMDI with $(\mchd,\mchdp)=(2,-2)$.
This model respects MSG $P4'/m'b'm$ and relevant symmetry operators are introduced in \sn~\ref{app:Tightbinding}B.

The main idea of spin-resolved topology is that a set of bands can be divided further into spin positive and negative sectors by appropriately defining spin projector, even with nonzero spin-orbit coupling.
Such spin projector $s(\bk)$ is defined in terms of the spin operator $S$ and the occupied-band projector $P_{\rm occ}(\bk) = \sum_{n=1}^{n_{\rm occ}} \ket{u_n(\bk)} \bra{u_n(\bk)}$:
\bg
s(\bk) = P_{\rm occ}(\bk) S P_{\rm occ}(\bk).
\eg
Note that the spin operator $s$ must commute with the orbital embedding matrix $V(\bk)_{\alpha \beta} = e^{-i \bk \cdot \bx_\alpha} \delta_{\alpha\beta}$ such that $[S,V(\bb G)]=0$ for a reciprocal lattice vector $\bb G$.
When a spectrum of the spin projector $s(\bk)$, i.e. $\{s_A(\bk)|A=1,\dots,n_{\rm occ}\} = {\rm Spec}[P_{\rm occ}(\bk) S P_{\rm occ}(\bk)]$ is gapped like \sfig~\ref{sfig:TB}(f), then the $S^{>0}$ and $S^{<0}$ sectors are defined according to the sign of $s_A(\bk)$.
Otherwise, it is not compatible with the periodic gauge in Eq.~\eqref{eqapp:pgauge}.
(For the full details of spin-resolved topology in time-reversal and inversion symmetric systems, we strongly recommend the reader to consult Ref.~\cite{lin2022spin}.)

Now, let us apply this idea to 3D TMDIs.
The bulk topological invariant of the Type-III TMDI is a pair of mirror Chern numbers $(\mchd,\mchdp)$, which are defined with respect to diagonal mirrors, $\mxybo$ and $\mxyo$.
Hence, spin-resolved topology in each spin sector can still be classified with $(\mchd,\mchdp)$, if a spin operator commutes with both diagonal mirrors.
However, there is no such spin operator: spin-rotation parts for $\mxybo$ and $\mxyo$ are given by $e^{-i \frac{\pi}{2} \frac{1}{\sqrt{2}} (1,-1,0) \cdot \bb \sg} = - \frac{i}{\sqrt{2}} (\sg_x - \sg_y)$ and $e^{-i \frac{\pi}{2} \frac{1}{\sqrt{2}} (1,-,0) \cdot \bb \sg} = - \frac{i}{\sqrt{2}} (\sg_x + \sg_y)$, respectively.
A generic spin operator $\bb n \cdot \bb \sg$ with a three-vector $\bb n \in \mathbb{R}^3$ (where we take a convention that spin eigenvalues are $\pm |\bb n|$) commutes only if $\bb n=0$.
Rather, we can fin a spin operator commuting with only one diagonal mirror, say $\mxybo$: $S_{x\cm{y}} = \frac{1}{\sqrt{2}} (\sg_x - \sg_y)$, which satisfies $[S_{x\cm{y}},U_{\mxybo}(\bk)]=0$ and $[S_{x\cm{y}},V(\bb G)]=0$.
The corresponding spin spectrum is shown in \sfig~\ref{sfig:TB}(e).
Note that there are gap closings in the BZ and thus spin sectors are not well-defined.

We resolve the obstruction of defining spin sectors by introducing a pseudo-spin.
By coupling the orbital degrees of freedom, we find a pseudo-spin operator $\td{S}_{x\cm{y}} = \mu_z \tau_z \sg_z$,
which commute with $V(\bb G)$, $U_{\mxybo}(\bk)$, and $U_{\mxyo}(\bk)$: $[\td{S}_{x\cm{y}},V(\bb G)]=[\td{S}_{x\cm{y}},U_{\mxybo}(\bk)]=[\td{S}_{x\cm{y}},U_{\mxyo}(\bk)]=0$.
Also note that $\{\td{S}_{x\cm{y}}, U_{TC_{4z}}^\mc{K}(\bk) \}=0$.
The corresponding pseudo-spin spectrum are shown in \sfig~\ref{sfig:TB}(f).
The pseudo-spin gap is finite and thus four occupied bands are resolved into two in the pseudo-spin positive ($\td{S}_{x\cm{y}}^{>0}$) sector and two in the negative ($\td{S}_{x\cm{y}}^{<0}$) sector.
For each pseudo-spin sector, the Wilson loop spectrum exhibits $(\mchd,\mchdp)=(1,-1)$, as shown in \sfigs~\ref{sfig:TB}(g) and (h).
This implies that a bulk response of TMDI with $\mchd=2$ can be contributed by two mirror TCIs with $\mchd=1$ but in opposite pseudo-spin sectors.
Finally, we comment that when a gap closing and reopening happens in pseudo-spin spectrum but not in bulk band structure, the mirror Chern numbers of each pseudo-spin sector can be changed.
We leave the detailed study of such topological phase transition in pseudo-spin spectrum for an interesting future study.
%%%%%%%%%%%%%%%%%%%%%%
\\

%%%%%%%%%%%%%%%%%%%%%%
\section{Topological phase transition in MSG 127.392 $P4'/m'b'm$}
\label{app:phasetransition}
\begin{figure*}[h!]
\centering
\includegraphics[width=0.33\textwidth]{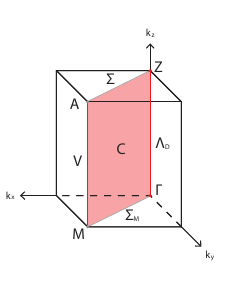}
\caption{
{\bf Topological phase transition of MSG 127.392 $P4'/m'b'm$.}
Gap closing can occur at red sections, i.e. $\Gamma$, $Z$, $\Lambda_D$, $V$, and $C$.
$\Delta \mchd = \pm 1$ at $\Gamma$ and $Z$; $\Delta \mchd = \pm 2$ or $0$ at $\Lambda_D$ and $V$, depending on the low-energy effective Hamiltonian; $\Delta \mchd = \pm 2$ at $C$.
}
\label{sfig:phase}
\end{figure*}

%%%%%%%%%%%%%%%%%%%%%%
In this section, we provide a detailed analysis on topological phase transition in DyB$_4$ with MSG 127.392 $P4'/m'b'm$, which is induced by the change of onsite Coulomb interaction $U$ and changes the mirror Chern number $\mchd$ (See \sfig~\ref{sfig:DFT+U:detail}).
For this, we discuss the possible gap closing and reopening on the mirror-invariant plane $k_x=k_y$ of MSG 127.392 $P4'/m'b'm$, in a general setting, and the corresponding change of mirror Chern number $\Delta \mchd$ of each case, based on $\bk \cdot \bb{p}$ theory.
For a topological phase transition to occur, a gap closing and reopening between the valence and conduction bands is necessary.
If such process is accompanied by a change in Chern number, we say that a band inversion has occurred.
Given that a band inversion at some $\bk$ point results in $\Delta \mchd = \pm 1$, a similar process must occur at symmetry related points, i.e. the star of $\bk$.
Depending on $\Delta \mchd$ at the star of $\bk$, the total change in mirror Chern number can be $0$ if they cancel out, $\pm n$ if they add up, assuming that the star is of multiplicity $n$.
For the rest of this section, we give a detailed analysis of the topological phase transition of MSG 127.392 $P4'/m'b'm$.
The results are summarized in \sfig~\ref{sfig:phase}.
%%%%%%%%%%%%%%%%%%%%%%
\\

%%%%%%%%%%%%%%%%%%%%%%
%%%%%%%%%%%%%%%%%%%%%%

%%%%%%%%%%%%%%%%%%%%%%
\tocless{\subsection{$\bk \cdot \bb{p}$ Hamiltonian construction}
\label{appsub:kptheory}}{}
%%%%%%%%%%%%%%%%%%%%%%
In this subsection, we give a brief review on the theory of constructing $\bk \cdot \bb{p}$ Hamiltonians, which can describe band inversion.
Let $\mc{G}$ be the space group that describes a system, where antiunitary symmetries are allowed.
For $\sg, \sg' \in \mc{G}$, where $\sg$ and $\sg'$ are unitary and antiunitary symmetries respectively, the Hamiltonian $H$ must be invariant under $\mc{G}$ such that $H = \sg H \sg^{-1}$ and $H = \sg' H \sg'^{-1}$.

By introducing a basis, $\ketr{\vph_\alpha (\bk)}$ ($\alpha=1,\dots, N_{\rm tot}$), where $N_{\rm tot}$ is the total number of bands, we can express the Hamiltonian as a $N_{\rm tot} \times N_{\rm tot}$ matrix, with $\mc{H} (\bk)_{\alpha\beta} = \brar{\vph_\alpha (\bk)} H \ketr{\vph_
\beta (\bk)}$.
We denote the action of symmetries as
\ba
\sg \ketr{\vph_\alpha (\bk)}
= \ketr{\vph_\beta (\sg \bk)} \mc{U}_{\sg} (\bk)_{\beta\alpha}, \quad
\sg' \ketr{\vph_\alpha (\bk)}
= \ketr{\vph_\beta (\sg' \bk)} \mc{U}_{\sg'} (\bk)_{\beta\alpha} \mc{K},
\label{eqapp:psiU}
\ea
where the symmetry constraints are embedded as
\ba
\mc{H} (\sg \bk)
= \mc{U}_{\sg} (\bk) \mc{H} (\bk) \mc{U}_{\sg} (\bk)^{-1}, \quad
\mc{H} (\sg' \bk)
= \mc{U}_{\sg'} (\bk) \cm{\mc{H} (\bk)} \mc{U}_{\sg'} (\bk)^{-1}.
\label{eqapp:projRep}
\ea
The set of $\mc{U}_{\sg} (\bk)$ and $\mc{U}_{\sg'} (\bk)$ is equivalent to a projective corepresentation of $\mc{G}$.
For a full account on the theory of corepresentations, see Ref.~\cite{bradley2010}.

A $\bk \cdot \bb{p}$ Hamiltonian $\mc{H}_{\bk^*} (\delta \bk)$ is an $N_{\rm eff} \times N_{\rm eff}$ matrix, which is a low-energy polynomial expansion of $\mc{H}(\bk)$ carried out in a suitable subspace $\ketr{\phi_A (\bk)}$ ($A=1,\dots,N_{\rm eff}$).
Here, we have defined $\bk = \bk^* + \delta \bk$, where $\bk^*$ is a (high symmetry) momentum around which the Hamiltonian is expanded.
The form of the $\bk \cdot \bb{p}$ Hamiltonian is both highly restricted, and fully determined, by the little group $\mc{G}_{\bk^*}$, which is the subgroup of $\mc{G}$ that leaves $\bk^*$ invariant up to reciprocal lattice translation.
In the effective subspace, the symmetries of the little group $\sg, \sg' \in \mc{G}_{\bk^*}$ act on the basis as
\ba
\sg \ketr{\phi_A (\bk)}
= \ketr{\phi_B (\sg \bk)} \rho_{r[\bk^*]} (\sg)_{BA}, \quad
\sg' \ketr{\phi_A (\bk)}
= \ketr{\phi_B (\sg' \bk)} \rho_{r[\bk^*]} (\sg')_{BA} \mc{K}.
\label{eqapp:phiRho}
\ea
Here, $\rho_{r[\bk^*]} (\sg)$ and $\rho_{r[\bk^*]} (\sg')$ are matrix representatives of the unitary and antiunitary symmetries of $\mc{G}_{\bk^*}$ for the representation $r[\bk^*]$.
By representation, we mean that the set of matrix representatives is unitarily equivalent, i.e. $U^{-1} \rho_{r[\bk^*]} (\sg) U$ and $U^{-1} \rho_{r[\bk^*]} (\sg') \cm{U}$ for a unitary matrix $U$, to a corepresentation of the little group $\mc{G}_{\bk^*}$.
The exhaustive list of irreducible corepresentations (irreps) at any high symmetry momentum can be found on the Bilbao Crystallographic Server~\cite{elcoro2021magnetic,xu2020high,bilbao_corep}.
Also, the basis choice of such irrep can be inferred by looking at any diagonal matrix representative.

The dimension of an irrep is the symmetry-protected degeneracy at a given $\bk$ point.
Since we are interested in band inversion, in which valence and conduction bands become degenerate, it is necessary to consider a model that can describe higher-fold degeneracies. 
Thus, a minimal model should be constructed from reducible representations of $\mc{G}_{\bk^*}$.
By definition, representative matrices of any reducible representation can be unitarily block diagonalized by a single matrix, to a direct sum of irreps~\cite{bradley2010}.
Therefore, every reducible representation, up to unitary equivalence, can be readily obtained by taking a direct sum of irreducible corepresentations listed in the Bilbao Crystallographic Server~\cite{bilbao_corep}.

Once the representative matrices have been defined, it is straightforward to construct a $\bk \cdot \bb{p}$ Hamiltonian by taking a matrix of polynomials, and imposing the symmetries of $\mc{G}_{\bk^*}$.
We define the $\bk \cdot \bb{p}$ Hamiltonian as a matrix of polynomials in $\delta \bk$, which obeys
\ba
\mc{H}_{\bk^*} (\sg \delta \bk)
= \rho_{r[\bk^*]} (\sg) \mc{H}_{\bk^*} (\delta \bk) \rho_{r[\bk^*]} (\sg)^{-1}, \quad
\mc{H}_{\bk^*} (\sg' \delta \bk)
= \rho_{r[\bk^*]} (\sg') \cm{\mc{H}_{\bk^*} (\delta \bk)} \rho_{r[\bk^*]} (\sg')^{-1}.
\label{eqapp:kpDef}
\ea
To describe the system in a different basis, one can simply apply a unitary transformation.
%
%%%%%%%%%%%%%%%%%%%%%%

%%%%%%%%%%%%%%%%%%%%%%
\tocless{\subsection{Gap closing points on the mirror-invariant plane $k_x = k_y$}
\label{appsub:DyB4gap}}{}
%%%%%%%%%%%%%%%%%%%%%%
A topological phase transition with $\Delta \mchd \neq 0$ necessarily involves gap closing between mirror eigenstates on the mirror-invariant plane, the $k_x=k_y$ plane.
In this subsection, we address the gap-closing points on the mirror-invariant plane by a so-called ``codimension analysis"~\cite{murakami2008}.
As an example, consider a two-band Hamiltonian in $d$ dimension, $H(m, \bk) = E(m, \bk) \sg_0 + \sum_{i=1}^{3} f_i (m, \bk) \sg_i$, where the $f_i$'s are independent functions, and $\bk = (k_1, ..., k_d)$.
Here, $m$ is an externally controlled variable parameter.
The number of functions that have to vanish in order to achieve a gapless phase is called the codimension $d_c$, which is $d_c=3$ ($f_1$, $f_2$, $f_3$), in this case.
On the other hand, the number of variable parameters $d_p$, is $d_p=1+d$.
If $d_p$ is smaller than the codimension $d_c$ ($d \leq 1$), gap closing cannot appear; if they are equal ($d=2$), a gap-closing point can occur, which is the case for band inversion.
Finally, if $d_p$ is larger than $d_c$, gap-closing points can form nodal lines or higher-dimensional manifolds in the momentum space~\cite{murakami2008}.
We present the results of codimension analysis using $\bk \cdot \bb{p}$ theory, conducted on MSG 127.392 $P4'/m'b'm$.
To this end, we first review the important symmetries of this system.

MSG 127.392, which is the symmetry group of DyB$_4$, is a Type-III space group with a primitive tetragonal unit cell.
The generators of this space group are $TC_{4z} = T\{c_{4z} | \bb{0}\}$, an antiunitary 4-fold rotation, $\mxybo = \{m_{x \bar{y}} | \hf, -\hf, 0\}$, an off-centered diagonal mirror whose action is $\mxybo (x, y, z) = (y + \hf, x - \hf, z)$, and $\cyo = \{c_{2y} | \hf, \hf, 0\}$, a 2-fold screw axis with $\cyo (x, y, z) = (-x + \hf, y + \hf, -z)$.
Note that the projective representations of the generators do not have $k_z$ dependence.
Thus, if the little group of two high symmetry points $\bk^*_1$ and $\bk^*_2$ are isomorphic, with $\bk^*_1 - \bk^*_2 = (0, 0, \Delta k_z)$, the $\bk \cdot \bb{p}$ Hamiltonians will have an identical form.

In addition to these symmetries, it is worth noting that the system has $TI = T \{i | \bb{0}\}$ symmetry, which commutes with $\mxybo$ on the mirror-invariant plane.
This implies that every band is doubly degenerate, and that Kramers pairs are formed with opposite mirror eigenvalues.
Due to Kramers' degeneracy, the matrix dimension of a minimal model describing band inversion is $4 \times 4$; at $A (\pi, \pi, \pi)$ and $M (\pi, \pi, 0)$, $8 \times 8$ due to the 4-fold Dirac fermion.

Now, we proceed to the codimension analysis, whose results are summarized in~\sfig~\ref{sfig:phase}.
Since the representation of $\mc{G}_{\bk^*}$ determine the $\bk \cdot \bb{p}$ Hamiltonian, we first classify high-symmetry momenta on the mirror-invariant plane by little groups.
Firstly, $\Gamma = (0, 0, 0)$, $Z = (0, 0, \pi)$, $A = (\pi, \pi, \pi)$, $M = (\pi, \pi, 0)$ have little groups isomorphic to the full space group.
Next, the little groups of $\Lambda_D = (0, 0, k_z)$ and $V = (\pi, \pi, k_z)$, with $(k_z \neq 0, \pi)$, are isomorphic to each other.
Finally, $C = (k_y, k_y, k_z)$, $\Sigma_M = (k_y, k_y, 0)$, and $\Sigma = (k_y, k_y, \pi)$, with $(k_y, k_z \neq 0, \pi)$, have identical little groups.
%%%%%%%%%%%%%%%%%%%%%%
\\

%%%%%%%%%%%%%%%%%%%%%%
\tocless{\subsubsection{$\Gamma$, $Z$, $A$, and $M$}
\label{appssub:hspoints}}{}
%%%%%%%%%%%%%%%%%%%%%%
The little group generators of high symmetry points $\Gamma$, $Z$, $A$, and $M$ are $TC_{4z}$, $\mxybo$, and $\cyo$.
The projective representation of these generators do not have $\bk$ dependence on the mirror-invariant plane.
However, the projective phase of $\cyo$, $e^{-i k_y}$, leads to a 4-fold degeneracy at $M (A)$, contrary to the 2-fold degeneracy at $V (\Gamma)$.
Therefore, the $\bk \cdot \bb{p}$ Hamiltonians can be further classified into those for $(\Gamma, Z)$ and $(A, M)$.

We first focus on $Z (\Gamma)$, which is also the gap-closing point in DyB$_4$ at $U$=6.338 eV (see \sfig~\ref{sfig:gap_closing}).
At $Z$, there are two irreducible corepresentations, $\bar{Z}_6$ and $\bar{Z}_7$.
A minimal model constructed with $Z_0 \equiv \bar{Z}_6 \oplus \bar{Z}_7$ yields $\mc{H}_{Z} (m) = E(m) \Gamma_{00} + f_1 (m) \Gamma_{30}$, with $\Gamma_{ij} \equiv \mu_i \otimes \tau_j$, where $\mu$ and $\tau$ are the Pauli matrices.
In this model, only $m$ is tunable, thus the number of tunable parameters is 1, i.e. $d_p=1$.
Also, the codimension is one ($d_c=1$) because a gap-closing condition requires $f_1(m)=0$.
Since $d_p$ and the codimension $d_c$ are equal to each other, gap closing can occur at $Z$ and $\Gamma$.
On the other hand, $\bk \cdot \bb{p}$ Hamiltonians constructed from $\bar{Z}_6 \oplus \bar{Z}_6$ or $\bar{Z}_7 \oplus \bar{Z}_7$ have codimension $d_c=2$, which is larger than $d_p=1$, so these representations cannot describe gap closing.

At $M$ and $A$, the matrix dimension of a minimal model that can describe gap closing is $8 \times 8$.
The $\bk \cdot \bb{p}$ Hamiltonian at $M (A)$ turns out to be $\mc{H}_{M} (m) = E(m) \Gamma_{000} + f_1 (m) \Gamma_{100} + f_2 (m) \Gamma_{230} + f_3 (m) \Gamma_{300}$, where $\Gamma_{abc} = \mu_a \otimes \tau_b \otimes \sg_c$.
Thus, $d_p=1$ and $d_c=3$.
Since the codimension $d_c$ is larger than $d_p$, gap closing cannot occur at $M (A)$.
%%%%%%%%%%%%%%%%%%%%%%
\\

%%%%%%%%%%%%%%%%%%%%%%
\tocless{\subsubsection{$\Lambda_D$ and $V$}
\label{appssub:hsline}}{}
%%%%%%%%%%%%%%%%%%%%%%
The little group of $\Lambda_D$ and $V$ is generated by $\mxybo$, $C_{2z} = \{c_{2z} | \bb{0}\}$, and $TI$.
Since the projective phase of these operators are identical at $\Lambda_D$ and $V$, the $\bk \cdot \bb{p}$ Hamiltonians have the same form at $\Lambda_D$ and $V$.
The minimal model obtained from $\Lambda_0 \equiv \Lambda_{D_5} \oplus \Lambda_{D_5}$ gives $\mc{H}_{\Lambda} (m, k_z) = E(m, k_z) \Gamma_{00} + f_1 (m, k_z) \Gamma_{10} + f_2 (m, k_z) \Gamma_{30}$, with $d_c=d_p=2$.
Since $d_c=d_p$, gap closing can occur at $\Lambda_D$ and $V$.
%%%%%%%%%%%%%%%%%%%%%%
\\

%%%%%%%%%%%%%%%%%%%%%%
\tocless{\subsubsection{$C$, $\Sigma_M$, and $\Sigma$}
\label{appssub:hsplane}}{}
%%%%%%%%%%%%%%%%%%%%%%
The little group of these high symmetry momenta is generated by $\mxybo$ and $TI$ at $C$, $\Sigma_M$, and $\Sigma$.
$\bk \cdot \bb{p}$ Hamiltonians of these momenta are unitarily equivalent, following the same reasoning presented in \sn~\ref{appssub:hsline}.
Constructing a $\bk \cdot \bb{p}$ Hamiltonian with $C_0 \equiv \bar{C}_3 \bar{C}_4 \oplus \bar{C}_3 \bar{C}_4$ gives $\mc{H}_C (m, \delta k_y, \delta k_z) = E(m, \delta k_y, \delta k_z) \Gamma_{00} + f_1 (m, \delta k_y, \delta k_z) \Gamma_{10} + f_2 (m, \delta k_y, \delta k_z) \Gamma_{23} + f_3 (m, \delta k_y, \delta k_z) \Gamma_{30} $, of which the codimension $d_c$ is 3.
Therefore, gap closing can occur at $C$, where $d_p=3$.
However, at $\Sigma$ and $\Sigma_M$, $d_p$ is reduced to 2 because $\delta k_z$ is fixed to 0; thus, band inversion is impossible.
%%%%%%%%%%%%%%%%%%%%%%
\\

%%%%%%%%%%%%%%%%%%%%%%
\tocless{\subsection{$\Delta \mchd$ for each type of band inversion}
\label{appsub:deltaCm}}{}
%%%%%%%%%%%%%%%%%%%%%%
For each type of band inversion, one can analytically calculate $\Delta \mchd$ using a linear-order $\bk \cdot \bb{p}$ Hamiltonian.
At $Z$, generators of the $Z_0$ representation defined in \sn~\ref{appssub:hspoints} can be chosen as~\cite{bilbao_corep}
\ba
\rho_{Z_0} (\mxybo)=
\bpm
0 & i & 0 & 0 \\
i & 0 & 0 & 0 \\
0 & 0 & 0 & i \\
0 & 0 & i & 0
\epm , \quad
\rho_{Z_0} (\cyo)=
\bpm
0 & \omega & 0 & 0 \\
\omega^3 & 0 & 0 & 0 \\
0 & 0 & 0 & \cm{\omega}^3 \\
0 & 0 & \cm{\omega} & 0
\epm, \quad
\rho_{Z_0} (TC_{4z})=
\bpm
0 & \cm{\omega}^3 & 0 & 0 \\
\cm{\omega} & 0 & 0 & 0 \\
0 & 0 & 0 & \omega \\
0 & 0 & \omega^3 & 0
\epm,
\label{eqapp:Zrep}
\ea
where $\omega=e^{-i \pi/4}$.
The first order $\bk \cdot \bb p$ Hamiltonian can be expressed as
\ba
\mc{H}_{Z} (\delta k_y, \delta k_z)
= t_1 \delta k_y \Gamma_{21} + t_2 \delta k_z \Gamma_{10} + m \Gamma_{30}.
\label{eqapp:hZfull}
\ea
Since $\mchd$ is the Chern number of the $+i$ mirror sector, we project the Hamiltonian into $+i$ eigenstates of $\rho_{Z_0} (\mxybo)$ in \eq{eqapp:Zrep}, $(0, 0, 1/\sqrt{2}, 1/\sqrt{2})^T$ and $(1/\sqrt{2}, 1/\sqrt{2}, 0, 0)^T$ to get
\ba
\mc{H}_{\rm proj} (\delta k_y, \delta k_z) = -t_1 \delta k_y \sg_2 + t_2 \delta k_z \sg_1 - m \sg_3.
\label{eqapp:hZproj}
\ea
Upon direct calculation, we get $\mc{C} = {\rm sign} (t_1 t_2 m)/2$.
Furthermore, no symmetry relates $Z$ to another point in the first BZ.
Thus, at $Z$, a topological phase transition with $\Delta \mchd=\pm 1$ can occur by tuning a single parameter $m$.
(Note that the same analysis can be applied for a band inversion at $\Gamma$.)
This is consistent with the topological phase transition of DyB$_4$, where a band inversion at $Z$ results in $\Delta \mchd = -1$.

Similar calculations show that topological phase transition can occur at $\Lambda_D$, $V$, and $C$.
A band inversion at $\Lambda_D$ or $V$ is accompanied with another at the $\cyo$-related momentum.
In this case, we find that $\Delta \mchd$ can be either $0$ or $\pm 2$ depending on the Dirac Hamiltonian.

Finally, a band inversion at $C$ simultaneously occurs at the $C_{2z}$ related point.
In this case, $\Delta \mchd$ can be calculated as follows.
First, $C_{2z}$ exchanges the $\pm i$ eigensectors due to $\{\mxybo, C_{2z}\}=0$ on the $k_x=k_y$ plane.
The Berry curvature at $C$ of the $+i$ eigensector and one at the $C_{2z}$-related point ($C_{2z}C$) of the $-i$ eigensector have the same magnitude but opposite signs.
(This can be inferred from the Berry curvature transformation formula introduced in \sn~\ref{appsub:mC_pcmm}.)
This implies that a band inversion at $C$, which gives rise to $\Delta \mchd$, is accompanied by another band inversion at $C_{2z}C$ with the same amount of change in mirror Chern numbers $\Delta \mchd$.
Hence, $\Delta \mchd = \pm 2$ in total.
%%%%%%%%%%%%%%%%%%%%%%
\\

%%%%%%%%%%%%%%%%%%%%%%
\section{Massive Dirac fermion and related topological phase transition via mirror breaking}
\label{app:mirrorbreaking}
%%%%%%%%%%%%%%%%%%%%%%
In this section, we study the topological phase transition induced by a mirror-breaking perturbation, which is mediated by the fourfold-degenerate Dirac fermion.
The fourfold degeneracy is protected by three MWGs, $p'_cmm$, $p'_c4mm$, and $p4'g'm$.
Therefore, any perturbation that reduces these MWGs to subgroups with lower symmetry can lift the fourfold degeneracy regardless of whether a Dirac fermion is a nodal structure in a purely 2D system or a surface state of a 3D TMDI.
(Note an exceptional case.
Since $p'_cmm$ is a subgroup of $p'_c4mm$, fourfold degeneracy is intact when $p'_c4mm$ is reduced to $p'_cmm$.)
However, there is a crucial difference between them.
In a 2D system, a symmetry-breaking perturbation not only lifts the fourfold degeneracy, but also gaps the system.
On the other hand, such perturbation must break diagonal mirrors in order to gap a surface band structure of a 3D TMDI.
Otherwise, the surface is still metallic because of nonzero mirror Chern numbers ($\mchd$, $\mchdp$) and mirror-protected surface states.

Below, we study the topology of gapped phases induced by a mirror-breaking perturbation, in order to describe a situation where the surface Dirac fermion of a TMDI becomes gapped.
The gapped phases and the topological phase transition which relates them can be systematically studied by constructing a $\bk \cdot \bb{p}$ Hamiltonian which describes such process.
%%%%%%%%%%%%%%%%%%%%%%

%%%%%%%%%%%%%%%%%%%%%%
\tocless{\subsection{Mirror breaking in MWG $p4'g'm$}}
%%%%%%%%%%%%%%%%%%%%%%
The MWG $p4'g'm$ is generated by $TC_{4z}=T\{c_{4z}|\bb 0\}$ and $TG_y=T\{m_y|\hf,\hf\}$.
There are two off-centered diagonal mirrors $\mxybo=\{m_{x\cm{y}}|\hf,-\hf\}$ and  $\mxyb=\{m_{xy}|\hf,\hf\}$.
Note that $\mxybo=TG_y (TC_{4z})^{-1}$ and $\mxybo \mxyo = \{E|\bb a_1\} (TC_{4z})^2$.
A fourfold degeneracy is protected at $M=(\pi,\pi)$ whose magnetic little group is $4'm'm$.
Following the $\bk \cdot \bb{p}$ method introduced in \sn~\ref{appsub:kptheory}, we now construct a $\bk \cdot \bb{p}$ Hamiltonian using the four-dimensional irreducible representation (irrep), in order to describe the fourfold degeneracy.
The matrix representatives of the four-dimensional irrep is given by~\cite{bilbao_corep}
\bg
\rho(C_{2z})=\bpm 0 & 1 & 0 & 0 \\ -1 & 0 & 0 & 0 \\ 0 & 0 & 0 & -1 \\ 0 & 0 & 1 & 0 \epm=i\tau_3 \sg_2, \quad
\rho(\mxybo)=\bpm 0 & i & 0 & 0 \\ i & 0 & 0 & 0 \\ 0 & 0 & -i & 0 \\ 0 & 0 & 0 & i \epm, \quad
\rho(TC_{4z})=\bpm 0 & 0 & 0 & i \\ 0 & 0 & i & 0 \\ i & 0 & 0 & 0 \\ 0 & -i & 0 & 0 \epm,
\eg
and $\rho(TG_y)=\rho(\mxybo) \rho(TC_{4z})$.
[Matrix representatives of other symmetry elements which are not written explicitly, such as $\mxyo$, $(TC_{4z})^{-1}$, and $TG_x$, can be obtained from group multiplications.]
According to \eq{eqapp:projRep}, the $\bk \cdot \bb{p}$ Hamiltonian at the $M$ point can be written as
\bg
\mc{H}_{p4'g'm}(\bk) = k_x(A_1 \Gamma_{10} + A_2 \Gamma_{20} + A_3 \Gamma_{31} + A_4 \Gamma_{33}) + k_y(A_3 \Gamma_{01} - A_4 \Gamma_{03} + A_2 \Gamma_{12} - A_1 \Gamma_{22}),
\eg
where $\Gamma_{ab}=\tau_a \otimes \sg_b$, $\bk$ is a small momentum expanded at $M$, and $A_{1,\dots,4}$ are real parameters.

Now, we introduce mirror-breaking terms, which break both $\mxybo$ and $\mxyo$.
Such terms reduce $p4'g'm$ to index-2 subgroups $p2g'g'$ and $p4'$.
This is because $\mxybo =TG_y (TC_{4z})^{-1}$ implies that $\mxybo$ must be broken with either $TC_{4z}$ or $TG_y$ (A similar constraint holds for $\mxyo$ breaking).
The former and latter cases correspond to $p2g'g'$ and $p4'$, respectively.
(Note that one can consider index-4 subgroups by breaking symmetries further.
For our purpose, we limit our discussion to index-2 subgroups.)

First, let us discuss the case where $p4'g'm$ is reduced to $p2g'g'$.
To obtain $p2g'g'$, we break $\mxybo$, $\mxyo$, $TC_{4z}$ and $(TC_{4z})^{-1}$, but preserve $C_{2z}$ and $TG_{x,y}$.
The resulting $\bk \cdot \bb{p}$ Hamiltonian is expressed as
\bg
\mc{H}_{p2g'g'}(\bk) = k_x(a_1 \Gamma_{10}+a_2 \Gamma_{20}+a_3 \Gamma_{31}+a_4 \Gamma_{33}) + k_y(a_5 \Gamma_{01}+a_6 \Gamma_{03}+a_7 \Gamma_{12}+a_8 \Gamma_{22}) + m \Gamma_{32}.
\label{eq:p2g'g'_kp}
\eg
Note that a single mass term is allowed, and it anticommutes with the kinetic terms.
Hence, $\mc{H}_{p2g'g'}(\bk)$ with $m>0$ and $m<0$ describe different topological phases.
This can be seen by choosing $a_3=a_6=1$ and 0 for other parameters in \eq{eq:p2g'g'_kp}.
We can do so without loss of generality, since one can adiabatically change the parameters without closing the gap for a sufficiently large $m$.
Then, \eq{eq:p2g'g'_kp} becomes $\mc{H}_{p2g'g'}(\bk)=k_x \tau_z \sg_x + k_y \sg_z + m \tau_z \sg_y$, which can be further block-diagonalized into eigensectors of $\tau_z$.
Each $\tau_z$ subspace contributes to the change of Chern number $\Delta \mc{C}=1$, when the mass term changes sign from $m<0$ to $m>0$ during the phase transition, resulting in $\Delta \mc{C}=2$ in total.
In other words, the fourfold-degenerate Dirac fermion in MWG $p4'g'm$ can be understood as a critical point between topological phases with Chern number difference $\Delta \mc{C}=2$ in $p2g'g'$-symmetric systems.

Now, we consider the mirror-breaking perturbation which reduces $p4'g'm$ to $p4'$. In this case, every symmetry except $TC_{4z}$ is broken.
The relevant $\bk \cdot \bb{p}$ Hamiltonian is given by
\ba
\mc{H}_{p4'}(\bk)
=& k_x(b_1 \Gamma_{01}+b_2 \Gamma_{03}+b_3 \Gamma_{10}+b_4 \Gamma_{12}+c_1 \Gamma_{31}-c_2 \Gamma_{33}+c_3 \Gamma_{22}+c_4 \Gamma_{20}) \nn
+& k_y(c_1 \Gamma_{01} + c_2 \Gamma_{03} + c_3 \Gamma_{10} + c_4 \Gamma_{12} - b_1 \Gamma_{31} + b_2 \Gamma_{33} - b_3 \Gamma_{22} - b_4 \Gamma_{20}) \nn
+& m_1 \Gamma_{02} + m_2 (\Gamma_{11}+\Gamma_{13}) + m_3 (\Gamma_{21}+\Gamma_{23}).
\ea
This Hamiltonian includes three anticommuting mass terms and therefore all the gapped phases are adiabatically connected.
Hence, it does not describe a topological phase transition.
%%%%%%%%%%%%%%%%%%%%%%

%%%%%%%%%%%%%%%%%%%%%%
\tocless{\subsection{Mirror breaking in MWG $p'_c4mm$}}
%%%%%%%%%%%%%%%%%%%%%%
Similar to the case of MWG $p4'g'm$, MWG $p'_c4mm$ protects a fourfold degeneracy at $M$.
The relevant matrix representatives of the four-dimensional irrep is given by~\cite{bilbao_corep}
\bg
\rho(\mxyb)=\bpm 0 & i & 0 & 0 \\ i & 0 & 0 & 0 \\ 0 & 0 & 0 & -i \\ 0 & 0 & -i & 0 \epm, \quad
\rho(M_x)=\bpm 0 & \om^{-1} & 0 & 0 \\ \om^{-3} & 0 & 0 & 0 \\ 0 & 0 & 0 & \om^{-3} \\ 0 & 0 & \om^{-1} & 0 \epm, \quad
\rho(T_G)=\bpm 0 & 0 & -1 & 0 \\ 0 & 0 & 0 & -1 \\ 1 & 0 & 0 & 0 \\ 0 & 1 & 0 & 0 \epm,
\eg
and $\rho(C_{4z})={\rm Diag}( \om^{-3}, \om^3, \om^{-1}, \om)$ where $\omega=e^{-i \pi/4}$.
Here, $\mxyb=\{M_{x\cm{y}}|\bb 0\}$, $M_x = \{m_x|\bb 0\}$, $T_G = T\{E|\hf,\hf\}$, and $C_{4z}=\{c_{4z}|\bb 0\}$.
Under such basis, the $\bk \cdot \bb{p}$ Hamiltonian at $M$ can be expressed as
\bg
\mc{H}_{p'_c4mm}(\bk) = (k_x+k_y)(A_1 \Gamma_{13} + A_2 \Gamma_{23} + A_3 \Gamma_{31}) + (k_x-k_y)(A_1 \Gamma_{23} - A_2 \Gamma_{20} - A_3 \Gamma_{02}).
\eg

Now we consider a perturbation that breaks diagonal mirror symmetries, $\mxyb$ and $\mxy=C_{2z}\mxyb$.
The relations $\mxyb = M_x C_{4z} = \cm{E} (T_G M_x)(T_G C_{4z})$ imply that there are two index-2 subgroups, $p'_cmm$ and $p4'mm'$.
We first consider the case where the symmetry is reduced to $p'_cmm$.
For this, we break $C_{4z}$.
In this case, the mirror-breaking perturbation breaks $C_{4z}$, $\mxyb$, and $\mxy$.
Since the MWG $p'_cmm$ is one of the three MWGs protecting the fourfold-degenerate Dirac fermion, the fourfold should be protected under such perturbation.
Indeed, the relevant $\bk \cdot \bb{p}$ Hamiltonian is given by
\ba
\mc{H}_{p'_cmm}(\bk)
=& k_x [ a_1(\Gamma_{02} - \Gamma_{31}) + a_2 (\Gamma_{10} - \Gamma_{23}) + a_3 (\Gamma_{13} + \Gamma_{20}) ] \nn
+& k_y [ b_1(\Gamma_{02} + \Gamma_{31}) + b_2 (\Gamma_{10} + \Gamma_{23}) + b_3 (\Gamma_{13}-\Gamma_{20})],
\ea
which has no mass term; thus, the fourfold degeneracy is protected.

As for the second case, we consider MWG $p4'mm'$ by breaking $T_G$, $T_G\mxy$, $T_G\mxyb$, $T_GC_{2z}$, and two diagonal mirrors $\mxy$ and $\mxyb$ as well.
Then, $T_GC_{4z}$, $C_{2z}$, $M_{x,y}$, and $T_G M_{xy,x\cm{y}}$ remain.
The relevant $\bk \cdot \bb{p}$ Hamiltonian has two anticommuting mass terms:
\bg
\mc{H}_{p4'mm'}(\bk) = (k_x+k_y) (c_1 \Gamma_{13} + c_2 \Gamma_{23} + c_3 \Gamma_{31} + c_4 \Gamma_{32}) \nn
+ (k_x-k_y) (b_1 \Gamma_{20} - b_2 \Gamma_{10} - b_3 \Gamma_{02} - b_4 \Gamma_{01}) \nn
+ m_1 \Gamma_{11} + m_2 \Gamma_{21}.
\eg
Hence, it does not describe topological phase transition.
%%%%%%%%%%%%%%%%%%%%%%

\end{document}